\documentclass[pdftex,twocolumn,epjc3]{svjour3}          
\journalname{Eur. Phys. J. C}
\usepackage{lipsum}  
\usepackage{cite}  
\DeclareMathAlphabet{\pazocal}{OMS}{zplm}{m}{n}
\newcommand{\Q}{\pazocal{Q}}

\usepackage{float}
\usepackage[utf8]{inputenc}
\usepackage{color}
\usepackage{comment}
\usepackage{lineno}
\usepackage{graphicx}
\usepackage{float} 
\usepackage[caption=false]{subfig} 
\usepackage{eqnarray,amsmath}

\usepackage{scalerel}
\usepackage{tikz}
\usetikzlibrary{svg.path}

\definecolor{orcidlogocol}{HTML}{A6CE39}
\tikzset{
  orcidlogo/.pic={
    \fill[orcidlogocol] svg{M256,128c0,70.7-57.3,128-128,128C57.3,256,0,198.7,0,128C0,57.3,57.3,0,128,0C198.7,0,256,57.3,256,128z};
    \fill[white] svg{M86.3,186.2H70.9V79.1h15.4v48.4V186.2z}
                 svg{M108.9,79.1h41.6c39.6,0,57,28.3,57,53.6c0,27.5-21.5,53.6-56.8,53.6h-41.8V79.1z M124.3,172.4h24.5c34.9,0,42.9-26.5,42.9-39.7c0-21.5-13.7-39.7-43.7-39.7h-23.7V172.4z}
                 svg{M88.7,56.8c0,5.5-4.5,10.1-10.1,10.1c-5.6,0-10.1-4.6-10.1-10.1c0-5.6,4.5-10.1,10.1-10.1C84.2,46.7,88.7,51.3,88.7,56.8z};
  }
}

\newcommand\orcidicon[1]{\href{https://orcid.org/#1}{\mbox{\scalerel*{
\begin{tikzpicture}[yscale=-1,transform shape]
\pic{orcidlogo};
\end{tikzpicture}
}{|}}}}

\def\ie{{\it i.e.}}
\def\eg{{\it e.g.}}
\def\etal{{\it et al.}}

\def\alphaS{\alpha_s}

\def\oSzo{{\bigl.^1\!S^{[1]}_0}}

\newcommand{\Br}{{\cal B}}

\usepackage{mathtools}
\usepackage{graphicx}
\usepackage{amsfonts}
\usepackage{slashed}

\newcommand{\overbar}[1]{\mkern 1.5mu\overline{\mkern-1.5mu#1\mkern-1.5mu}\mkern 1.5mu}

\usepackage[toc,page]{appendix}
\usepackage{pdfpages}

\usepackage{makecell}

\newcommand{\eqs}[1]{\begin{equation} \begin{split} #1\end{split} \end{equation} }
\def\be{\begin{equation*}}
\def\ee{\end{equation*}}
\def\bsp#1\esp{\begin{split}#1\end{split}} 
\def\bpm{\begin{pmatrix}}
\def\epm{\end{pmatrix}}
\newcommand{\ce}[1]{Eq.~(\ref{#1})}
\newcommand{\cf}[1]{{Fig.~\ref{#1}}}
\newcommand{\ct}[1]{{Table~\ref{#1}}}
\newcommand{\msbar} {\overline{\text{MS}}}

\usepackage{txfonts}

 \usepackage{ifpdf}
 \ifpdf
 \usepackage[pdftex]{hyperref}
 \else
 \usepackage[hypertex]{hyperref}
 \fi

 \hypersetup{
   pdftitle={Curing the unphysical behaviour of NLO quarkonium production at the LHC and its relevance to constrain
gluon PDF at low scales},%
   pdfauthor={},%
   pdfsubject={},%
   pdfkeywords={},%
   pdfstartview={},%
   bookmarksopen=true, breaklinks=true, debug=true, %
   colorlinks=true, linkcolor=red, citecolor=blue, urlcolor=blue
 }

\begin{document} 

\title{Curing the unphysical behaviour of NLO quarkonium production at the LHC and its relevance to constrain the
gluon PDF at low scales}

\author{
Jean-Philippe Lansberg\thanksref{addr1}\protect\orcidicon{0000-0003-2746-5986}
\and 
Melih A. Ozcelik\thanksref{addr1}\protect\orcidicon{0000-0002-8312-7116}
} 

\institute{
Universit\'e Paris-Saclay, CNRS, IJCLab, 91405 Orsay, France \label{addr1}
\\~ \\
\email{Jean-Philippe.Lansberg@in2p3.fr \& Melih.Ozcelik@ijclab.in2p3.fr}
}

\date{Version of \today}

\maketitle
\begin{abstract}
We address the unphysical energy dependence of quarkonium-hadroproduction cross sections at Next-to-Leading Order (NLO) in $\alphaS$ which we attribute to an over-subtraction in the factorisation of the collinear singularities inside the PDFs in the $\msbar$ scheme. Such over- or under-subtractions have a limited phenomenological relevance in most of the scattering processes in particle physics. On the contrary, it is particularly harmful for $P_T$-integrated charmonium hadroproduction which renders a wide class of NLO results essentially unusable. Indeed, in such processes, $\alphaS$ is not so small, the PDFs are not evolved much and can be rather flat
for the corresponding momentum fractions and, finally, some process-dependent NLO pieces are either too small or too large. We propose a scale-fixing criterion which avoids such an over-subtraction. We demonstrate its efficiency for $\eta_{c,b}$ but also for a fictitious light elementary scalar boson. Having provided stable NLO predictions for $\eta_{c,b}$ $P_T$-integrated cross sections, $\sigma^{\rm NLO}_{\eta_Q}$, and discussed the options to study $\eta_{b}$ hadroproduction, we argue that their measurement at the LHC can help better determine the gluon PDF at low scales and tell whether the local minimum in conventional NLO gluon PDFs around $x=0.001$ at scales below 2 GeV  is physical or not.
\end{abstract}


\section{Introduction}
The production of quarkonia ($\Q$) in inclusive proton-proton and electron-proton collisions when the protons break apart
is one of most often studied process at high-energy colliders. Yet one still does not agree on how these heavy quark-antiquark bound states are produced. The interested reader will find it useful to consult the following reviews ~\cite{Kramer:2001hh,Brambilla:2004wf,Lansberg:2006dh,Brambilla:2010cs} addressing HERA and Tevatron results
and more recent ones~\cite{Andronic:2015wma,Lansberg:2019adr} as what regards the recent advances in the field
with the RHIC and LHC. 
Besides probing QCD at the interplay between its perturbative and nonpertubative regimes, quarkonium production --once theoretically understood-- should in principle allow us to probe the proton gluon content in terms of PDFs (see \eg~\cite{Halzen:1984rq,Martin:1987ww,Martin:1987vw,Jung:1992uj}) or TMDs  (see \eg~\cite{Boer:2012bt,Dunnen:2014eta,Boer:2016bfj,Lansberg:2017dzg,Lansberg:2017tlc,Lansberg:2018fwx,Bacchetta:2018ivt,DAlesio:2019qpk,Kishore:2019fzb,Scarpa:2019fol,Boer:2020bbd}).

What we have learnt in the recent years with the advent of NLO computations of $P_T$-differential cross sections, $d\sigma^{\rm NLO}/dP_T$, of $J/\psi$ and $\Upsilon$~\cite{Kramer:1995nb,Artoisenet:2008fc,Lansberg:2008gk,Lansberg:2009db,Gong:2012ah,Lansberg:2013qka,Lansberg:2014swa} is that
the inclusion of NLO corrections in any data-theory comparison is absolutely mandatory to extract qualitatively reliable statements. 
This is particularly true in two of the three most used approaches, the Colour-Singlet Model (CSM)~\cite{Chang:1979nn,Berger:1980ni,Baier:1983va} and  Non-Relativistic QCD (NRQCD)~\cite{Bodwin:1994jh}, where $P_T$-enhanced NLO contributions notably affect observables like $d\sigma/dP_T$ and the yield polarisation as a function of $P_T$. In fact, for $S$-wave quarkonia, the CSM  is the leading NRQCD contribution in the heavy-quark velocity, $v$. For the Colour-Evaporation Model~\cite{Fritzsch:1977ay,Halzen:1977rs}, the impact of NLO corrections~\cite{Lansberg:2016rcx,Lansberg:2020rft} to $d\sigma/dP_T$ is limited. In the latter model, all the spin and colour contributions of the heavy-quark pair are summed over and
the possible additional gluon radiations at NLO do not open new production channels at variance with the CSM and NRQCD.

When one integrates over $P_T$, these NLO channels, which are $P_T$-enhanced in the CSM and NRQCD and which are precisely responsible for the large impact of the NLO corrections at mid and large $P_T$, are just suppressed by one power of $\alphaS$ without any $P_T$-enhancement factor. In this context, in 2015, we studied~\cite{Feng:2015cba} the energy dependence of the $J/\psi$ and $\Upsilon$ $P_T$-integrated cross section at NLO, $\sigma^{\rm NLO}$,  in NRQCD to verify the coherence with the NRQCD predictions for $d\sigma^{\rm NLO}/dP_T$. 

Beside the confirmation of a possible breakdown of NRQCD universality\footnote{The LDME values obtained by fitting $d\sigma/dP_T$ are ten times larger than those fit from $\sigma$.}--as first claimed~\cite{Maltoni:2006yp} by F. Maltoni \etal\ based on a partial NLO study--, we found out that, for all the NRQCD contributions\footnote{Both the Colour-Singlet (CS) and Colour-Octet (CO) contributions.}, the energy dependence became unphysical once the $\alphaS$ corrections were added. The same observation was made for the $\eta_c$ which is at the centre of this study.
More precisely, the charmonium cross sections would become negative at increasing energies for a wide class of factorisation and renormalisation scales. 
In the  $J/\psi$ case, the NLO corrections significantly reduce the predicted yields close to RHIC energies\cite{Brodsky:2009cf} and $\sigma^{\rm NLO}_{J/\psi}$ already becomes negative at a couple of hundred GeV at central rapidities, $y$, for $\mu_F \geq M_\psi$~\cite{Feng:2015cba}.  Such observations were already made in the 1990's regarding the $\eta_c$ independently by Schuler~\cite{Schuler:1994hy} and then by Mangano \& Petrelli~\cite{Mangano:1996kg} but were then essentially forgotten, see \eg\ \cite{Ozcelik:2019qze}. 

For bottomonia and for some --small-- $\mu_F$ scale choices for charmonia (see~\cite{Feng:2015cba} for details), $\sigma^{\rm NLO}$ would not become negative but the $K^{\rm NLO}$ factor, defined as $\sigma^{\rm NLO}/\sigma^{\rm LO}$, would steadily deviate from unity for increasing $\sqrt{s}$. As discussed above, large $K^{\rm NLO}$ factors have already been observed in quarkonium production at  finite $P_T$ but they can then be explained by kinematical factors scaling like $P_T/m_Q$. These are absent when $P_T$ is integrated over. As we noted such an intriguing behaviour can already be observed for $\sqrt{s}$ on $200 \sim 300$ GeV~\cite{Feng:2015cba}, so not necessarily at very high energies where large logarithms of the colliding gluon momentum fraction $x$ should be accounted for. Indeed, such energies typically corresponds to $x=0.01$ and even above.

In this article, we propose a solution to this issue which we attribute to an over-subtraction in the factorisation of the collinear singularities inside the PDF in the $\msbar$ scheme. As such, it may appear in any NLO computations once a couple of unfavourable factors combine. In general, such over-subtractions indeed have a limited phenomenological relevance. It is clearly not the case for charmonium production which therefore offers a neat study case. As we will discuss, we propose a simple solution which consists in a factorisation-scale choice based on the high-energy limit of the partonic cross section and we demonstrate how well it works for $\eta_c$ and $\eta_b$ production and for the production of a fictitious elementary light boson, whose production mechanism is at odds with the production of a non-relativistic pair of heavy quarks then forming a quarkonium.

Having proposed a way to get sound NLO perturbative results, we discuss the interplay between the behaviour of the gluon PDFs at low scales and, in particular, the $\eta_c$ production cross sections. This motivates us to encourage a vigorous experimental effort to measure it and, before data are available to fit them, we are tempted to suggest experts in PDF fits to analyse how degraded global fits would be if $x g(x;\mu_F)$ at NLO is required to be monotonous for $x < 0.01$ at $\mu_F \sim m_c$.

The structure of the article is as follow. In section 2, we outline the structure of the NLO $\eta_Q$ production cross sections and explain how to reproduce the existing results. On the way, we provide analytical expressions in terms of the partonic cross section and of the partonic luminosities needed to compute the rapidity differential cross section, $d\sigma^{\rm NLO}/dy$, 
which are not available in the literature. In section 3, we make a brief historical survey of the past phenomenology of NLO $\eta_Q$ hadroproduction and of the attempts to identify the origin of these negative NLO cross sections  and we explain that they come from the subtraction procedure in the factorisation of the collinear singularities in the $\msbar$ scheme. Section 4 is devoted
to our factorisation-scale choice.  Section 5 gathers our resulting cross sections for $\eta_c$ and $\eta_b$. We first demonstrate that our proposal works by discussing the behaviour of the $K^{\rm NLO}$ factors for $\eta_Q$ and elementary scalar bosons. Then, we discuss the interplay between the gluon luminosity and our obtained cross sections and finally we present what we believe to be the best possible NLO predictions.  Section 6 gathers our conclusions and an outlook at other quarkonium-production processes.

\section{$\eta_Q$ production up to NLO in the collinear and NRQCD factorisations} 
\subsection{CSM, NRQCD and collinear factorisation}
The present study essentially bears on collinear factorisation~\cite{Brock:1993sz} whereby the hadronic cross-section to produce
a quarkonium $\Q$ is factorised
into a convolution of PDFs and a partonic cross section, $\hat{\sigma}(\Q)$. Through NRQCD factorisation~\cite{Bodwin:1994jh}, the latter is further factorised
into short-distance perturbative parts, computable with Feynman graphs, and long distance non-perturbative parts.
As a result, one starts for the production of a quarkonium $\Q$ in a collision of two hadrons $A$ and $B$ from:
\eqs{
   d\sigma_{AB}=\sum_{ab}\int\!\! dx_1 dx_2 f_{a/A}(x_1;\mu_F) f_{b/B}(x_2;\mu_F)\, \times
    \\
    \underbrace{\sum_{n}{d\hat{\sigma}_{ab}\Big(Q\bar{Q}\left[n\right]+\{k\}\Big)\,(\mu_R,\mu_F,\mu_\Lambda)
    \langle {\cal O}_{\Q}^n\rangle_{\mu_\Lambda}},}_{\displaystyle d\hat{\sigma}_{ab}(\Q)}
    \label{eq:colfactorisation}
}
where \eg\ $f_{a/A}$ is the PDF of the parton $a$ inside the hadron $A$, $d\hat{\sigma}_{ab}(Q\bar{Q}\left[n\right]+\{k\})$ 
are proportional to the partonic differential cross-section 
to produce a $Q\bar{Q}$ pair in the (spin and colour) quantum number $n$, possibly with 
other particles $\{k\}$ from the scattering of the partons $ab$ and
 $\langle {\cal O}_{\Q}^n)\rangle$ is an NRQCD Long-Distance Matrix Element (LDME) 
 for the non-perturbative hadronisation of the pair in the state $n$ into the quarkonium $\Q$. NRQCD factorisation stems from an expansion in the relative velocity $v$ between the $Q\bar{Q}$ pair in the quarkonium rest frame. In this work, we focus on the terms leading in $v$ and sub-leading in $\alphaS$.
 As such, we only need to consider the colour singlet $\oSzo$ state for pseudo-scalar quarkonia, which is thus equivalent to the CSM. In such a case, 
  the sum over $n$ in $d\hat{\sigma}_{ab}(\Q)$ reduces to a single term.
 The purpose of the next sections is to explain how $d\hat{\sigma}_{ab}(\Q)$
can computed up to NLO accuracy in order to explain the appearance of
negative cross sections in past computations.
 
\subsection{$\eta_Q$ hadroproduction at LO}

At LO ($\alphaS^2$), $\eta_Q$ hadroproduction proceeds through gluon fusion, $g(k_1)+g(k_2)\rightarrow \eta_Q (P)$, which can be computed via Feynman diagrams like \cf{fig:gg-etaQ}. 
In the CSM~\cite{Chang:1979nn,Berger:1980ni,Baier:1983va}, the matrix element to create
a $^1S_0$ pseudoscalar  quarkonium $\eta_Q$ with a momentum $P$,  possibly accompanied by other partons, noted $\{k\}$, 
is obtained from the product of the amplitude to create the corresponding heavy-quark pair, ${\cal M}(ab \to Q \bar Q+\{k\})$, a spin
 projector, $N(Ps.| s_1,s_2)$ and $R(0)$, the $\eta_Q$ radial wave function at the origin in the configuration space. 
The CSM being the leading $v$ contribution to NRQCD, $R(0)$ can naturally be related to
a NRQCD LDME as follows:
\eqs{\langle \mathcal{O}_{\eta_Q}^{^1S_0^{[1]}} \rangle= \frac{2 (2J+1) N_c\lvert R(0)\rvert^2}{ 4 \pi}.} By virtue
of heavy-quark spin symmetry, $R(0)$ is identical for the $\eta_c$ and $J/\psi$ for instance up to $v^2$ corrections. It  can 
then be obtained from the well measured leptonic width of the $J/\psi$ computed in the CSM/NRQCD 
or from potential models. In what follows, we will use $|R_{\eta_c}(0)|^2=1\text{ GeV}^3$ and $|R_{\eta_b}(0)|^2=7.5\text{ GeV}^3$~\cite{Brodsky:2009cf}.

Overall, one has
\eqs{ \label{eq:CSM_generic}
&{\cal M}(ab \to {\Q}(P)+\{k\})=\!\!\sum_{s_1,s_2,i,i'}\!\!\frac{N(Ps.| s_1,s_2)}{ \sqrt{m_Q}} \times \\ 
&\frac{\delta^{ii'}}{\sqrt{N_c}} 
\frac{R(0)}{\sqrt{4 \pi}} 
{\cal M}(ab \to Q^{s_1}_i \bar Q^{s_2}_{i'}(\mathbf{p}=\mathbf{0})  + \{k\}),
}
where $P=p_Q+p_{\bar Q}$, $p=(p_Q-p_{\bar{Q}})/2$, $s_1$ and $s_2$ are the heavy-quark spins, and $\delta^{ii'}/\sqrt{N_c}$ 
is the projector onto a CS state. For $v\to0$, the spin projector on a pseudoscalar state,
$N(Ps.| s_i,s_j)= \frac{1}{2 \sqrt{2} m_{Q} } \bar{v} (\frac{\mathbf{P}}{2},s_j) \gamma_5 u (\frac{\mathbf{P}}{2},s_i)$.
After one sums over the quark spin, one obtains traces which can be evaluated in a standard way.

\begin{figure}[hbt!]
\centering
\subfloat[]{\includegraphics[scale=0.4]{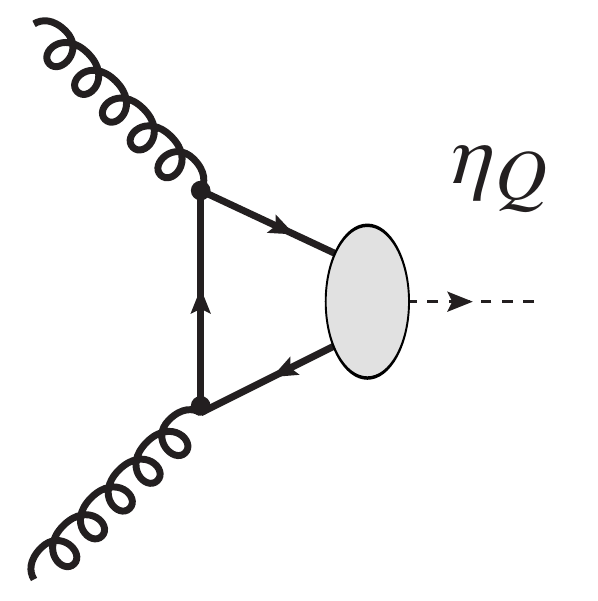}\label{fig:gg-etaQ}}
\subfloat[]{\includegraphics[scale=0.4]{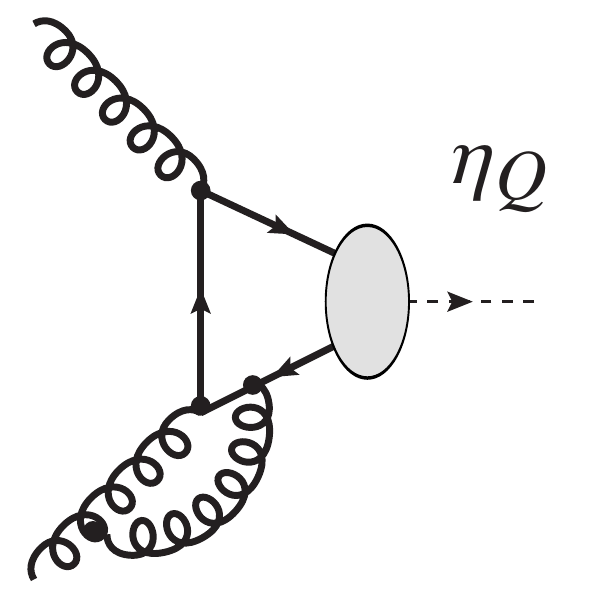}\label{fig:gg-etaQ-OneLoop}}
\subfloat[]{\includegraphics[scale=0.4]{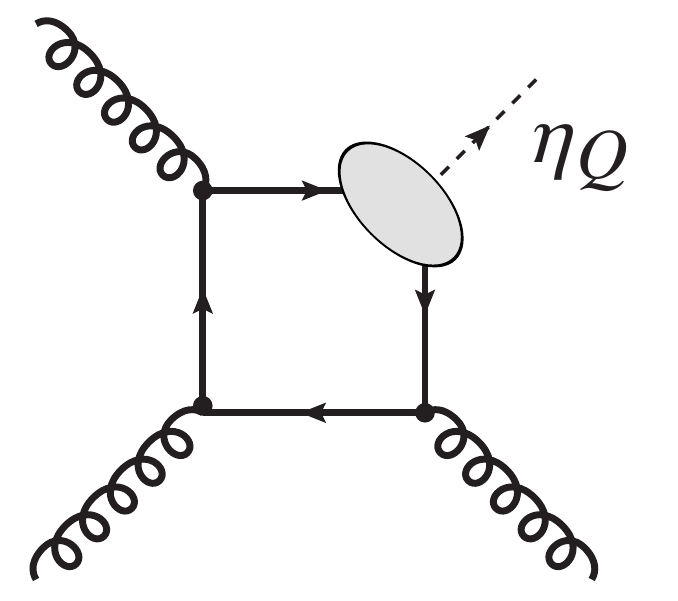}\label{fig:gg-etaQg-PT8}}\\
\subfloat[]{\includegraphics[scale=0.4]{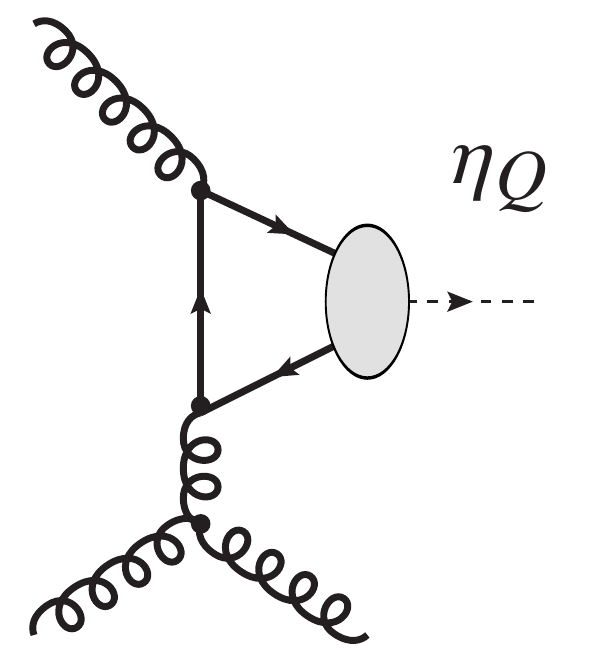}\label{fig:gg-etaQg-PT6}}
\subfloat[]{\includegraphics[scale=0.4]{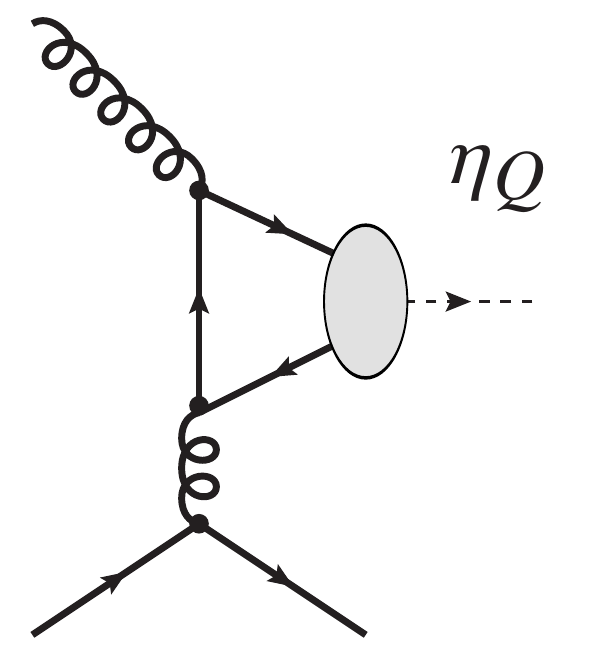}\label{fig:gq-etaQq-PT6}}
\subfloat[]{\includegraphics[scale=0.4]{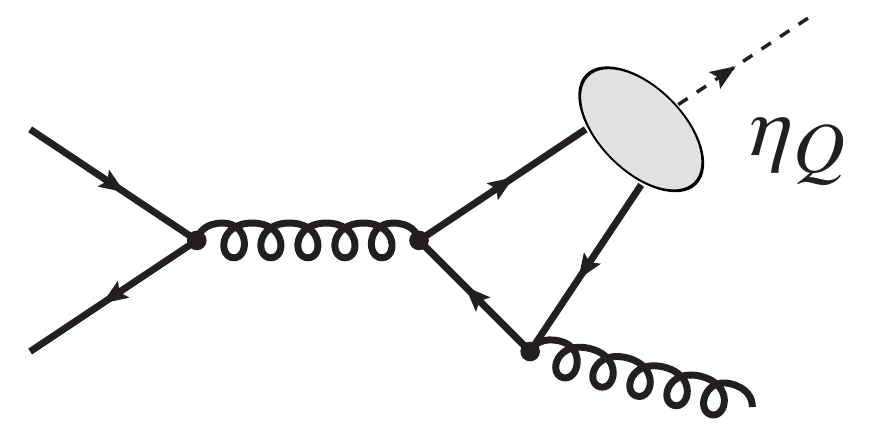}\label{fig:qq-etaQg}}
\caption{Representative diagrams contributing to $\eta_Q$ hadroproduction via 
CS channels at orders $\alphaS^2$ (a), $\alphaS^3$ (b,c,d,e,f).
The quark and antiquark attached to the ellipsis are taken as on-shell
and their relative velocity $v$ is set to zero.}
\label{diagrams-CS-etac}
\end{figure}

However we note here the explicit appearance of $\gamma_5$ in $N(Ps.| s_i,s_j)$ which can cause issues within the framework of dimensional regularisation. Here we employ the standard 't Hooft-Veltman scheme to deal with $\gamma_5$ in $D=4-2\epsilon$-dimensions~\cite{tHooft:1972tcz}. We obtain for the LO matrix element squared~\cite{Kuhn:1992qw},
\begin{equation}
    |\mathcal{M}|^2=\frac{N_c^2-1}{N_c}\frac{16 \alpha_s^2 \pi |R_0|^2}{M_\Q}\mu_R^{4\epsilon}\left(1-\epsilon\right)\left(1-2\epsilon\right),
\end{equation}
and for the partonic cross section (where we have set $M_{\Q}=2m_{\Q}$ as expected within NRQCD and $N_c=3$),
\eqs{
    \hat{\sigma}^{\text{LO}}_{gg}&=\frac{\pi}{\hat{s}^2}\frac{1}{8^2}\frac{1}{\left(2-2\epsilon\right)^2}|\mathcal{M}|^2 \delta\Big(1-\frac{M^2_{\Q}}{\hat{s}}\Big)
    \\
    &=\underbrace{\frac{\pi}{M_{\Q}^2}\frac{1}{8^2}\frac{1}{\left(2-2\epsilon\right)^2}|\mathcal{M}|^2}_{\displaystyle \hat{\sigma}_0^{\text{LO}}} \delta\Big(1-z\Big)
}
where we have defined $z=M^2_{\Q}/\hat{s}$. 

The hadronic section  then reads with $\tau_0=M^2_{\Q}/s=4m^2_Q/s$ and $\tau=\tau_0/z$,
\eqs{\label{eq:dsidy-LO}
&\frac{d\sigma^{\text{LO}}}{dy}(\sqrt{s},y; \mu_F)=\int d\tau \, \frac{\partial\mathcal{L}_{gg}}{\partial y\partial \tau}
\, \hat{\sigma}^{\text{LO}}_{0}\,\tau\,\delta\Big(\tau-\tau_0\Big), \\
&\sigma^{\text{LO}}(\sqrt{s}; \mu_F)=\int d\tau \, \frac{\partial\mathcal{L}_{gg}}{\partial \tau}\,
\hat{\sigma}^{\text{LO}}_{0}\,\tau\,\delta\Big(\tau-\tau_0\Big),
}
in terms of the following differential gluon luminosities: 
\eqs{
&\frac{\partial\mathcal{L}_{gg}}{\partial y\partial \tau}(\tau,y; \mu_F)=f_{g}(\sqrt{\tau}e^y,\mu_F)f_{g}(\sqrt{\tau}e^{-y};\mu_F),\\
&\frac{\partial\mathcal{L}_{gg}}{\partial \tau}(\tau; \mu_F)=\int_{1/2 \log{\tau}}^{-1/2\log{\tau}} dy\, f_{g}(\sqrt{\tau}e^y;\mu_F)f_{g}(\sqrt{\tau}e^{-y};\mu_F).
}
These fully encapsulate the energy and rapidity dependences of the $\eta_Q$ yields at LO. 

\subsection{$\eta_Q$ hadroproduction at NLO}
\label{subsec:NLOpart}

Let us now outline how to compute the $\eta_Q$ cross section up to NLO accuracy~\cite{Kuhn:1992qw,Schuler:1994hy,Petrelli:1997ge} which we will then use throughout our study.  
 
NLO contributions involve both virtual(-emission) and real(-emission) corrections via $gg$ fusion that can be represented by diagrams in \cf{fig:gg-etaQ-OneLoop}, \cf{fig:gg-etaQg-PT8} and \cf{fig:gg-etaQg-PT6}. In addition to the $gg$ fusion, $qg$ and $q\bar{q}$ channels contribute at order $\alpha_s^3$ as shown in \cf{fig:gq-etaQq-PT6} and \cf{fig:qq-etaQg}. Both real and virtual contributions individually exhibit singularities. In order to deal with these singularities, we employ dimensional regularisation where we define $D=4-2\epsilon$.

As usual, the virtual contributions exhibit both Ultra-Violet (UV) and Infra-Red (IR) divergences. The former are removed via the renormalisation procedure. To do so, we apply the on-shell (OS) renormalisation scheme for the gluon/quark wave functions 
and the {heavy-quark} mass counter-term 
while, for the coupling, we perform the renormalisation $\delta Z_g^{\msbar}$ within the $\msbar$-scheme and we take~\cite{Klasen:2004tz}
\begin{equation}
\begin{split}
    \delta Z_2^\text{OS}&=-C_F \frac{\alpha_s}{4\pi}\left(\frac{1}{\epsilon_{\text{UV}}}+\frac{2}{\epsilon_{\text{IR}}}+3\log{\left(\frac{\mu_R^2}{m_Q^2}\right)}+4\right),
    \\
    \delta Z_3^\text{OS}&=\frac{\alpha_s}{4\pi}\left(\Big(\beta_0-2C_A\Big)\left(\frac{1}{\epsilon_{\text{UV}}}-\frac{1}{\epsilon_{\text{IR}}}\right)\right),
    \\
    \delta Z_m^\text{OS}&=-3C_F\frac{\alpha_s}{4\pi}\left(\frac{1}{\epsilon_{\text{UV}}}+\log{\left(\frac{\mu_R^2}{m_Q^2}\right)}+\frac{4}{3}\right),
    \\
    \delta Z_g^{\msbar}&=-\frac{\beta_0}{2}\frac{\alpha_s}{4\pi}\left(\frac{1}{\epsilon_{\text{UV}}}\right),
\end{split}
\end{equation}
where $\beta_0=\frac{11}{3}C_A-\frac{4}{3}T_F n_f$ with $n_f$ being the number of active light flavours. We have above made a distinction between $\epsilon_{\text{UV}}$ and $\epsilon_{\text{IR}}$ to label the poles coming from UV and IR divergences respectively. In the following we will only label the $\epsilon$ poles to show their UV/IR character but not the $\epsilon$ appearing in exponents. The $\epsilon$ with and without labels ultimately originate from the regulator in $D=4-2\epsilon$.
We have also absorbed a global factor of $e^{-{\epsilon} \gamma_{E}}\left(4\pi\right)^{{\epsilon}}$ inside the $\msbar$-renormalised $\alphaS$ coupling.

As what regards the virtual corrections, we are thus only left with soft IR divergences. In contrast to the virtual contributions where the singularities are already manifest in the $\eta_c$--gluon form factor\footnote{That is the contribution proportional to $\delta(1-z)$.}, the divergences in the real-emission part only reveal themselves after taking the phase-space integration\footnote{$z\rightarrow 1$ for soft and $\hat{t}, \hat{u}\rightarrow 0$ for collinear divergences.}. 

For ${d\hat{\sigma}}/{dy}$, the phase-space integration is slightly less straightforward to be performed analytically than for $\hat{\sigma}$ where one can just integrate over the full phase-space without separating out the rapidity $y$ and the transverse momentum $P_T$.

For both $\hat{\sigma}$ and ${d\hat{\sigma}}/{dy}$, after combining the virtual with the real corrections, the soft singularities vanish\footnote{But for a soft singularity proportional to $\beta_0$ that arises through renormalisation. This factor will be absorbed inside the PDFs, see later.} and we are left, as usual, with the initial-state collinear divergences which originate from diagrams such as in \cf{fig:gg-etaQg-PT6} and \cf{fig:gq-etaQq-PT6}. The occurrence of these divergences is a consequence of the fact that the initial states are fixed by the kinematics and therefore not integrated over.

Under the collinear factorisation, the divergences arising from such collinear emissions from specific initial partons are subtracted in the factorised PDFs via the corresponding Altarelli-Parisi (AP) Counter-Terms (CTs) which introduce the factorisation scale $\mu_F$  in the partonic cross section \cite{Altarelli:1977zs}.
In the $\msbar$ scheme, the AP CT for $\bar{\sigma}_{ag}$ reads,
\begin{equation}
    \bar{\sigma}_{ag}^{\text{(AP-CT)}}=\frac{1}{\epsilon_{\text{IR}}}\frac{\alphaS^{\text{bare}}}{2\pi}\left(\frac{4\pi\mu_R^2}{\mu_F^2}\right)^{\epsilon}\Gamma\left[1+\epsilon\right]\hat{\sigma}_{0}^{\text{LO}}z P_{ag}\left(z\right),
    \label{eq:APcounter}
\end{equation}
where $\Gamma\left[1+\epsilon\right]$ is the Gamma function and $P_{ag}(z)$ are the splitting functions between parton $a$ and a gluon. We have given their expressions in \ref{sec:split}.

Using these standard procedures, we have reproduced the expressions of the partonic cross sections up to $\alphaS^3$~\cite{Petrelli:1997ge,Kuhn:1992qw,Schuler:1994hy}. It does not generate any specific complications to fold these with PDFs. We have collected in \ref{sec:sigma} the final expressions for the integrated cross section $\sigma^{\rm NLO}_{\eta_Q}$ in terms of the partonic luminosities.  On the contrary, the analytical expressions needed to
obtain ${d\sigma^{\rm NLO}_{\eta_Q}}/{dy}$ are absent in the literature. We have gathered them in terms of the partonic luminosities for the three channels $gg$, $qg$ and $q\bar{q}$ in \ref{sec:dsigdy}. The codes which we have derived from these expressions and which we have used to generate the results presented later have been successfully cross-checked versus the semi-automatic code FDC~\cite{Wang:2004du}.

\section{On the origin of unphysical $\eta_Q$ cross section at high energies}
\subsection{The NLO partonic cross section and its HE behaviour}
\label{subsec:HE-behaviour}
In this section, we focus on the partonic high-energy (HE) limit ($\hat{s}\to \infty$ or equally $z \to 0$) and show how this limit can help us understand the origin of the unphysical cross-section results which we referred to in the introduction. 

The first NLO computation for pseudo-scalar quarkonium production was done~\cite{Kuhn:1992qw} by K\"uhn \& Mirkes in 1992 for toponium. At the time, it was not known that a toponium state could not bind. Their NLO results were confirmed by G.~Schuler~\cite{Schuler:1994hy} two years later who performed the first phenomenological application for charmonia. He was the first to report negative cross sections for $\eta_c$ production at $\sqrt{s}$ just above 1 TeV for the central scale choice. He explained this unphysical behaviour by the fact that the partonic $gg$ cross section was approaching a negative constant for $\mu_F=M_{\Q}$ at high $\hat{s}$. When folded with PDFs, such negative contributions coming from real emissions would become larger than the Born contributions for too flat low-$x$ gluon PDFs. However, as what regarded the reason why the $gg$-partonic cross section was approaching a negative constant at high $\hat{s}$, he did not provide any explanation, only a suggestion of a possible side effect of the restriction in the heavy-quark kinematics for them to be at threshold to form a non-relativistic bound state like a quarkonium.

 In 1996, while presenting preliminary NLO cross-section results within NRQCD, Mangano \& Petrelli discussed in a proceedings contribution~\cite{Mangano:1996kg} similar issues;
 they then attributed these negative cross sections to a possible over-subtraction of the collinear divergences inside the PDFs, thus rendering the partonic cross section negative in the HE limit.
 Quoting them, ``there is nothing wrong in principle with these [partonic] cross sections turning negative in the small-[$z$] region, as what is subtracted is partly returned to the gluon density via the evolution equations". They however also noted that, for processes like charmonium production occurring at scales near where the PDF evolution is initiated, it is insufficient in practice -- hence the negative hadronic cross sections then observed by Schuler. 

An important observation we would like to make here for our following reasoning is that the magnitude of these negative partonic cross sections for $z\to 0$ is process dependent. As such, the universal PDF evolution, for a given scale, cannot thus possibly fix the issue in a global manner. 

To further assess this, let us indeed focus on the small-$z$ limit of $\hat{\sigma}_{ab}$ which we obtained in the previous section, as done by Schuler, Mangano and Petrelli. Taking this limit in \ce{eq:sig-NLO-gg} \& (\ref{eq:sig-NLO-gq})\footnote{The following discussion applies to both $\hat{\sigma}$ and $\frac{d\hat{\sigma}}{dy}$.}, one gets
\begin{equation}
   \lim_{z\rightarrow 0} \hat{\sigma}^{\text{NLO}}_{ag}(z) = C_{a} \,\frac{\alpha_s}{\pi}\,\hat{\sigma}^{\text{LO}}_{0}\,\left(\log{\frac{M_\Q^2}{\mu_F^2}}+A_{a}\right),
   \label{eq:highenergylimit}
\end{equation}
where $C_{g}=2C_A$, $C_{q}=C_F$ and $M_\Q$ is the mass of the produced quarkonium (or $2m_Q$). Equivalent expressions were obtained for $P$-wave quarkonium~\cite{Schuler:1994hy,Mangano:1996kg}. 

One can also consider such a limit~\cite{Harlander:2009my} for the production of the Brout-Englert-Higgs (BEH) scalar boson $H^0$ using NLO expressions~\cite{Dawson:1990zj,Graudenz:1992pv,Spira:1995rr} or a fictitious elementary scalar boson, dubbed $\tilde{H}^0$, whose coupling to gluons also occurs through a loop of heavy quarks and get a similar limit. In all these cases, we stress, since it will be essential for our forthcoming discussion, that this limit~\cite{Harlander:2009my} in fact yields $A_{g}=A_{q}$.  

We further note the presence of the factorisation scale $\mu_F$ inside $\log{\frac{M^2}{\mu_F^2}}$ in these limits. This term is in fact universal and \textit{process-independent} as it originates from the AP CT (see~\ce{eq:APcounter}) to subtract the initial-state collinear divergences.  On the other side, $A_{a}$ is clearly \textit{process-dependent} as \ct{tab:Aval} illustrates it. 

\begin{table}[hbt!]\renewcommand{\arraystretch}{1.6}
    \centering 
    \begin{tabular}{c||c|c}
     & $A_{a}$ & $\hat{\mu}_F$ \\
    \hline  \hline 
    $\eta_{Q}^{[1,8]}$ & $-1$ & $\frac{M_\Q}{\sqrt{e}}=0.607 M_{\Q}$ \\
    \hline
    $\chi_{Q, \,J=0}^{[1,8]}$ & $-{43}/{27}$ & $0.451 M_{\Q}$ \\
    \hline
    $\chi_{Q, \,J=2}^{[1,8]}$ & $-{53}/{36}$ & $0.479 M_{\Q}$ \\
    \hline\hline
    \makecell{Fictitious $\tilde{H}^0$\\ ($2m_Q/m_{\tilde{H}}=1$)} & $-0.147$ & $0.93M_{\tilde{H}^ 0}$\\
    \hline
    \makecell{Fictitious $\tilde{H}^0$\\ ($m_Q/m_{\tilde{H}}=1$)} & $1.61$ & $2.43M_{\tilde{H}^ 0}$\\
    \hline
    \makecell{Real $H^0$ \\ ($2m_t/m_H=2.76$)} & $2.28$ & $3.12M_{H^0}$\\
        \end{tabular}
    \caption{The process-dependent constants $A_{a}$ [For quarkonia, $A_{a}$ is identical for CS and CO states] along with the
     $\mu_F$ value cancelling the HE limit. For the quarkonia, $A_{a}$ is identical for CS and CO states. For the scalar
particles, these value haves been derived for the HE expressions of~\cite{Harlander:2009my}. [$M$ represents here, and in what follows, 
the mass of the produced particle, be it a quarkonium or a, elementary scalar boson.]
}
    \label{tab:Aval}
\end{table}

As a consequence, if $A_{a}<0$, the HE limit of the {\it partonic} cross section thus gets negative for the natural scale choice $\mu_F=M_{\Q}$ and above. Whether this can make the {\it hadronic} cross section turn negative is then a matter of a complex interplay between the hadronic energy, the PDFs, $\mu_R$ and the size of the (process-dependent) virtual corrections.

Before discussing this interplay, let us however go back to the notion of over-subtraction to trace back the origin of these negative limits. Away from $\hat{s}=M^2$, only the real emissions contribute. In fact, at large $\hat{s}$, the sole $\hat{t}$-channel gluon-exchange topologies depicted by \cf{fig:gg-etaQg-PT6} \& \ref{fig:gq-etaQq-PT6} contribute; the other real-emission graphs depicted by \cf{fig:gg-etaQg-PT8} \& \ref{fig:qq-etaQg}--which are not divergent in the collinear region-- are suppressed at least one power of $M^2_\Q/\hat{s}$. At this stage, the amplitude square  can only be positive-definite by construction as it is a full Hermitian square for $\hat{s}\neq M^2$.\footnote{For $\hat{s}=M^2$, we note that the virtual contributions are not squared as their square contributes at $\alphaS^4$.}

When integrating over $\hat{t}$, one will encounter the aforementioned collinear divergences which are to be absorbed in the PDF via the AP CT. 
Anticipating this subtraction, we can exhibit the corresponding divergence and recast the unrenormalised cross section $\bar{\sigma}$ as
\eqs{\bar{\sigma}_{ag}^{\text{NLO},z\neq 1}=& \int d\hat{t}\,  \frac{d\bar{\sigma}_{ag}^{\text{NLO}, z\neq 1}}{d\hat{t}}
\\
=&- \frac{1}{\epsilon_{\text{IR}}}\frac{\alphaS^{\text{bare}}}{2\pi}\left(\frac{4\pi\mu_R^2}{M^2_\Q}\right)^{\epsilon}\Gamma\left[1+\epsilon\right]\hat{\sigma}_{0}^{\text{LO}}z P_{ag}\left(z\right)D_{a}
\\
&+ \frac{\alphaS^{\text{bare}}}{\pi} \hat{\sigma}_{0}^{\text{LO}} C_{a} \bar A_{a}\left(z\right),
}
where above we have split the collinear part from $\bar{A}_{a}\left(z\right)$\footnote{We remark at this stage that if the form factor of the Born cross section is resolved, \ie\ considering the top-quark loop with a finite mass in the case of $H^0$ production via gluon fusion, $A_{a}=\bar{A}_{a}(z=0)$ is a constant. On the contrary, \ie\ the coupling is tree-level type-like as in the Higgs EFT with $m_t\rightarrow \infty$, then we have an additional $\log{z}$ dependence and a different off set for the $gg$ and $qg$ channel.  It is not very surprising as for $z\to 0$, $m_t$ and $\hat{s}$ are both large and HEFT cannot be applied.} which is free of divergences for any $0\leq z< 1$. We have multiplied the first term by a factor $D_{a}=\left(1+\delta_{ag}\right)$ to account for the fact that one has collinear singularities for each gluon in the $gg$ channel. Therefore one would need to take $2\bar{\sigma}_{gg}^{\text{(AP-CT)}}$, \ie\ for each parton, to eliminate the poles. From the equation above, it follows that $\bar{\sigma}_{ag}^{\text{NLO},z\neq 1}$ is positive-definite due to the fact that the first term evaluates to positive infinite\footnote{For IR poles, one has that $\epsilon_{\text{IR}}<0$, while for UV poles $\epsilon_{\text{UV}}>0$.} as $\epsilon_{\text{IR}}\rightarrow 0^-$ irrespective of $\bar{A}_{a}\left(z\right)$.

Clearly, other schemes to absorb these collinear divergences inside the PDFs would yield different $\bar A_{a}\left(z\right)$\footnote{In principle, one could thus look for a scheme where the partonic cross sections simply do not become negative. This is left for future investigations as it would entail refitting the PDFs with different evolution equations.}.  In the DIS scheme for instance, $\bar{A}_{a}\left(z\right)$~\cite{Kuhn:1992qw} exhibits a $\log{z}$ dependence, which does not create any issue once integrated over $z$ and this different $z$ dependence should in principle be compensated by a different evolution of the PDFs. Yet, $\bar{A}_{a}\left(z\neq0\right)$ should remain finite. 

What we wish to argue here is that, since this subtraction is the only possible source of negative numbers at $z\neq 1$,
if $\bar A_a(z)$ happens to be negative in a given scheme where PDFs are supposedly positive (see \cite{Candido:2020yat} for $\msbar$), this signals that the AP CT have likely over-subtracted some collinear contributions from the real-emission contributions, and this can yield the observed negative hadronic cross sections. This is indeed what happens for quarkonia since the NLO threshold contributions ($\hat{s}=M^2$) are found to be positive-definite for $\eta_\Q$ and several other states at least for $\mu_F=\mu_R$\footnote{This is also the case for $H^0$ and $\tilde{H^0}$.}. Note that  $\sigma^{\rm NLO}_{\eta_\Q}$ also goes negative at large $\sqrt{s}$ for $\mu_F=\mu_R$.

Let us re-iterate at this stage that contributions of type full square $|\mathcal{M}|^2$ like the real emissions are always positive-definite by construction at any kinematical point $z$. The only way to render them negative is the over-subtraction via the AP-CT inside the PDFs. We agree that evolved PDFs can reduce the weight of these regions in $z$ where the partonic cross sections are negative, and eventually avoid negative hadronic cross sections. Yet, it is hard to believe that they would do so for all possible processes where this can occur as the coefficients $A_{a}$ are \textit{process-dependent} while the DGLAP evolution is \textit{process-independent}.

\subsection{From negative partonic cross sections to negative (or positive) hadronic cross sections}

Having now identified the origin of the negative cross sections, we can discuss their relevance to the past phenomenology which we recalled in the previous subsection. 

First, we note that the $\eta_b$ phenomenology, for which $\sigma^{\text{NLO}}_{\eta_b}$ remains positive in the LHC range, is less pathological. We have indeed found out~\cite{Feng:2015cba} that $\sigma^{\text{NLO}}_{\eta_b}$ only slightly deviates from $\sigma^{\text{LO}}_{\eta_b}$ in the LHC range. It thus seems that it is less sensitive to the limit  of~\ce{eq:highenergylimit}. Both charmonia and bottomonia have the same partonic cross section but for three changes: the mass shift and a trivial rescaling of the LDME and
$n_f$ which plays a minor role here. This mass shift however has three immediate effects : 
(i) a given $z=M_\Q^2/\hat{s}$ value for bottomonia corresponds to 3 times larger $\sqrt{\hat{s}}$. Considering the rescaling on the integration bounds, $[M_\Q^2/{s},1]$, when convoluted with PDFs, this effectively corresponds to a 3 times larger $\sqrt{{s}}$,  which is thus easily outside the range of past studies, 
(ii) however, even at fixed $z$, the results would differ since $\alphaS(\mu_R\simeq M_\Q)$ is smaller and this reduces the impact of $\alphaS^3$ contributions compared to the (positive) Born ones at $\alphaS^2$, 
(iii) the evolved gluon PDFs up to a larger $\mu_F$ become steeper which reduces the relative importance of the small-$z$ domain compared to the threshold contribution at $z=1$ which remains positive. Taken together, these 3 points explain very well why the charmonium case, at low scales, is the most pathological one and that the issue of a possible over-subtraction of the collinear divergences is usually considered to be rather academical with a limited impact on other hadronic cross sections.

It is however legitimate to wonder if further aspects specific to the modelling of quarkonium production renders its phenomenology particular. Our answer tends to be negative. Indeed, as \ct{tab:Aval} shows, 
CO and CS states are equally affected, in agreement with the past phenomenology~\cite{Feng:2015cba}. 
This confirms that neglecting such higher-order $v$ corrections is not the source the issue. Since CS and CO are both computed 
in the non-relativistic limit, one may wonder whether that this limit is also a source of issues, as suggested by Schuler. 
Yet, we anticipate that the phenomenology of a $\tilde{H}^0$ with $m_Q=M_{\tilde{H}^0}/2$ should also be affected since $A_{a}$ is also negative.
As our numerical results will show, it is indeed the case and this will thus confirm that this is not a quarkonium issue {\it per se}.

\section{A scale choice as solution}

We now come to our proposal to solve this unnatural behaviour of the cross section. In fact, it simply amounts to set the factorisation scale $\mu_F$ such that partonic cross-section vanish at large $\hat{s}$, instead of risking it to become negative. Of course, such a scale choice is only possible provided that it is the same for all the partonic channels. Dubbing our scale choice $\hat{\mu}_F$, we just define it as 
\begin{equation}
\hat{\mu}_F= M e^{{{A}_{a}}/{2}},
\end{equation}
having in mind that that $A_q=A_g$.
It is clear, from our definition that, since ${A}_{a}$ is a \textit{process-dependent} quantity,  $\hat{\mu}_F$ will be \textit{process-dependent}. We have listed some values of $\hat{\mu}_F$ in~\ct{tab:Aval} for the different particles we considered. It is important to note that the $\hat{\mu}_F$ values we have found are within or close to the usual ranges of values anyway taken in phenomenological studies.

Let us now turn to the physical picture of our reasoning. Our motivation is clear as it amounts to avoid negative cross sections which we attribute to an over-subtraction in the $\msbar$ factorisation scheme. Our scale choice avoids that $\hat \sigma_{ab}$ be negative at small $z$. It makes sense to base its construction from this limit as it becomes more and more relevant at large $s$, precisely where $\sigma^{\rm NLO}_{\eta_Q}$ can become negative.

Even if $\bar{A}_{a}(z)$ becomes more negative than its limiting value, ${A}_{a}$, our results will show that cancelling $\hat \sigma_{ab}(z\to0)$ with $\mu_F=\hat{\mu}_F$ will be sufficient to get much more sound results, in particular to avoid $\sigma^{\rm NLO}_{\eta_Q}<0$. Going further, we stress that $\bar{A}_{g}(z)$ also
contains real emissions from the heavy-quark line (see \cf{fig:gg-etaQg-PT8}) and thus differs from $\bar{A}_{q}(z)$.
Working at finite $z$ where  $\bar{A}_{g}(z)\neq \bar{A}_{q}(z)$ does not allow us to derive an equally simple gauge-invariant solution based on a scale choice. In the quarkonium case, the latter contributions to $\hat{\sigma}_{gg}$ are relatively suppressed by $M^2_\Q/\hat{s}=z$ and thus their effect disappear at small $z$.

Another reason to focus on the small-$z$ limit is that when folding $\hat \sigma_{ab}(z)$ with the PDFs (see \ce{eq:colfactorisation} or \ce{eq:sig-NLO-gg}), the Jacobian to transform the integration measure from $dx_1 dx_2$ to a measure involving $dz$ will comprise a multiplicative factor $1/{z^2}$. As a result, the impact of the small-$z$ region certainly depends much on whether $\hat \sigma_{ab}(z\to0)$ is zero or not, even though the $z$ range has a lower bound set by $\tau_0=M^2/s$. Indeed, for nonzero $\hat \sigma_{ab}(z\to0)$, the PDFs are the key element regulating the integral. By virtue of evolution, they should become steep enough as to essentially damp down the contribution of the small-$z$ region. However, at low $\mu_F$, the PDFs can be rather flat. This can give a large weight to this small-$z$ region where the real-emission contributions are negative for large $\mu_F/M$, hence the possibility that $\sigma^{\rm NLO}_{\eta_Q}<0$. Now, if $\hat \sigma_{ab}(z\to0)=0$ as our $\mu_F$ choice entails, the PDF shape at low scales is suddenly much less crucial to damp down a contribution which should not be the leading one in any case.

Physics wise, our scale choice essentially amounts, in the partonic HE limit, to reshuffle the entirety of the real emissions inside the PDFs\footnote{Understanding the possible connection with a recent study of the positivity of the PDFs in the $\msbar$ scheme~\cite{Candido:2020yat} and the collinear factorisation scheme~\cite{Maltoni:2007tc} is left for a future study.}. From a HE viewpoint, such contributions are expected to be important, in particular at small $\hat{t}$ since they are supposed to be enhanced by logarithms of $\hat{s}$, which should eventually be resummed. 
In fact, as our discussion has illustrated, the prominent effect of such contributions is a source of issues in a fixed-order computation as it jeopardises the convergence of the perturbative series with NLO contributions being more important than the Born ones.
In this sense, our scale setting amounts to include all these possible HE effects in the PDFs. This makes sense
as the PDFs are ultimately determined by fitting data -- containing all type of higher order corrections. In fact, recent PDF analyses have been made taking into account HE effects in their evolution~\cite{Ball:2017otu,Abdolmaleki:2018jln}.

\section{Results and discussion}

\subsection{A word on our  PDF choice}
Given the importance of the PDF shape at low scales in the previous discussions, we have employed on purpose, thanks to LHAPDF6~\cite{Buckley:2014ana}, 3 NLO sets which show rather different features:
: 
\begin{enumerate}
\item a  representative\footnote{Our choice has mainly been driven by technical reasons. For instance, at low scales, 
as can be seen on~\cf{fig:PDF-comparison} (a) NNPDF31\_nlo\_as\_0118 seems to suddenly saturate at $x=5\times10^{-5}$ whereas CT14nlo (like CT18nlo) comprises two outstanding eigensets (one low and one high) which are extremely different from the others. We stress that our forthcoming physical conclusions would not
be affected if we made other choices.} set
of the conventional NLO PDFs, PDF4LHC15\_nlo\_30~\cite{Butterworth:2015oua},
\item a dynamical PDF set, JR14NLO08VF~\cite{Jimenez-Delgado:2014twa}, where gluons are radiatively generated from a valence-like positive input distributions at a low scale which is optimally chosen, and 
\item a set taking into account HE effects in the evolution, NNPDF31sx\_nlonllx\_as\_0118~\cite{Ball:2017otu},
\end{enumerate}
in order to perform our NLO cross-section evaluations.

These are plotted on~\cf{fig:PDF-comparison} along with CT14nlo~\cite{Dulat:2015mca}, MMHT14nlo~\cite{Harland-Lang:2014zoa},
NNPDF31\_nlo\_as\_0118~\cite{Ball:2017nwa} for comparison, for two scale choices, 1.55 GeV and 3 GeV. We note that $x\, g(x)$ from PDF4LHC15\_nlo\_30, MMHT14nlo, CT14nlo and NNPDF31\_nlo\_as\_0118 all show a maximum around 0.02 and then a local mininum below 0.001.

\begin{figure}[h!]
\centering
\subfloat[$\mu_F=1.55$~GeV]{\includegraphics[width=0.9\columnwidth]{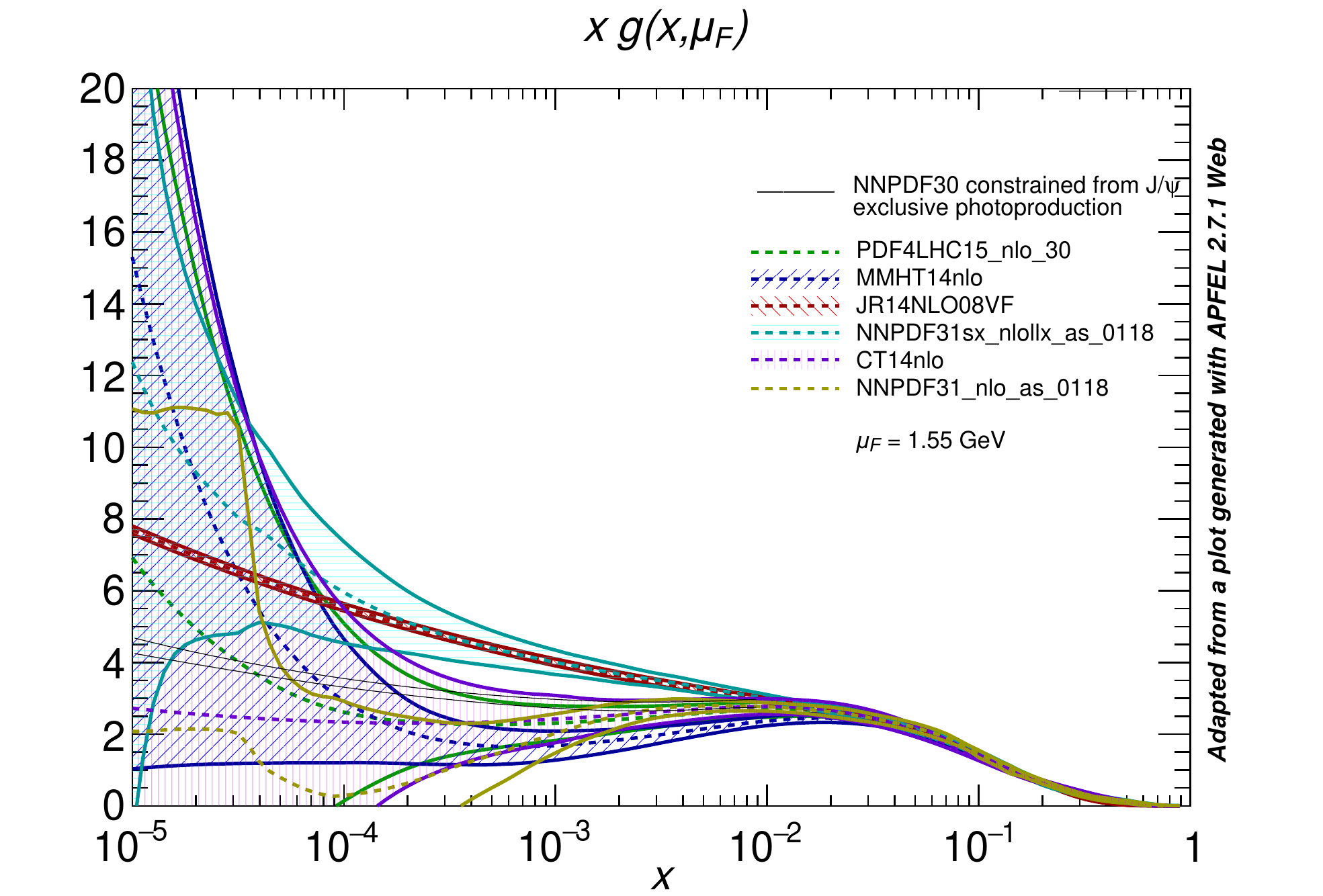}\label{fig:PDF-comparision-1_5}}\\
\subfloat[$\mu_F=3.0$~GeV]{\includegraphics[width=0.9\columnwidth,]{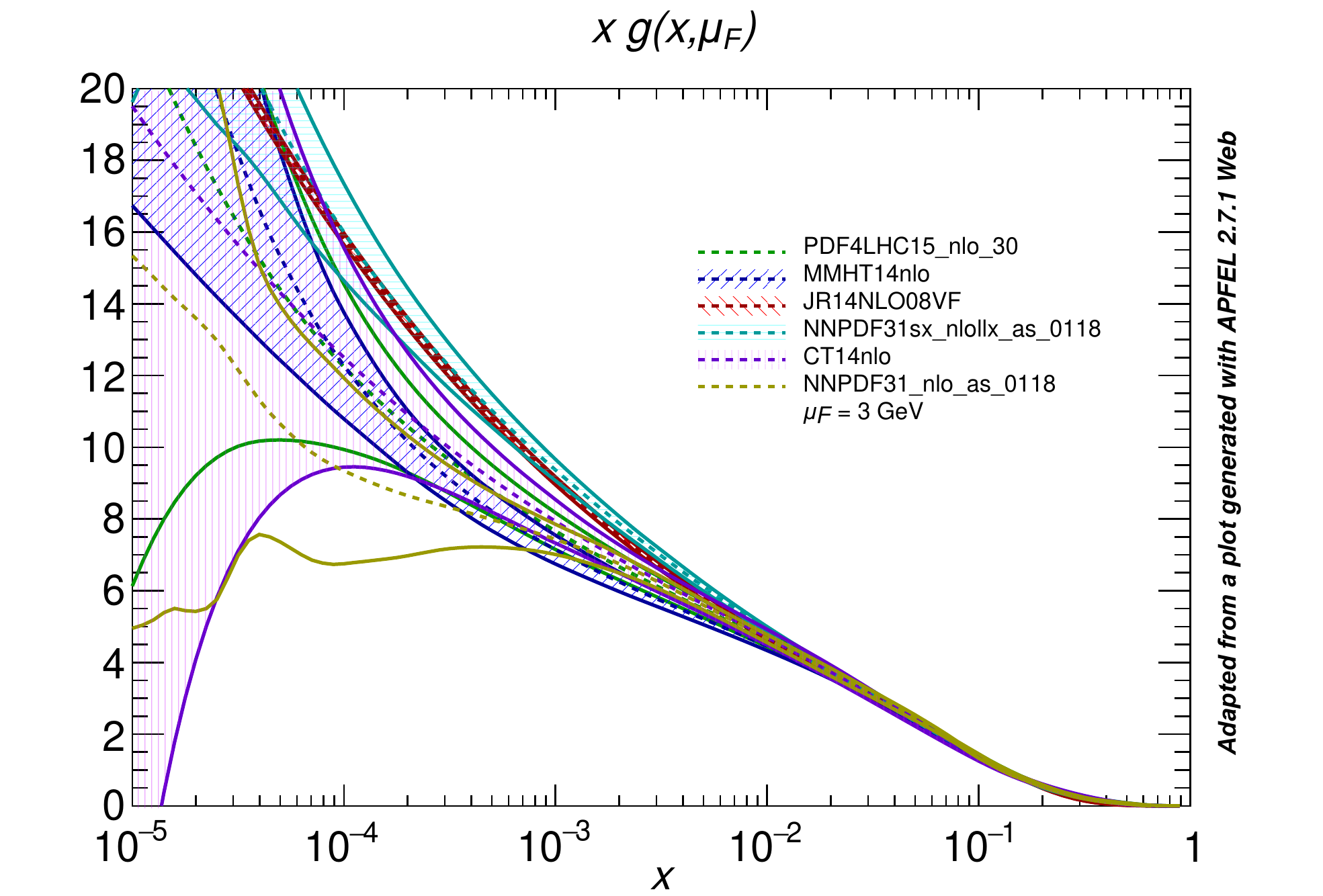}\label{fig:PDF-comparision-3}}
\caption{Gluon PDFs as encoded in PDF4LHC15\_nlo\_30~\cite{Butterworth:2015oua}, JR14NLO08VF~\cite{Jimenez-Delgado:2014twa},
NNPDF31sx\_nlonllx\_as\_0118~\cite{Ball:2017otu}, CT14nlo~\cite{Dulat:2015mca}, MMHT14nlo~\cite{Harland-Lang:2014zoa},
NNPDF31\_nlo\_as\_0118~\cite{Ball:2017nwa}  for two scale values  : (a) 1.55~GeV and (b) 3~GeV. In addition,
we have added on (a) (solid black lines) the resulting constraints on NNPDF3.0 obtained by Flett~\etal\ under some asumptions~\cite{Flett:2020duk}~from $J/\psi$ exclusive photoproduction. [These plots have been adapted from plots generated by APFEL web~\cite{Bertone:2013vaa,Carrazza:2014gfa}].}\label{fig:PDF-comparison}
\vspace*{-0.5cm}
\end{figure}

Such features are absent in both JR14NLO08VF and NNPDF31sx\_nlonllx\_as\_0118 whereas they have a significant impact on the phenomenology as we will show later on. However, we stress that the local minimum has already disappeared once the gluon PDFs are evolved up to 3 GeV, where the 3 sets we have used display similar features but for the size of the uncertainties.  At $\mu_F= 1.55$~GeV, we note that both for PDF4LHC15\_nlo\_30 and NNPDF31sx\_nlonllx\_as\_0118, the shape can be very different within the uncertainty spanned by their PDF eigensets. At $\mu_F= 3$~GeV, this only remains the case for PDF4LHC15\_nlo\_30. These different behaviours will in fact be very useful to study the interplay between the scale and the PDF choices.

We further note that a recent study by Flett~\etal\ has shown that one could extract, under specific assumptions~\cite{Flett:2020duk}, actual constraints on the gluon PDF at low scales from $J/\psi$ exclusive production data. These are represented by the solid black lines in~\cf{fig:PDF-comparison} (a) when applied to NNPDF30\_nlo. Under such assumptions, the presence of a local minimum below 0.001 is unlikely as it would yield to a decrease in the $J/\psi$ exclusive production cross section, which is absent in the data. At this stage, since it is not an actual PDF fit and since the exclusive cross sections are not directly related
to the PDFs, we consider this finding as a guidance, yet a very interesting one anticipating our results.

\subsection{Assessing the perturbative convergence with $\hat{\mu}_F$ using the $K^{\rm NLO}$ factors }

We have found so far that significant NLO contributions to $\eta_Q$ production are expected
to appear if the hadronic cross section becomes sensitive to $z=M_{Q}^2/\hat{s}$ values far away from threshold.
This would result in a significant $\sqrt{s}$ dependence of the NLO/LO hadronic cross-section ratio ($K^{\rm NLO}$). As we explained, it is due to an over subtraction, in the $\msbar$ scheme, of collinear contributions from the real-emission NLO contributions inside the PDFs. A relative constant offset is however expected from the virtual corrections at $z=1$, like for the decay widths, in particular for reactions where $\alphaS$ is not very small. 

\begin{figure*}[hbt!]
\centering
\subfloat[]{\includegraphics[width=0.66\columnwidth,height=5cm]{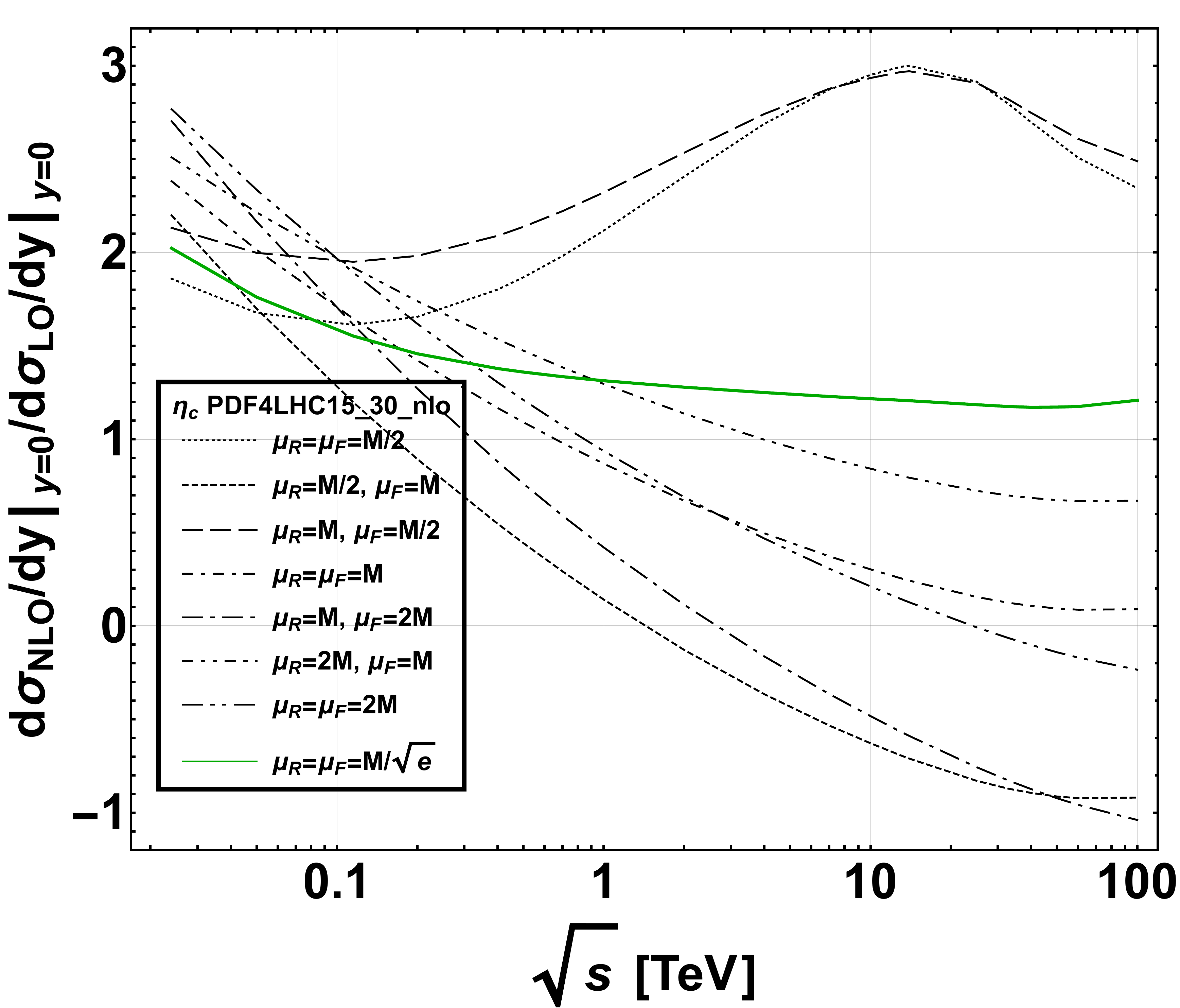}\label{fig:PDF4LHC_K_etac_EnEvo}}
\subfloat[]{\includegraphics[width=0.66\columnwidth,height=5cm]{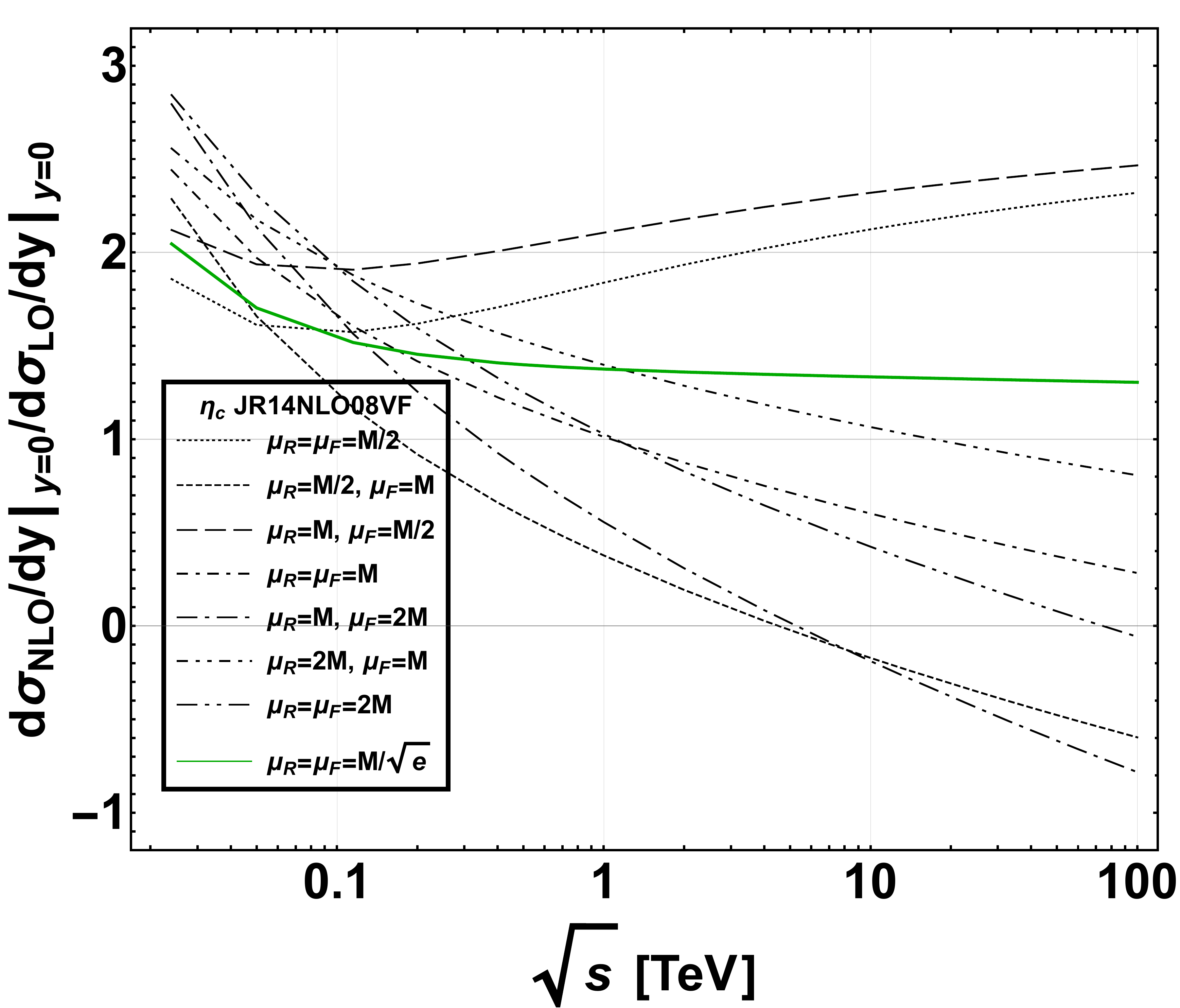}\label{fig:JR14NLO08VF_K_etac_EnEvo}}
\subfloat[]{\includegraphics[width=0.66\columnwidth,height=5cm]{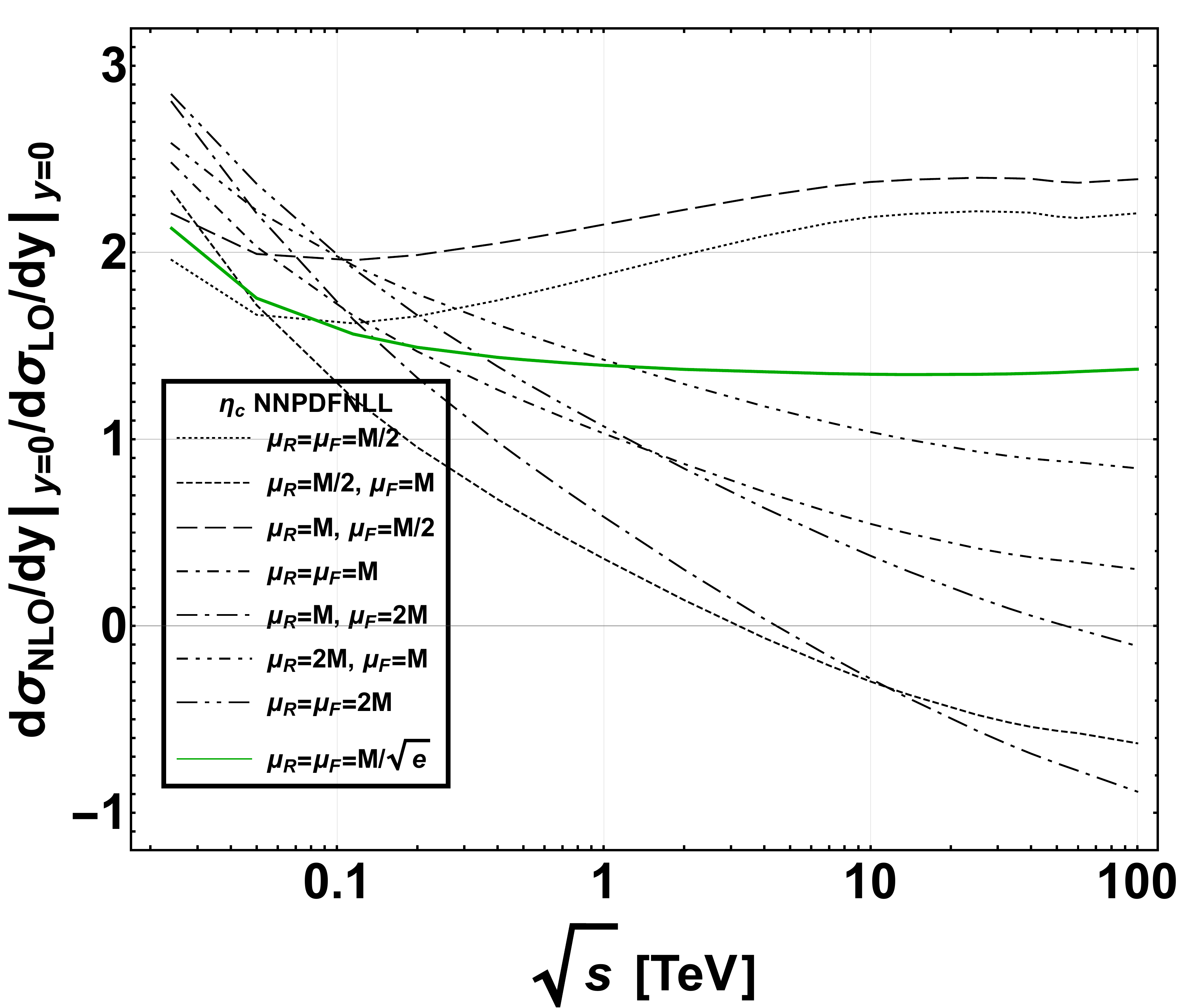}\label{fig:NNPDF31sxNLONLL_K_etac_EnEvo}}\\
\subfloat[]{\includegraphics[width=0.66\columnwidth,height=5cm]{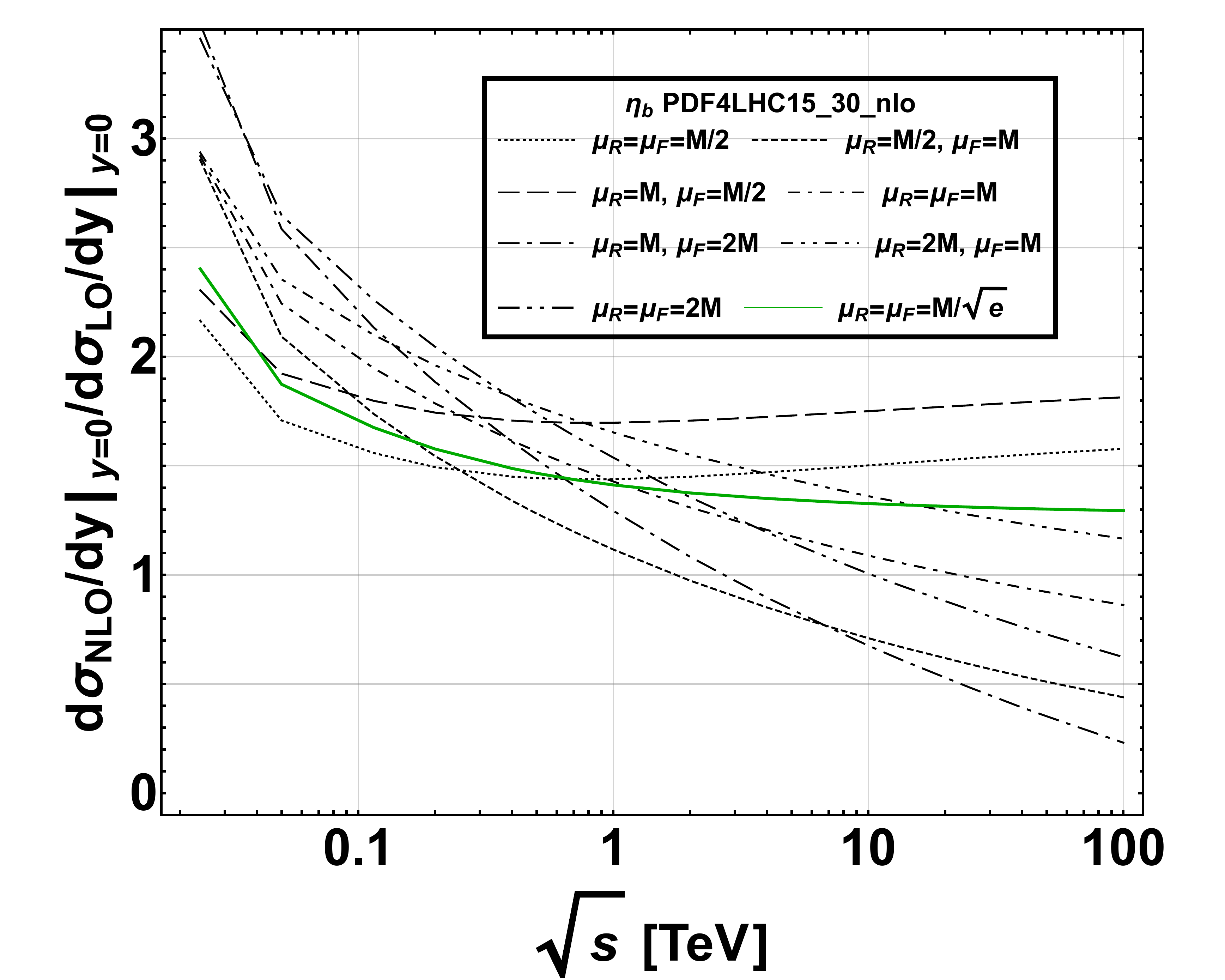}\label{fig:PDF4LHC_K_etab_EnEvo}}
\subfloat[]{\includegraphics[width=0.66\columnwidth,height=5cm]{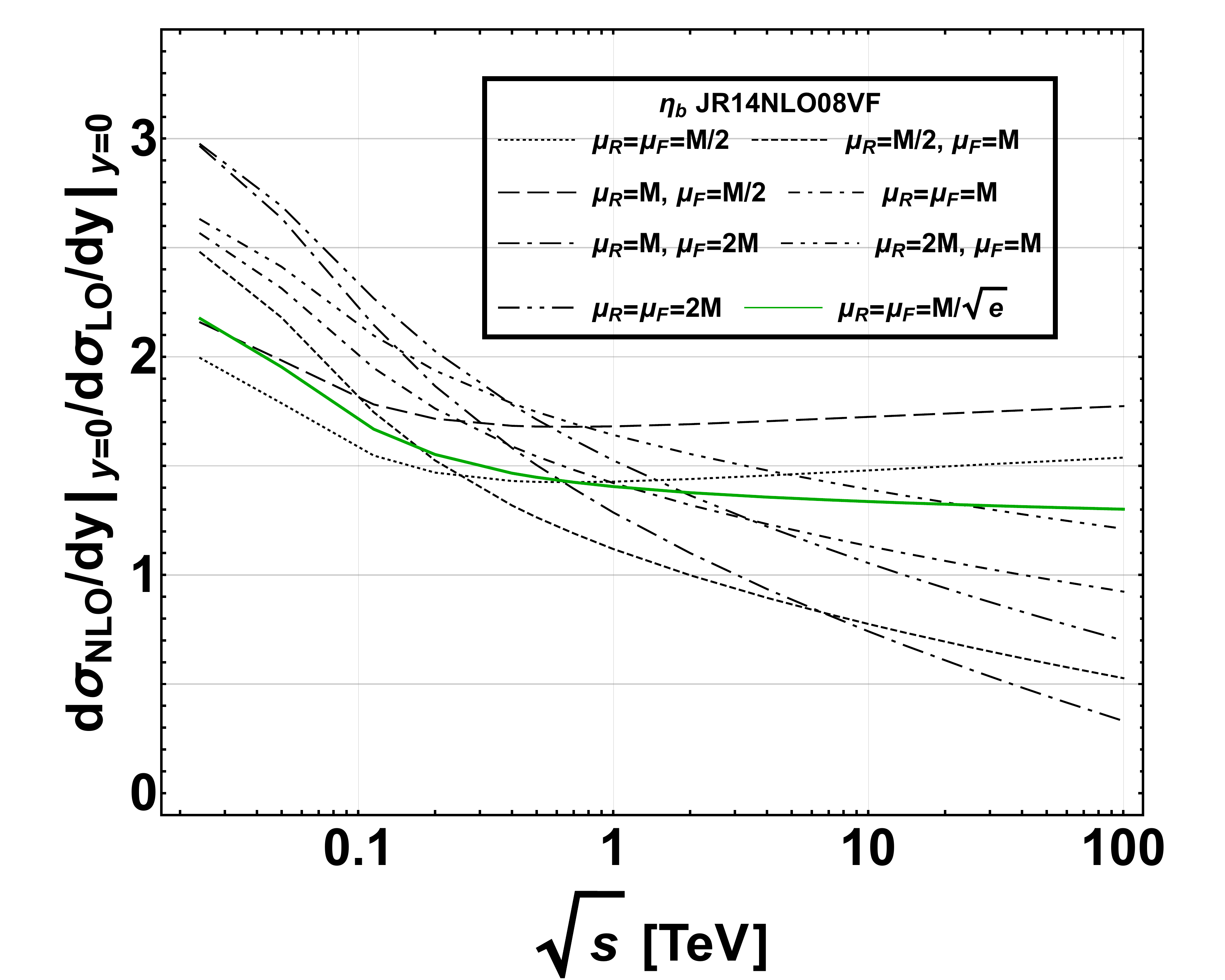}\label{fig:JR14NLO08VF_K_etab_EnEvo}}
\subfloat[]{\includegraphics[width=0.66\columnwidth,height=5cm]{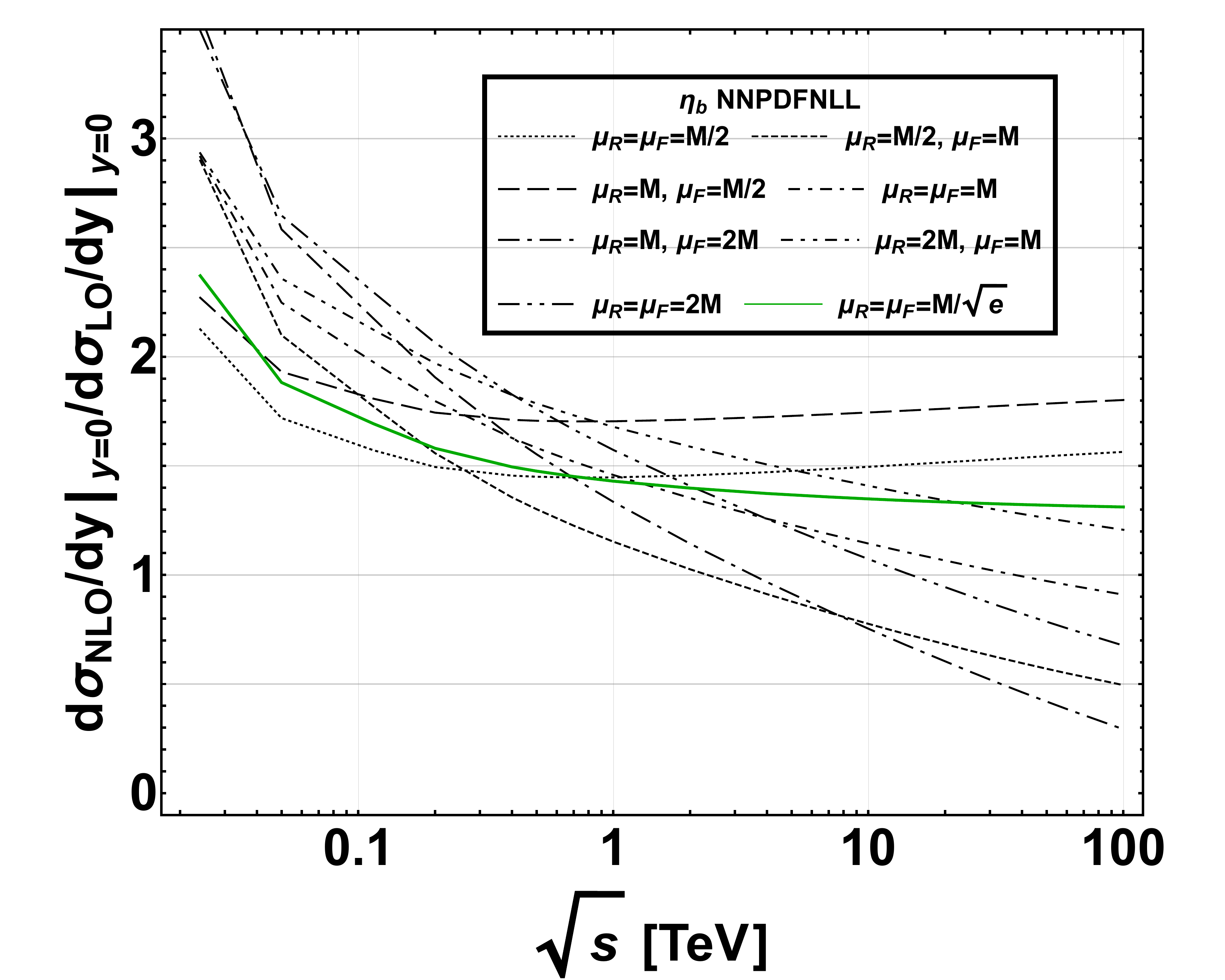}\label{fig:NNPDF31sxNLONLL_K_etab_EnEvo}}
\caption{$K^{\rm NLO}|_{y=0}$ for $\eta_c$ (top) and  $\eta_b$ (bottom) (for PDF4LHC15\_nlo\_30 (left), JR14NLO08VF (middle) and NNPDF31sx\_nlonllx\_as\_0118 (right)) as a function of $\sqrt{s}$ for the usual 7-point scale choices and our $\hat\mu_F$ scale with $\mu_R=\mu_F$.}\label{fig:K-etaQ} 
\vspace*{-0.5cm}
\end{figure*}

To mitigate this fixed-order treatment shortcoming, we have thus proposed a specific scale choice which corresponds to the inclusion, in the $z\to0$ limit, of the entirety of such NLO contributions in the PDFs.  The logic behind is that PDFs are fit to data which incorporate all such emissions. This is probably not a perfect solution but, beside of corresponding to perfectly acceptable $\mu_F$ values, it indeed avoids erratically varying $K^{\rm NLO}$ factors and negative and unphysical hadronic cross section, as the results of this section show. 

Before discussing our results for $K^{\rm NLO}$, let us describe our set-up. We have evaluated them at $y=0$ using \ce{eq:dsidy-LO} for LO and \ce{eq:dsidy-NLO-gg}, (\ref{eq:dsidy-NLO-qg}) \& (\ref{eq:dsidy-NLO-qq}) for NLO. The same PDF have been used for both. We have set $m_c=1.5$ GeV  for the $\eta_c$ and $m_b=4.75$ GeV for $\eta_b$. We have used the $\alphaS$ corresponding to our PDF choice thanks to LHAPDF6.

As for a fictitious $\tilde{H^0}$, we have set its mass at 3 GeV, close to that of $\eta_c$. Having at our disposal, 
the small-$z$ limit for different $M_{\tilde{H^0}}/m_Q$ ratio, we have chosen three values for the mass of the heavy-quark active in the loop, namely $0.5 \times M_{\tilde{H^0}}$, $M_{\tilde{H^0}}$ and $(m_t/m_{H^0}) \times M_{\tilde{H^0}}$. As can be seen from \ct{tab:Aval}, $m_Q=0.5 \times M_{\tilde{H^0}}$ renders ${A}_{a}$ slightly negative, $-0.147$, whereas it is large and positive, 2.28, for the SM $H^0$ with $m_H=125$~GeV and $m_t=173$~GeV. {The plotted $K^{\rm NLO}$ factors have been computed with the publicly available code {\tt ggHiggs} by Bonvini~\cite{Ball:2013bra,Bonvini:2014jma,Bonvini:2016frm} based on~\cite{Bonciani:2007ex,Harlander:2009my} for the NLO result with a finite heavy-quark mass in the loop. Other than this, we have run with its default setup.}

Let us first discuss the $\eta_Q$ results. \cf{fig:K-etaQ} gathers our result for the $K^{\rm NLO}$ factor computed at $y=0$ for $\eta_c$ (top) and  $\eta_b$ (down) and for the central eigenset of our 3 NLO PDF sets, namely PDF4LHC15\_nlo\_30 (left), JR14NLO08VF (middle) and NNPDF31sx\_nlonllx\_as\_0118 (right). We have used the conventional 7-point scale-choice values obtained by independently varying $\mu_R$ and $\mu_F$ by a factor of 2 about a default value which we simply chose here to be the mass of the quarkonium, $M_\Q$. We stress that LO cross sections used to compute $K^{\rm NLO}$ were obtained with the same PDF and  scale as those used for the NLO cross sections.  In addition, we have plotted $K^{\rm NLO}$ for $\mu_F=\hat{\mu}_F$ which we expect to provide the best behaviour. We have only plotted it for $\mu_R={\mu}_F$. 

We now discuss the qualitative features of the results. First, we note that, for PDF4LHC15\_nlo\_30, negative cross sections ($K^{\rm NLO}<0$) appear as expected as early as 1 TeV. This happens first for $\mu_F=M_\Q$ and $\mu_R=0.5 M_\Q$, then for $\mu_F=2 M_\Q$ and $\mu_R=M_\Q$, and then for $\mu_F=2 M_\Q$ and $\mu_R=2 M_\Q$, while $K^{\rm NLO}$ essentially converges to 0 for $\mu_F=2 M_\Q$ and $\mu_R=M_\Q$, which is also not acceptable. In short, all the results with $\mu_F$ equal or larger than the default choice are pathological and the situation is worsened by a lower value of $\mu_R$ which comes along with a larger size of $\alphaS$. On the other hand, for $\mu_F= 0.5 M_\Q$, $K^{\rm NLO}$ does not get negative, neither particularly small, but shows a peak at the top LHC energies which is related to the peak in the low scale gluon distribution at low scales encoded in PDF4LHC15\_nlo\_30.
So far, these results confirm the aforementioned past phenomenology.

\begin{figure*}[hbt!]
\centering
\subfloat[]{\includegraphics[width=0.66\columnwidth]{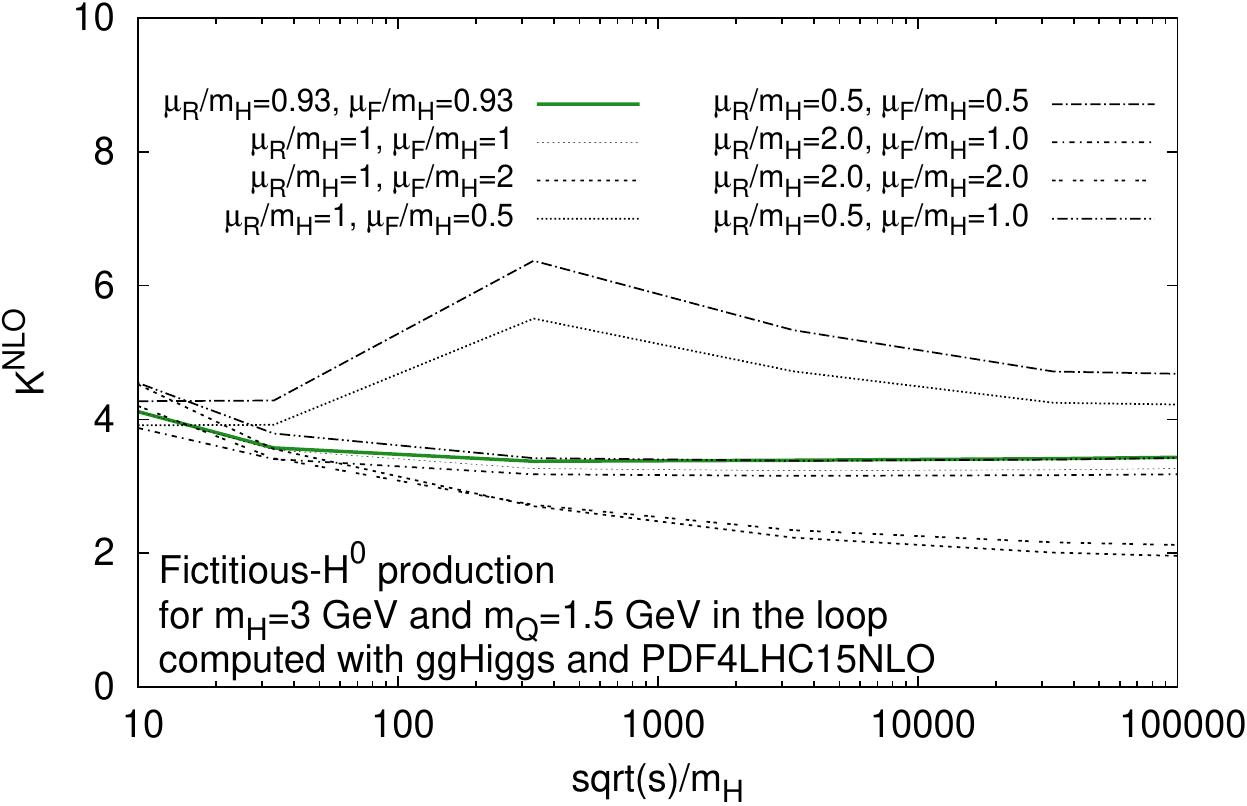}\label{fig:K-H0-3-1_51-PDF4LHC15_nlo_30}}
\subfloat[]{\includegraphics[width=0.66\columnwidth]{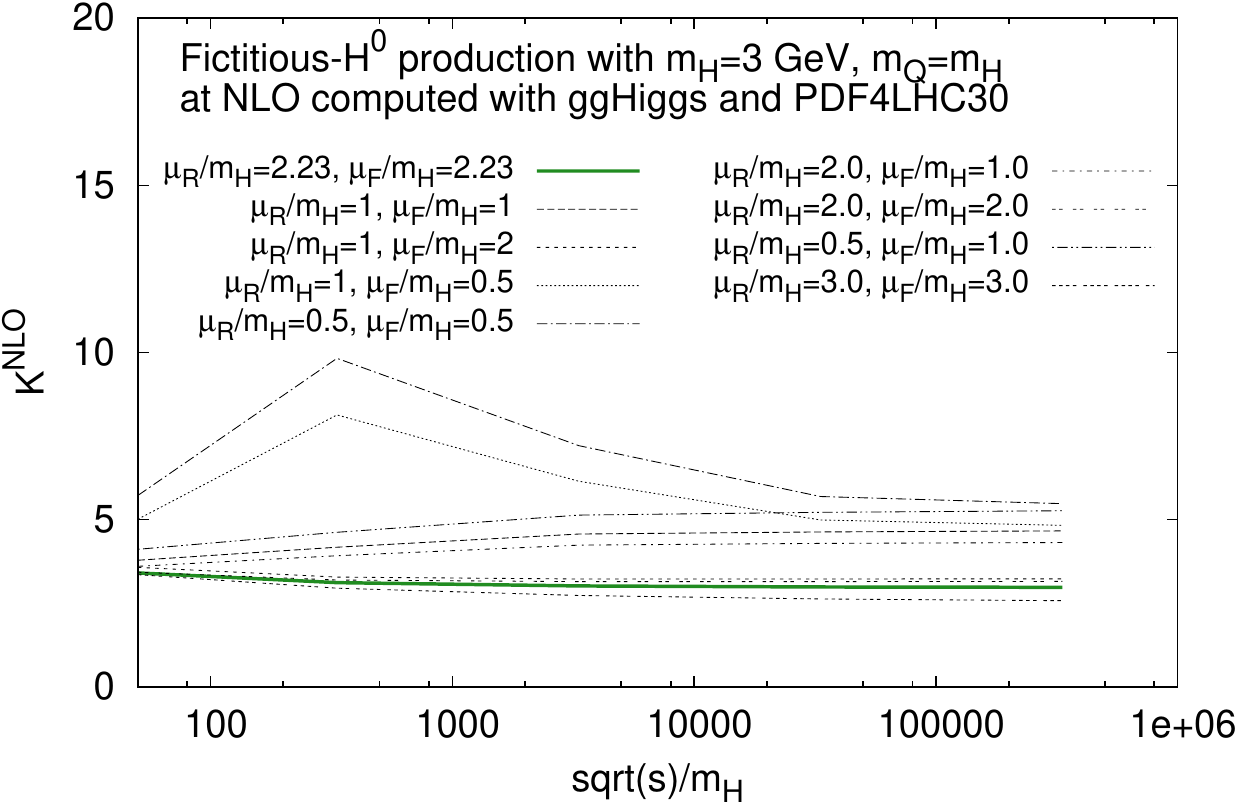}\label{fig:K-H0-3-3-PDF4LHC15_nlo_30}}
\subfloat[]{\includegraphics[width=0.66\columnwidth]{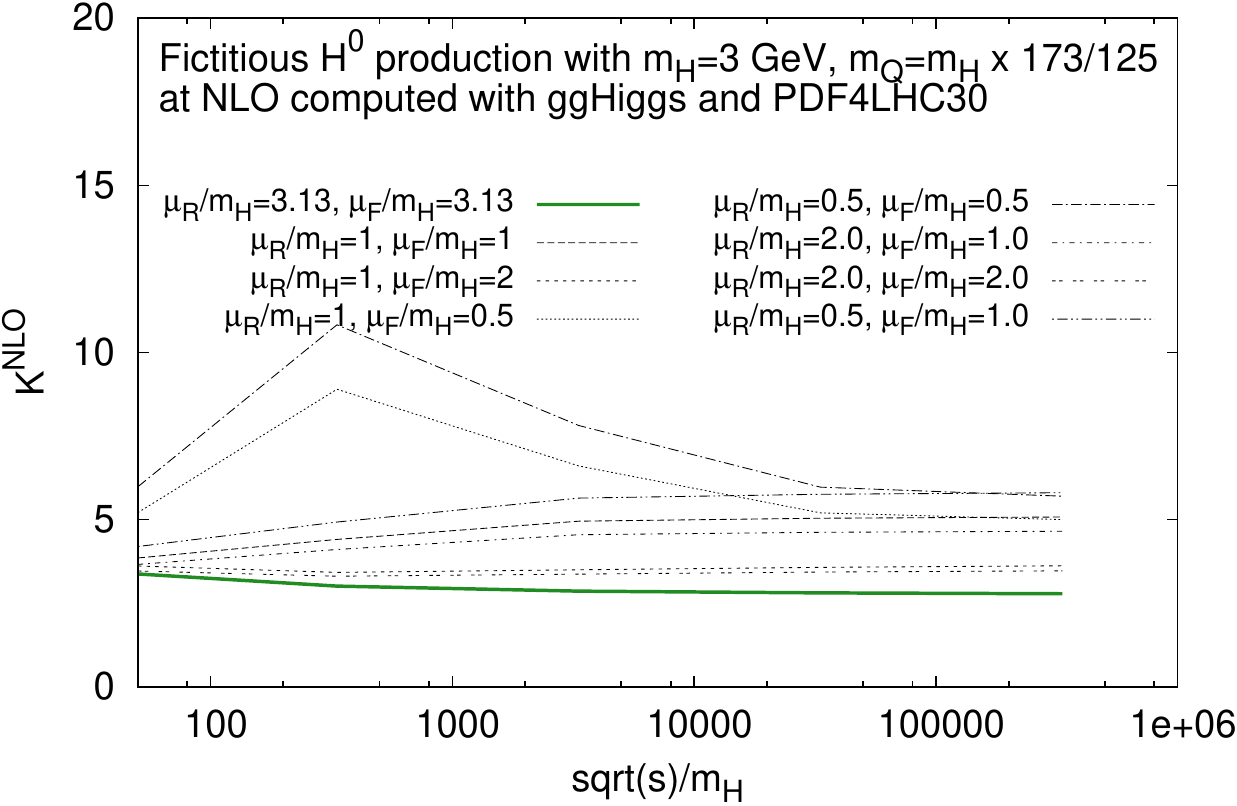}\label{fig:K-H0-3-4_15-PDF4LHC15_nlo_30}}
\\
\subfloat[]{\includegraphics[width=0.66\columnwidth]{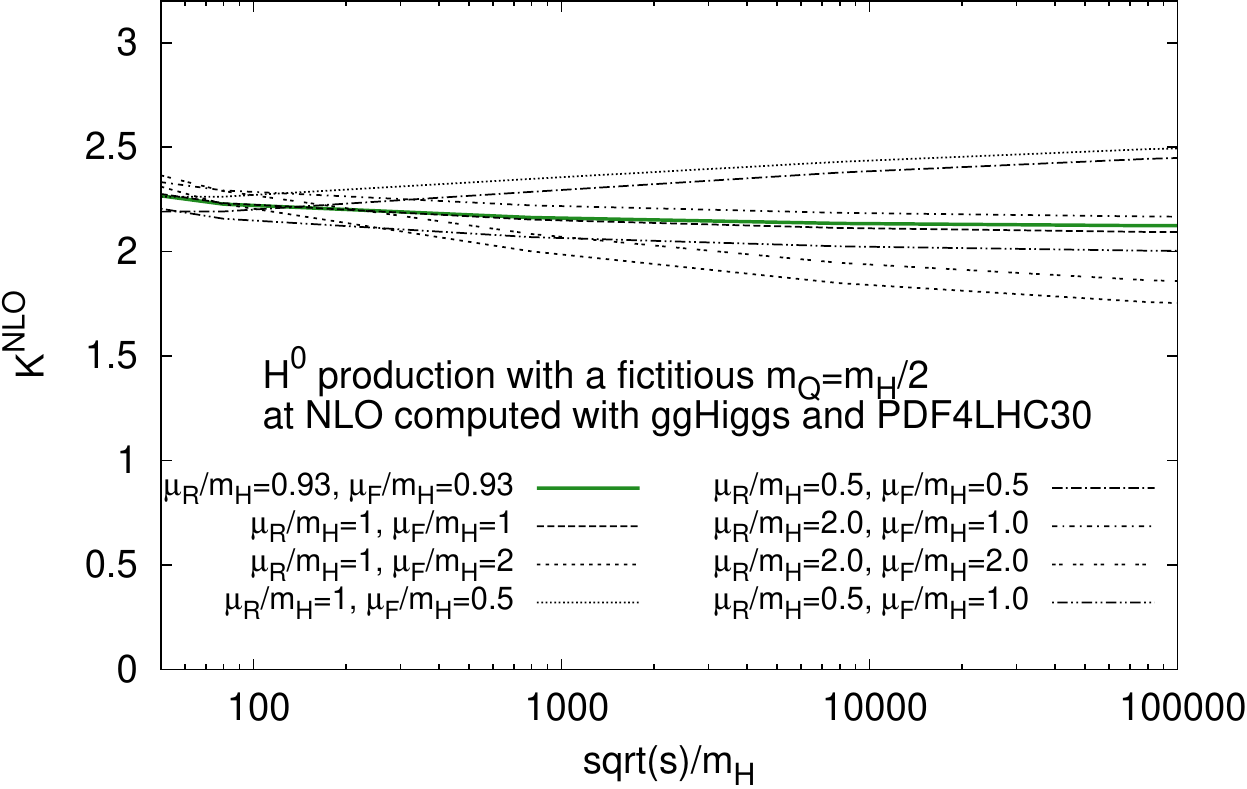}\label{fig:K-H0-125-63-PDF4LHC15_nlo_30}}
\subfloat[]{\includegraphics[width=0.66\columnwidth]{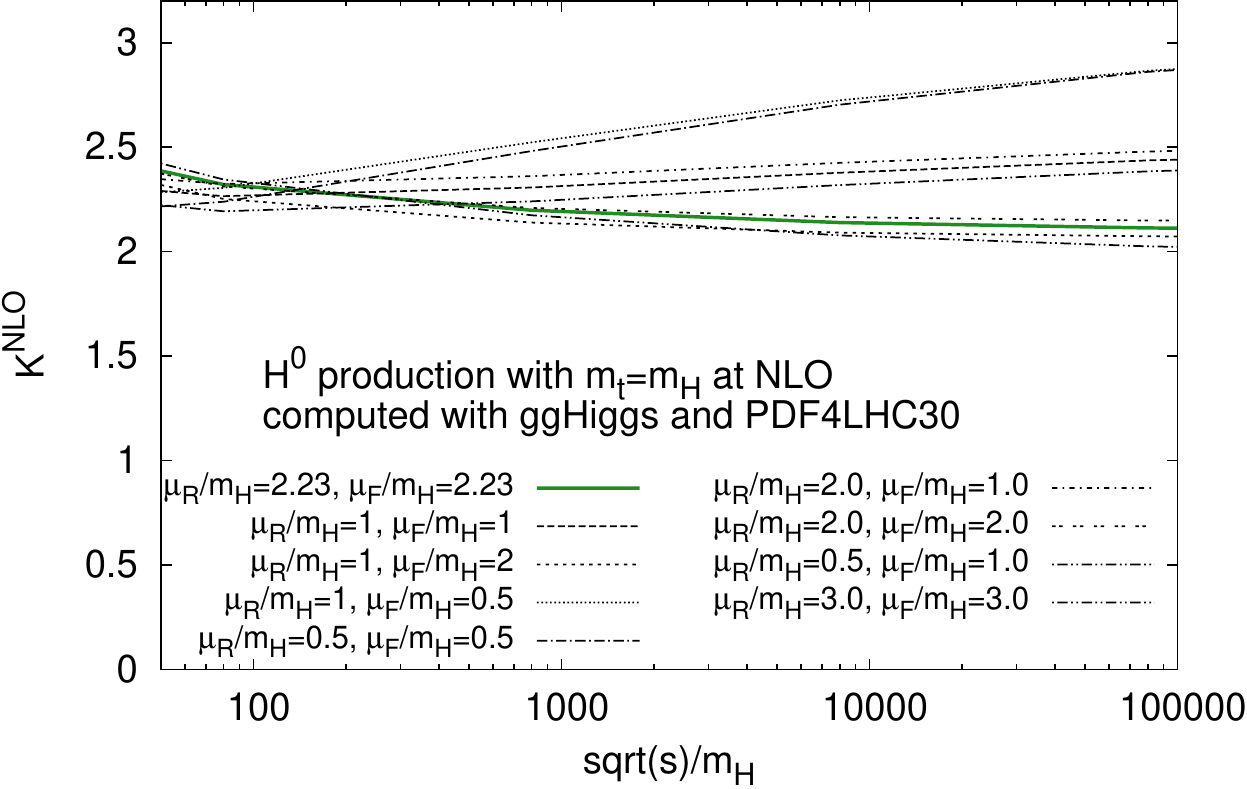}\label{fig:K-H0-125-125-PDF4LHC15_nlo_30}}
\subfloat[]{\includegraphics[width=0.66\columnwidth]{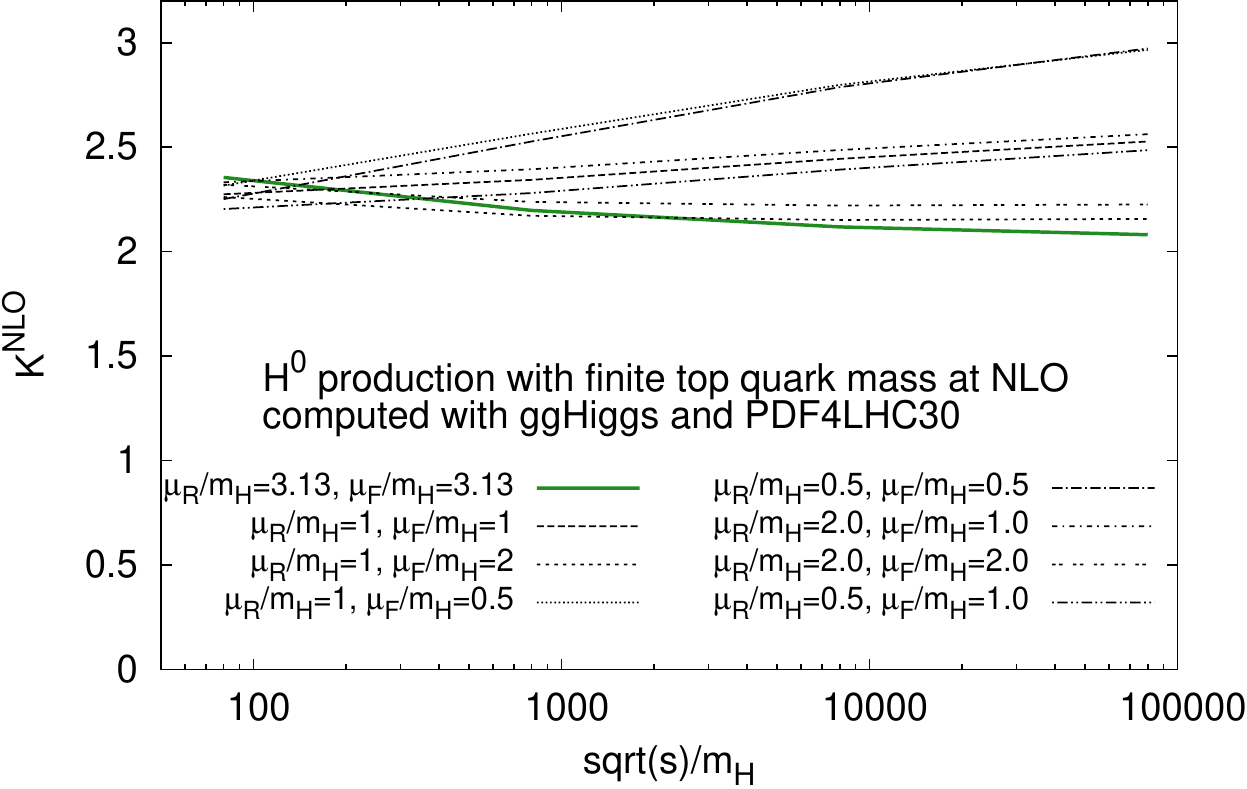}\label{fig:K-H0-125-173-PDF4LHC15_nlo_30}}
\caption{$K^{\rm NLO}$ for (top) fictitious $\tilde{H}^0$ and (bottom)  $H^0$ with different (fictitious) heavy-quark masses for PDF4LHC15\_nlo\_30 as a function of $\sqrt{s}/M_H$ for the usual 7-point scale choices and our $\hat\mu_F$ scale with $\mu_R=\mu_F$.
[Only the case (f) is realistic, all the other are academical examples.]}\label{fig:K-H0-PDF4LHC15_nlo_30}
\vspace*{-0.5cm}
\end{figure*}

On the other hand, when adopting our scale choice, $\mu_F=\hat{\mu}_F(=\mu_R)$, the behaviour is smooth and seems to slowly converge towards a constant value slightly above the unity. In fact, we have checked that varying $\mu_R$ at fixed $\mu_F=\hat{\mu}_F$ would simply shift the curve without changing its shape. This is connected to the fact that the limiting values is also driven by the threshold contribution at $z=1$ where the virtual corrections which are sensitive on $\mu_R$ sit. Now, summarising our result for a conventional PDF like PDF4LHC15\_nlo\_30, we can claim that the scale choice which we advocate provides a very simple solution to avoid pathological behaviour of the $P_T$-integrated $\eta_c$ cross sections at NLO. 

All the above observations can be made again for $\eta_b$ (see \cf{fig:PDF4LHC_K_etab_EnEvo}) but for the fact that the $K^{\rm NLO}$ factor does not get negative. Nonetheless, it gets so small for the large scale choices that the results remain meaningless. Presumably at $\sqrt{s}$ above those of a FCC, the cross section for $\mu_F=2 M_\Q$ and $\mu_R=M_\Q$ would turn negative. However, at such $\sqrt{s}$, we admit that it is rather an academic example. Yet, we stress that $K^{\rm NLO}$ varying by a factor of 10 from fixed-target energies to FCC energies is the sign of a bad convergence of the NLO computation for such scales.  Besides, we do not observe any more a peak $\mu_F= 0.5 M_\Q$, for which the energy dependence of $K^{\rm NLO}$ starts to be acceptable. Choosing $\mu_F=\hat{\mu}_F$ gives the best trend with a quasi constant value, close to 1, for 1 TeV and above. Such a choice completely stabilises the $K^{\rm NLO}$ energy dependence as it results that going to higher energies does not give an artificial importance to the $\alphaS^3$ corrections.

From the early studies of Schuler, Mangano and Petrelli one expects a strong sensitivity of the PDF shape on the impact of the NLO corrections (see also \cite{Ozcelik:2019qze}). We have checked that the PDF uncertainty on $K^{\rm NLO}$ derived from the 30 PDF4LHC15\_nlo\_30 eigensets is indeed smaller for $\mu_F=\hat{\mu}_F$ than for larger scales, despite the fact that the PDF uncertainty themselves usually decrease for growing scales. Actually to assess the PDF-shape sensitivity, it can more insightful to compare the trend with the central set of JR14NLO08VF  and NNPDF31sx\_nlonllx which show a clear different shape in particular close to 1.5 GeV (see \cf{fig:PDF-comparison}). These are respectively shown on \cf{fig:JR14NLO08VF_K_etac_EnEvo} \& \ref{fig:JR14NLO08VF_K_etab_EnEvo} and \cf{fig:NNPDF31sxNLONLL_K_etac_EnEvo} \& \ref{fig:NNPDF31sxNLONLL_K_etab_EnEvo}.  For $\eta_c$, the trend is very similar compared to what we obtained with PDF4LHC15\_nlo\_30 except for the absence of the peak for $\mu_F= 0.5 M_\Q$. As we wrote above, such a peak resulted from the local maximum and minimum  in the central PDF4LHC15\_nlo\_30 eigenset\footnote{Two effects can come into play here. First, the average momentum fraction of the gluons in the NLO contributions is slightly larger than for the LO one. As such, if the gluon PDF oscillates, it could happen that the PDFs product multiplying the $gg$ NLO partonic cross section could be larger than the LO one. Second, as we previously discussed, with flatter PDFs, ${\sigma}$ is in principle more sensitive to the large $\hat{s}$ behaviour of $\hat{\sigma}_{gg}$. For the considered scale, $\mu_F= 0.5 M_\Q$, this limit is positive, thus $K^{\rm NLO}$ is expected to get larger, precisely right after the bump in the PDF luminosity. It is likely that the latter effect actually dominates. Indeed, for $\mu_F=\hat{\mu}_F$, the limiting value of $\hat{\sigma}$ is set to 0 while the bump in $K^{\rm NLO}$ has nearly disappeared although there is still a slight bump in the PDF.}.

In conclusion, even with {\it a priori} the steepest possible gluon PDFs compatible with a global NLO PDF analysis, one gets negative or strongly suppressed NLO $\eta_c$ cross sections for a majority of the conventional scale choices (5 out of 7), whereas that obtained
with our scale choice $\mu_F=\hat{\mu}_F$ is remarkably stable. For $\eta_b$, the 3 PDFs essentially yield the same $K^{\rm NLO}$ factors which also shows the most stable behaviour for $\mu_F=\hat{\mu}_F$.

Having demonstrated the efficiency of our scale choice to avoid anomalously large NLO corrections to pseudoscalar quarkonium production attributed to an over-subtraction of the collinear divergence inside the PDFs, let us now investigate whether it works for elementary scalar bosons coupling to gluons via heavy quarks. If our argumentation is correct, $K^{\rm NLO}$ should, first, be rather $\mu_F$- and $\sqrt{s}$-dependent for a scalar boson of similar mass than the $\eta_c$ and, second, become stable for $\mu_F=\hat{\mu}_F$. 

{Our results, shown on \cf{fig:K-H0-3-1_51-PDF4LHC15_nlo_30}, exactly confirms our expectation for $\tilde{H}^0$ with $M_{\tilde{H}^0}=3$~GeV and $m_Q=1.5$~GeV. If we were to work at even larger $\sqrt{s}$, the larger $\mu_F$ choices would eventually yield very small $K^{\rm NLO}$. They would probably not become negative but we recall that $|{A}_{a}|$ is smaller for $\tilde{H}^0$ than for $\eta_c$ rendering the HE limit slightly less harmful. On the other hand, too small scales yield strongly growing $K^{\rm NLO}$ at large $\sqrt{s}$. Finally, setting $\mu_F=\hat{\mu}_F$, or close to it  with $\mu_F=M_{\tilde{H}^0}$ since $\hat{\mu}_F = 0.93 M_{\tilde{H}^0}$, gives remarkably stable $K^{\rm NLO}$. We consider this to be a confirmation that the instabilities in NLO computations of quarkonium production are not connected to the modelling of quarkonium production.} The situation is equally good if we set consider, as an academic example,  $M_{{H}^0}=125$~GeV and $m_Q=M_{\tilde{H}^0}/2$. Indeed, $\mu_F=\hat{\mu}_F$
yields the most stable results on \cf{fig:K-H0-125-63-PDF4LHC15_nlo_30}.

\begin{figure*}[h!]
\centering
\subfloat[]{\includegraphics[width=0.66\columnwidth]{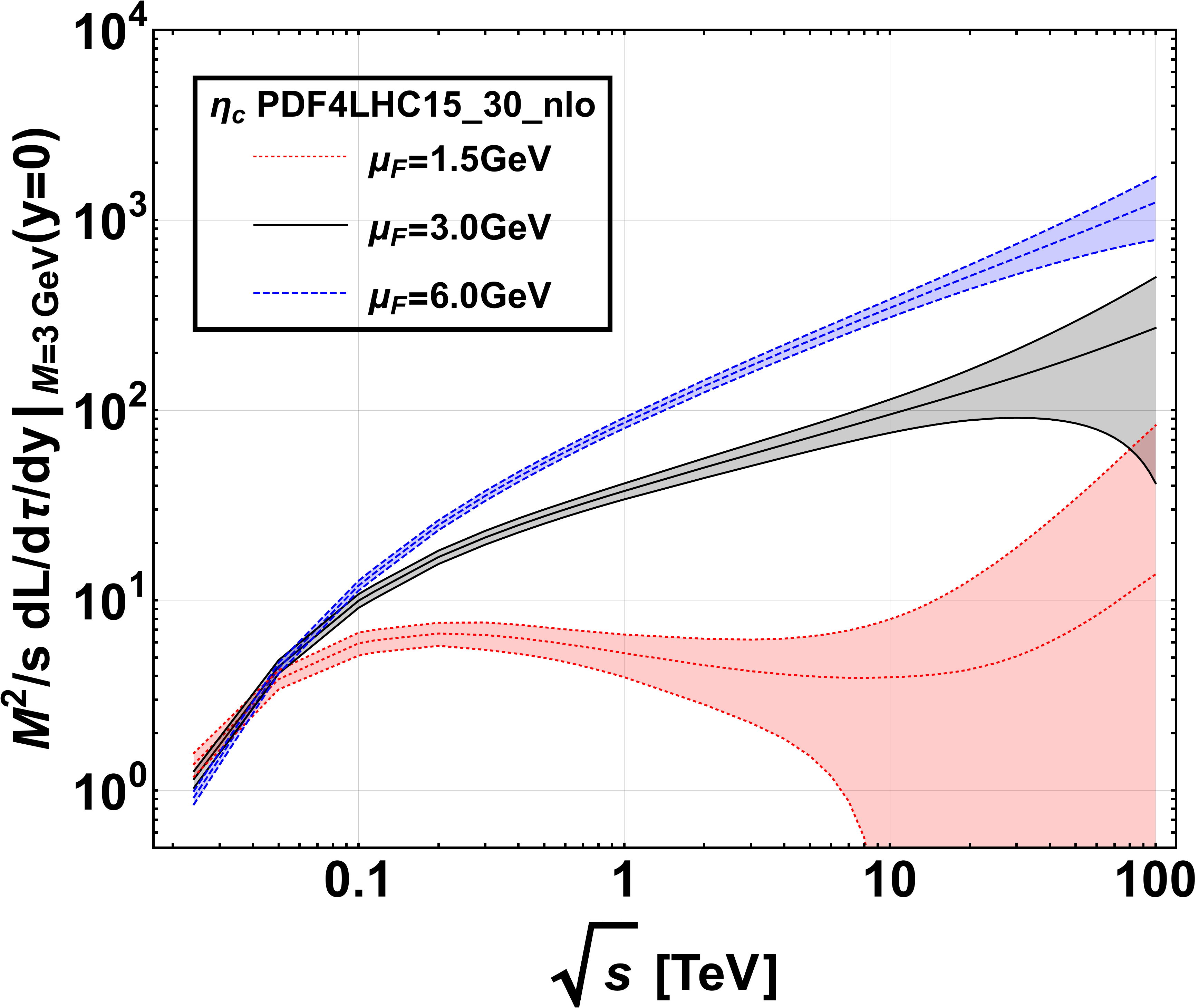}\label{fig:PDF4LHC_dLdtau_etac_EnEvo_Uncertainty}}
\subfloat[]{\includegraphics[width=0.66\columnwidth]{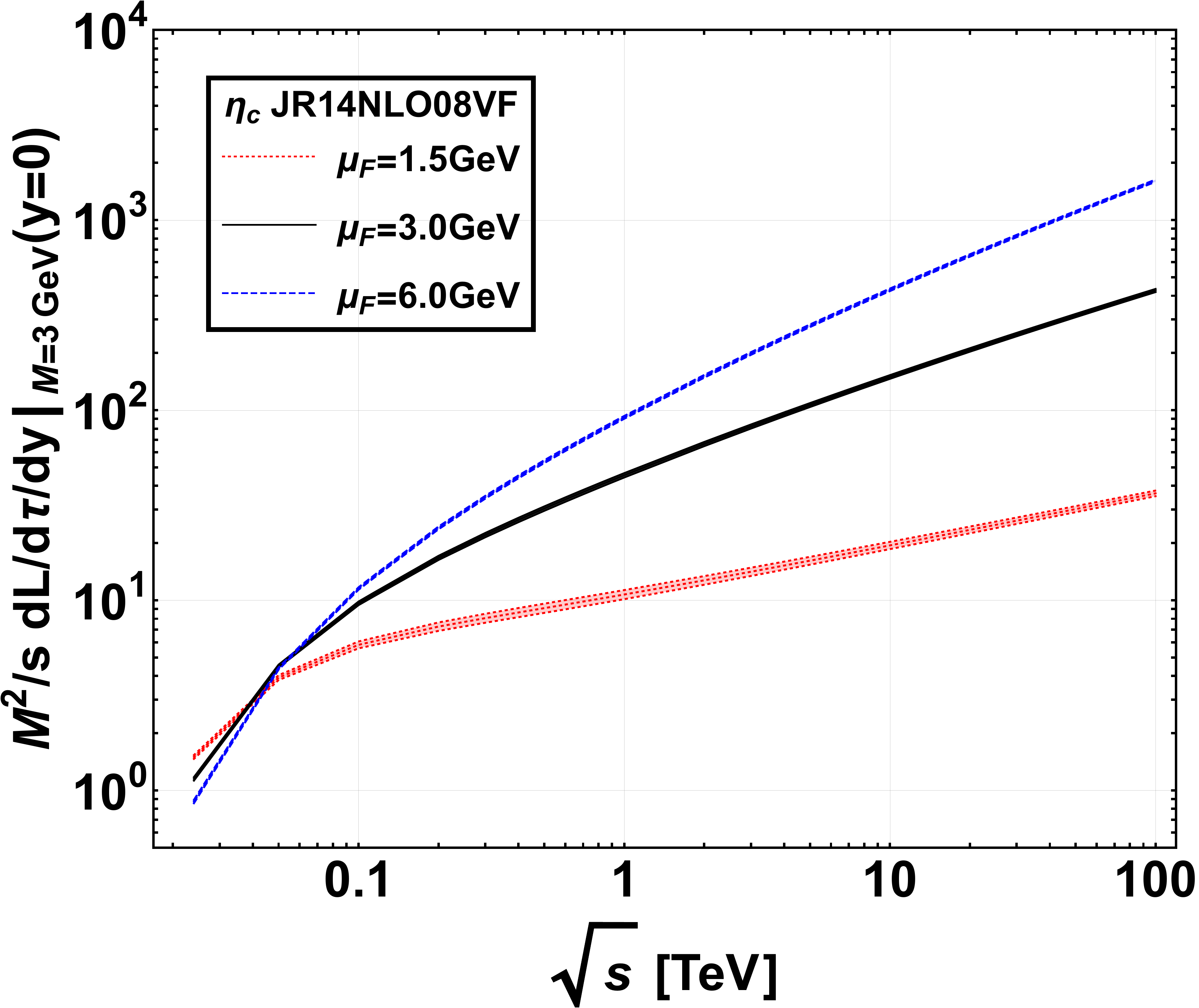}\label{fig:JR14NLO08VF_dLdtau_etac_EnEvo_Uncertainty}}
\subfloat[]{\includegraphics[width=0.66\columnwidth]{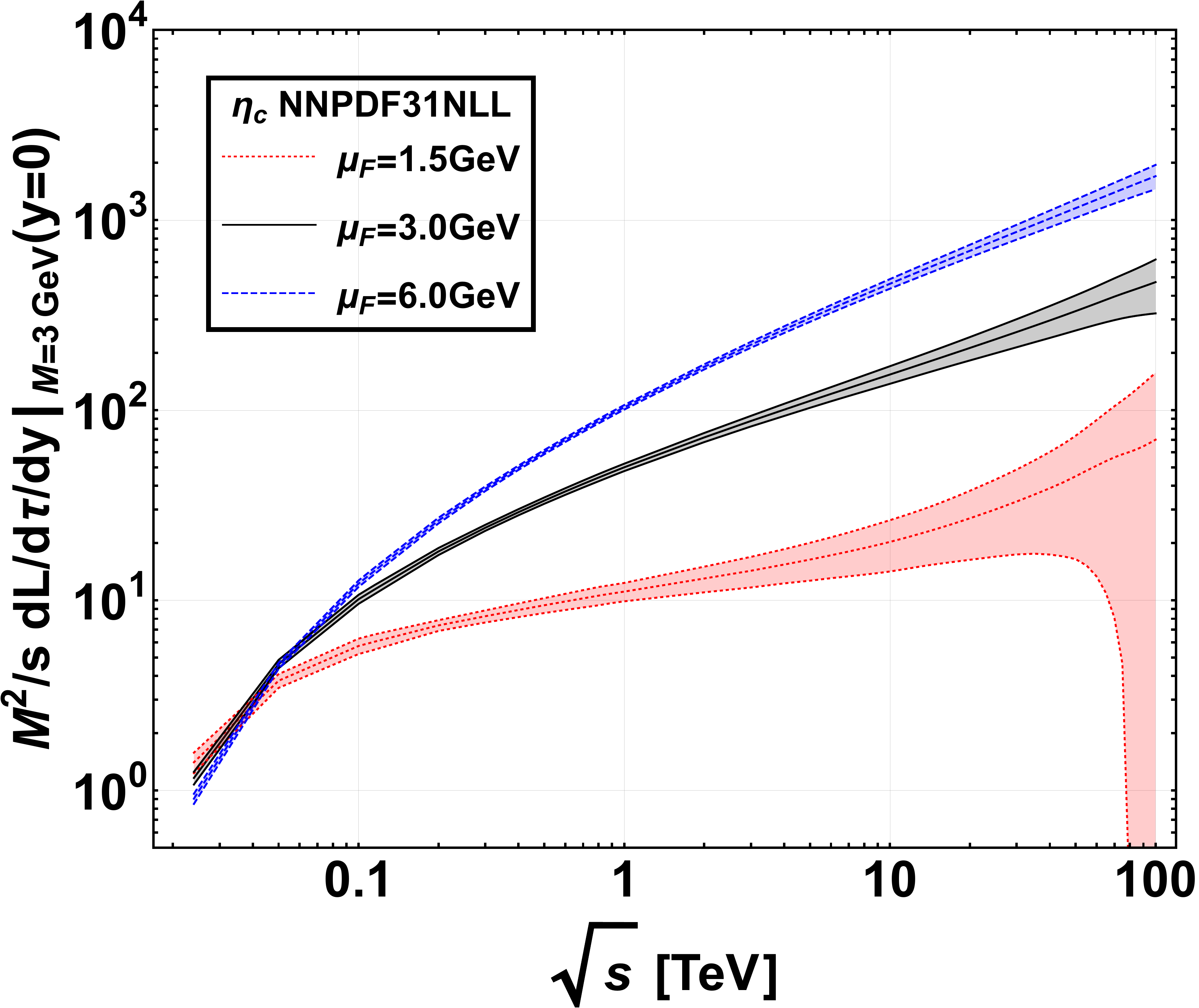}\label{fig:NNPDF31sxNLONLL_dLdtau_etac_EnEvo_Uncertainty}}\vspace*{-0.5cm}
\\
\subfloat[]{\includegraphics[width=0.66\columnwidth]{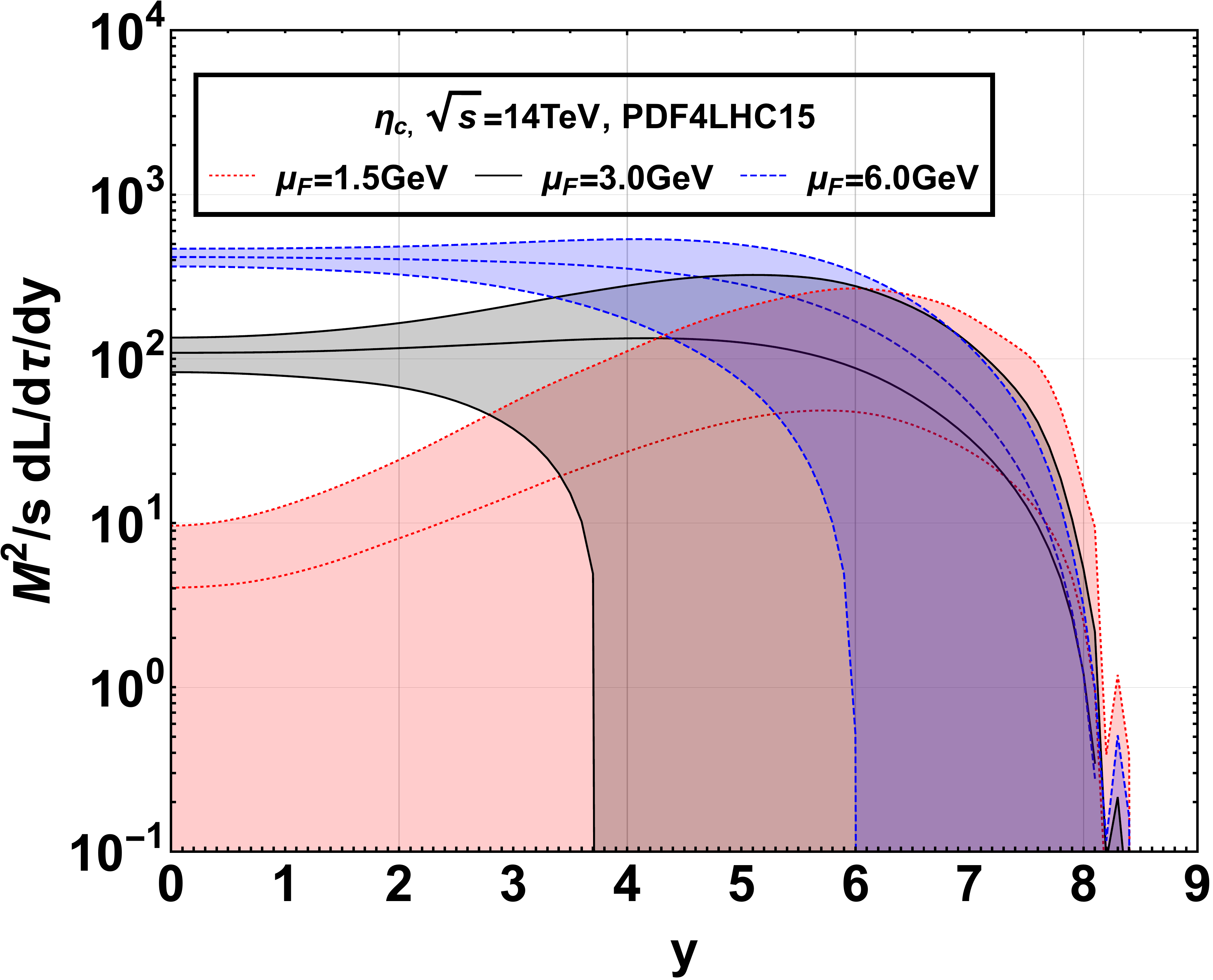}\label{fig:PDF4LHCplots/PDF4LHC_dLdy_etac_Rap14TeV_Uncertainty}}
\subfloat[]{\includegraphics[width=0.66\columnwidth]{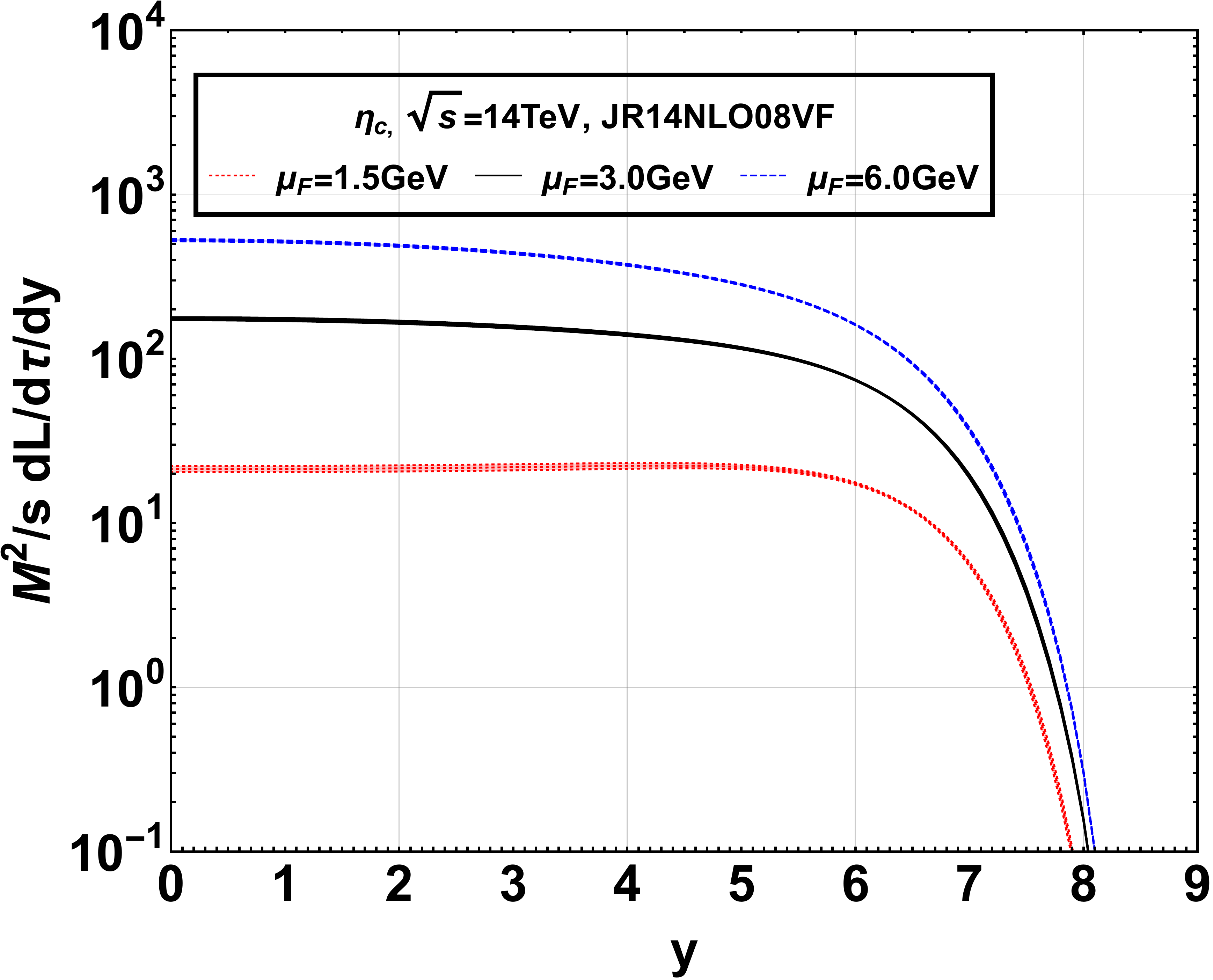}\label{fig:JR14NLO08VF_dLdy_etac_Rap14TeV_Uncertainty}}
\subfloat[]{\includegraphics[width=0.66\columnwidth]{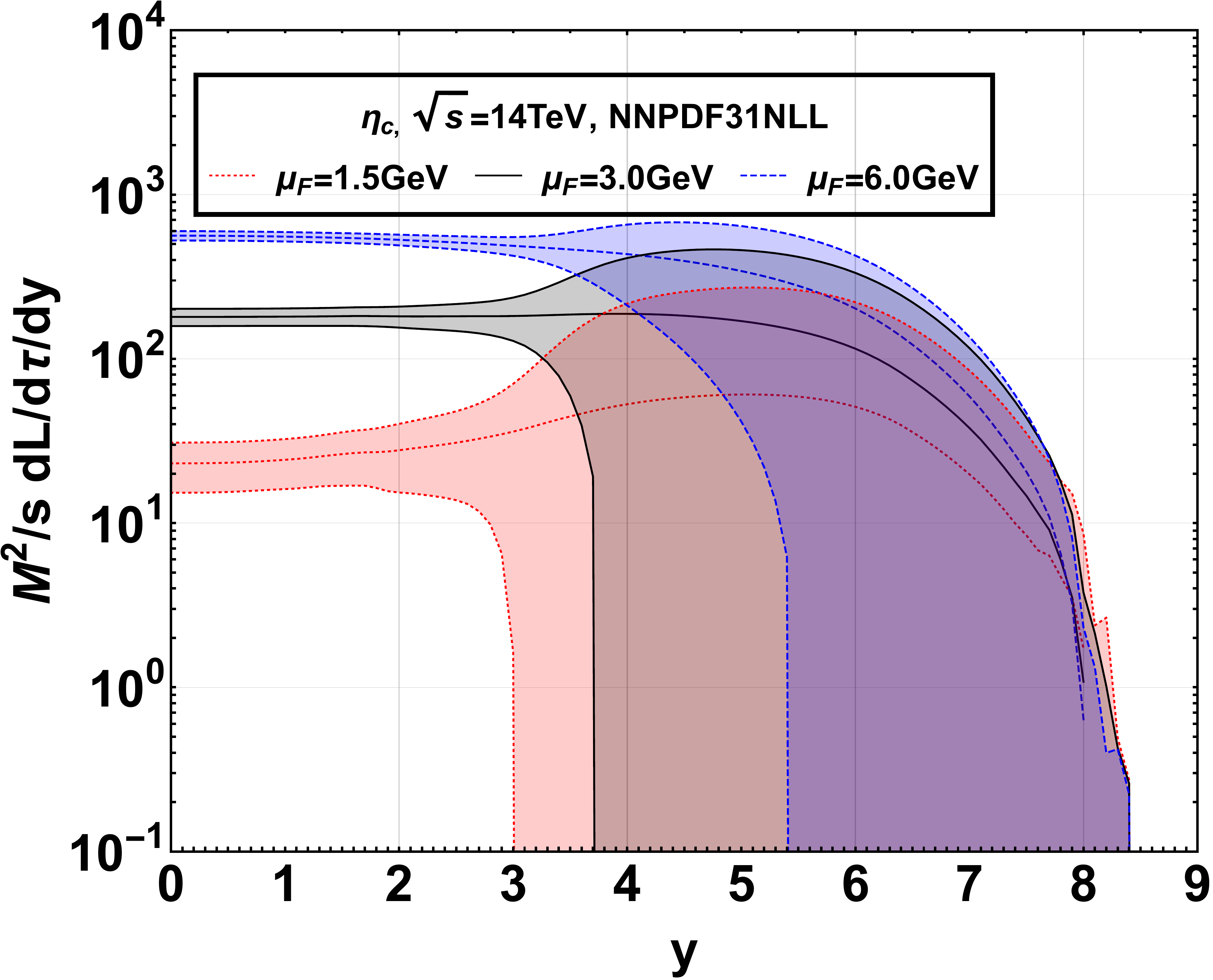}\label{fig:NNPDF31sxNLONLL_dLdy_etac_Rap14TeV_Uncertainty}}
\caption{$\tau_0 \frac{d\mathcal{L}}{d\tau dy}$ as function of energy $\sqrt{s}$ and at $y=0$ (top) and $\frac{d\mathcal{L}}{d\tau dy}$ as function of $y$ at $\sqrt{s}=14$ TeV (bottom) for $M=3$~GeV  (for PDF4LHC15\_nlo\_30 (left), JR14NLO08VF (middle) and NNPDF31sx\_nlonllx\_as\_0118 (right))  for 3 $\mu_F$ values ($0.5 M$, $M$ and $2 M$).}\label{fig:dLdtau_etac_Uncertainty}
\vspace*{-0.5cm}
\end{figure*}

{
On the other side, for the real $H^0$ case with $m_t=173$~GeV --for which the situation is of course not problematic--, we observe on \cf{fig:K-H0-125-173-PDF4LHC15_nlo_30} that $K^{\rm NLO}$ tends to clearly increase with $\sqrt{s}$ for the smaller scale choices, like ${\mu}_F = 0.5 M_{{H}^0}$. Choosing $\hat{\mu}_F = 2 M_{{H}^0}$ or ${\mu}_F = \hat{\mu}_F = 3.3 M_{{H}^0}$ yield to more stable trends. In principle, we would expect ${\mu}_F = \hat{\mu}_F$ to yield the most stable curve. More investigations are needed to understand why ${\mu}_F = 2 M_{{H}^0}$ shows the best trend. One observes the same for $M_{\tilde{H}^0}=3$~GeV and the same $m_Q/M_{\tilde{H}^0}$ ratio on \cf{fig:K-H0-3-4_15-PDF4LHC15_nlo_30}. \cf{fig:K-H0-3-3-PDF4LHC15_nlo_30} and \cf{fig:K-H0-125-125-PDF4LHC15_nlo_30}, in between both cases, hint at an effect which would scale like $m_Q$. A possible explanation could come from contributions of the box diagrams which would yield $\bar{A}_g(z)\neq \bar{A}_g(z)$ down to very low $z$.

Yet, the phenomenology of $H^0$ being now  made at N$^3$LO accuracy~\cite{Anastasiou:2015ema,Anastasiou:2016cez} in the infinite top-quark-mass limit, it is rather an academical question. For the validation of our scale proposal, the success of the light $\tilde{H}^0$ case with $M_{\tilde{H}^0}=3$~GeV and $m_Q=1.5$~GeV, with strongly decreasing $K^{\rm NLO}$ for large $\mu_F$ and
a very stable one for $\mu_F=\hat{\mu}_F$ is much more telling since it perfectly confirms what we observed with the $\eta_c$.}


\subsection{A word on gluon luminosities at NLO and low scales }
\label{sec:Gluon-luminosities}

\begin{figure*}[hbt!]
\centering
\subfloat[]{\includegraphics[width=0.66\columnwidth]{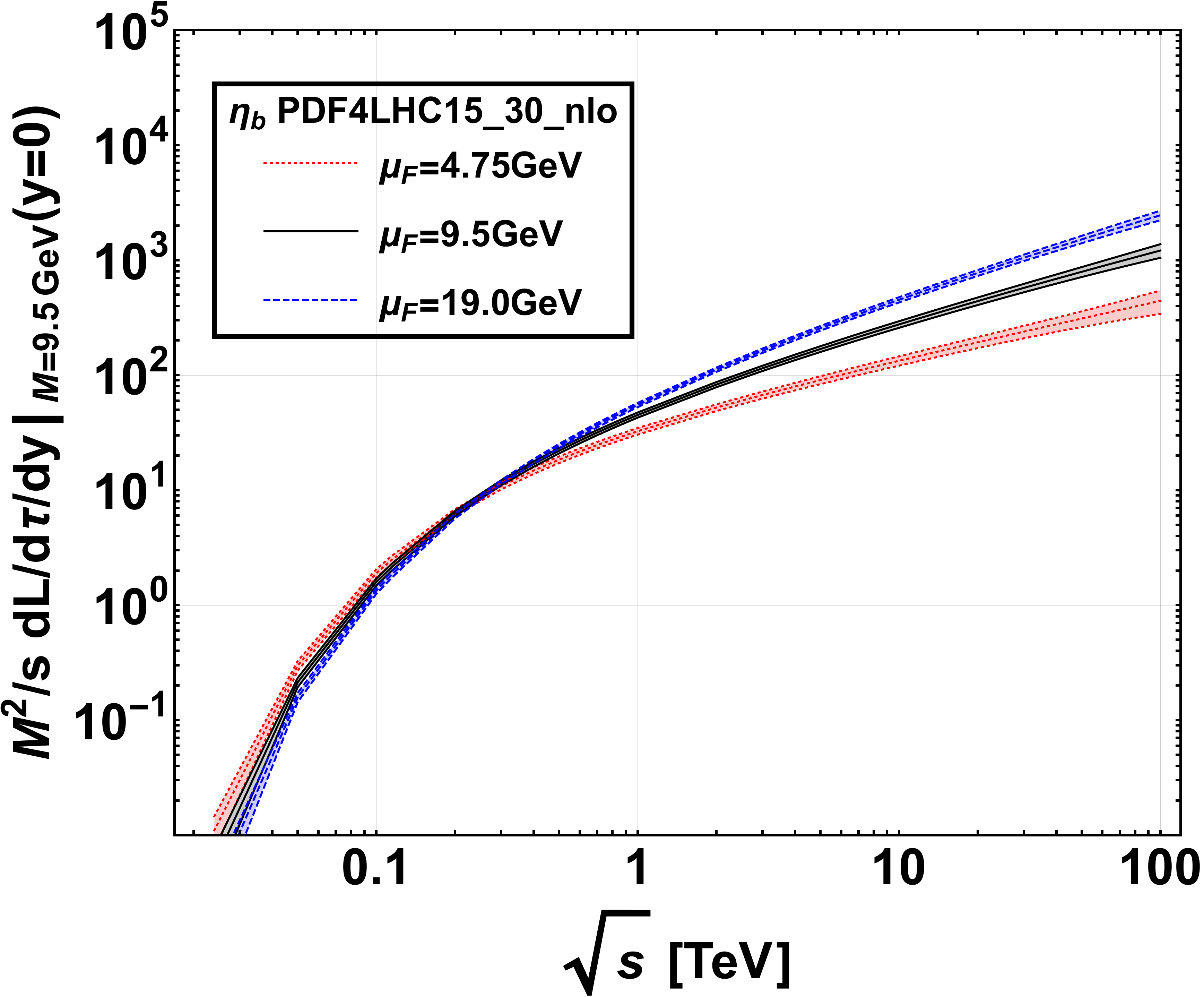}\label{fig:PDF4LHC_dLdtau_etab_EnEvo_Uncertainty}}
\subfloat[]{\includegraphics[width=0.66\columnwidth]{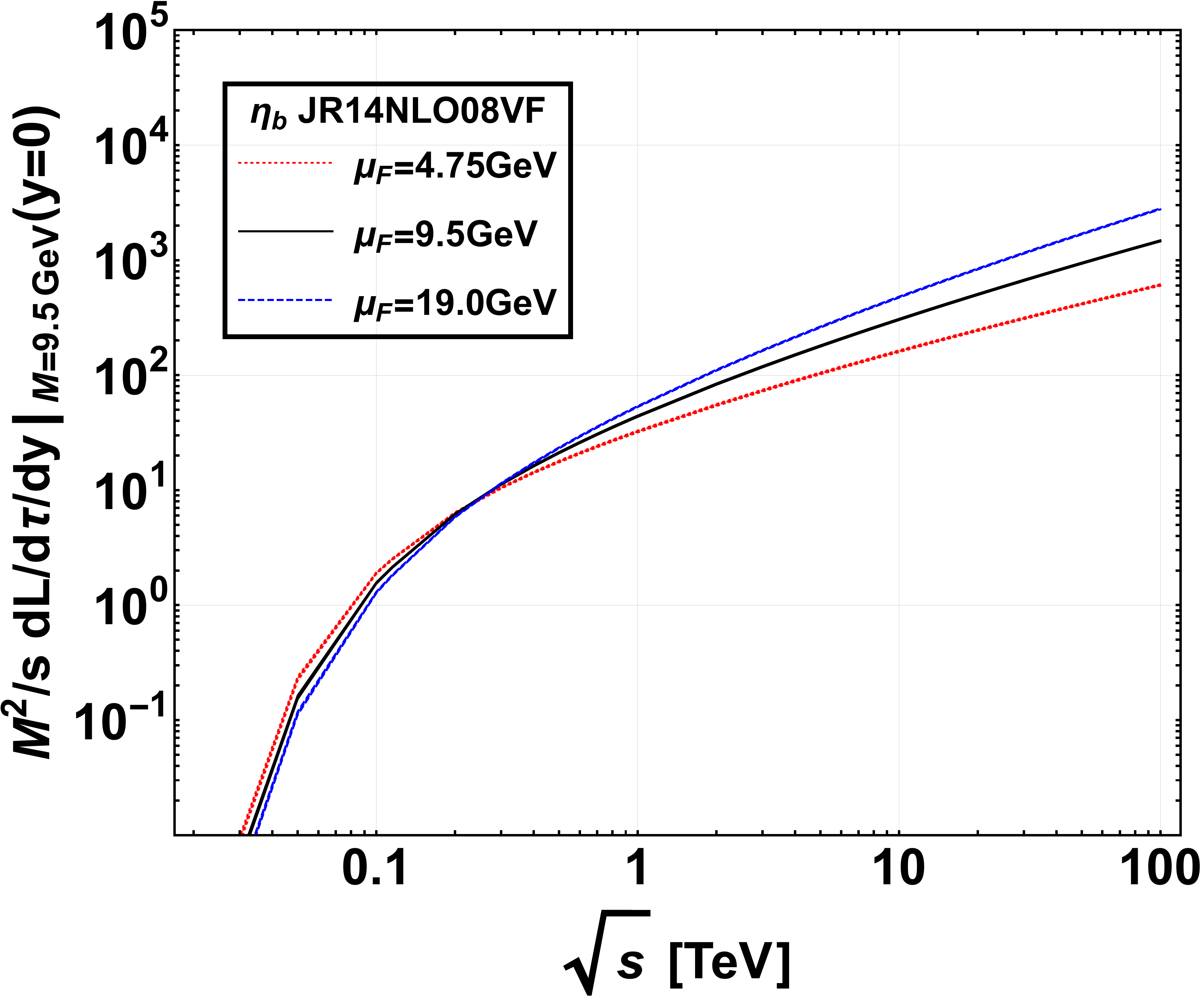}\label{fig:JR14NLO08VF_dLdtau_etab_EnEvo_Uncertainty}}
\subfloat[]{\includegraphics[width=0.66\columnwidth]{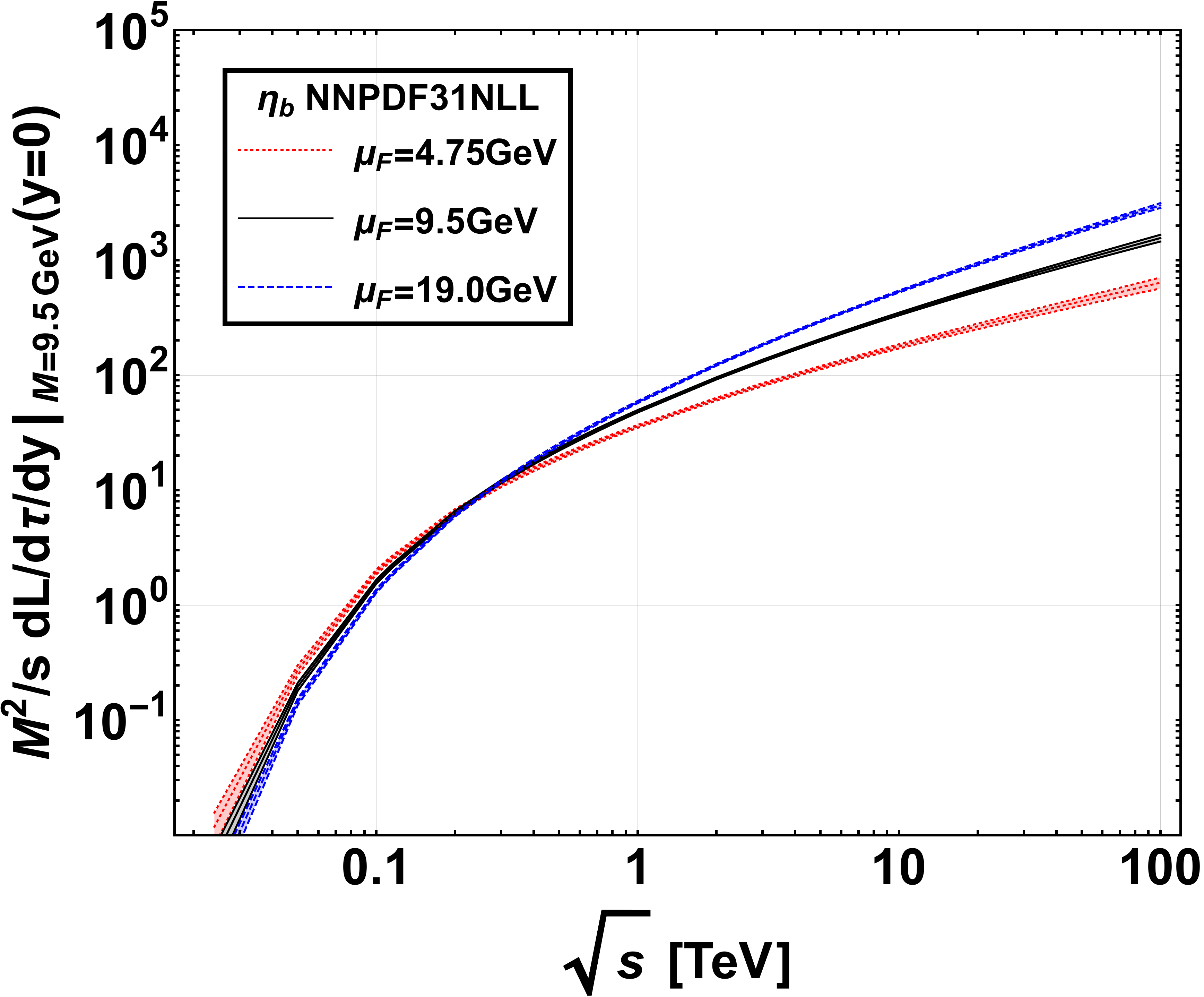}\label{fig:NNPDF31sxNLONLL_dLdtau_etab_EnEvo_Uncertainty}}\vspace*{-0.5cm}
\\
\subfloat[]{\includegraphics[width=0.66\columnwidth]{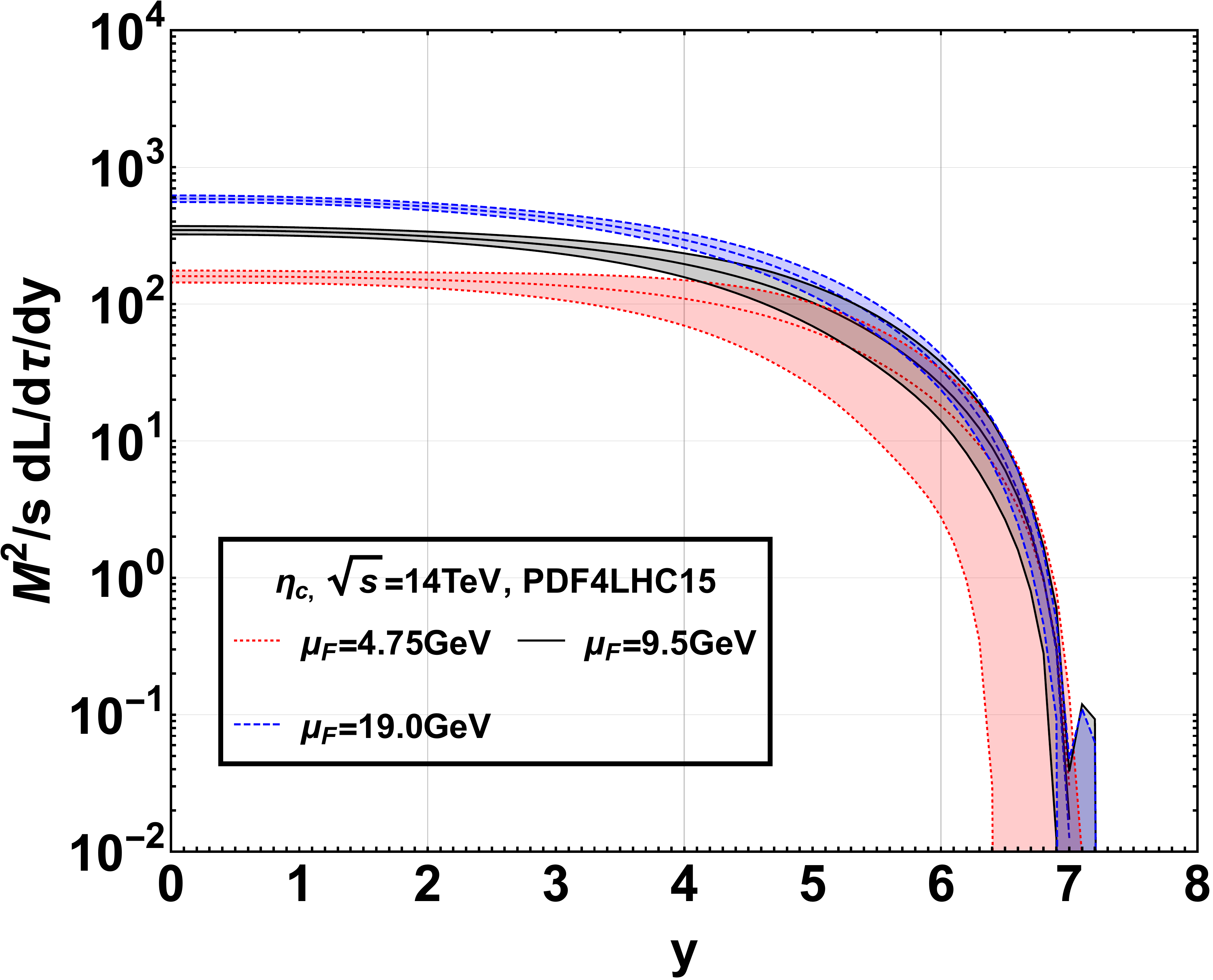}\label{fig:PDF4LHCplots/PDF4LHC_dLdy_etab_Rap14TeV_Uncertainty}}
\subfloat[]{\includegraphics[width=0.66\columnwidth]{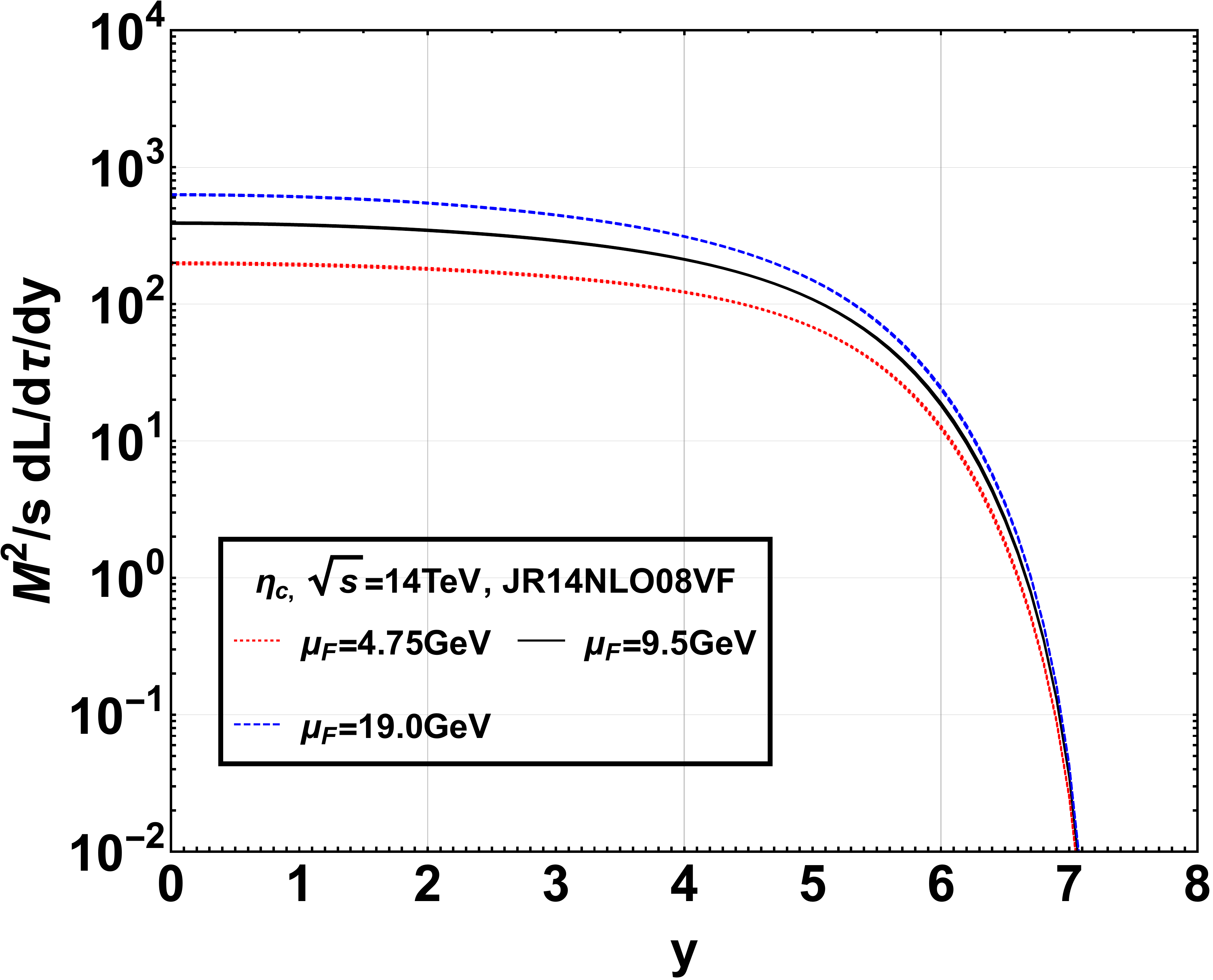}\label{fig:JR14NLO08VF_dLdy_etab_Rap14TeV_Uncertainty}}
\subfloat[]{\includegraphics[width=0.66\columnwidth]{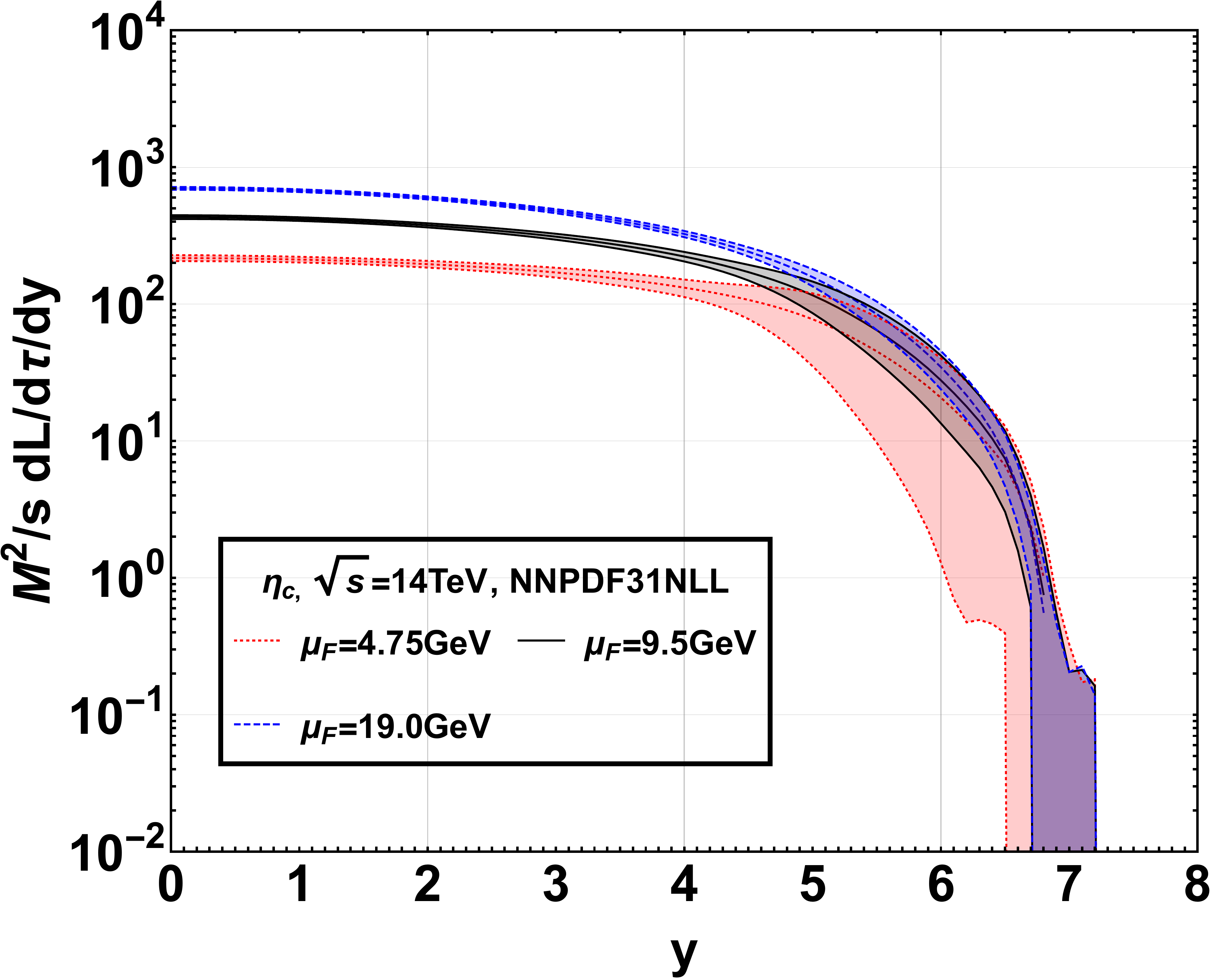}\label{fig:NNPDF31sxNLONLL_dLdy_etab_Rap14TeV_Uncertainty}}
\caption{$\tau_0 \frac{d\mathcal{L}}{d\tau dy}$ as function of energy $\sqrt{s}$ and at $y=0$ (top) and $\frac{d\mathcal{L}}{d\tau dy}$ as function of $y$ at $\sqrt{s}=14$~TeV (bottom) for $M=9.5$~GeV  (for PDF4LHC15\_nlo\_30 (left), JR14NLO08VF (middle) and  NNPDF31sx\_nlonllx\_as\_0118 (right)) for 3 $\mu_F$ values ($0.5 M$, $M$ and $2 M$).).}\label{fig:dLdtau_etab_Uncertainty}
\vspace*{-0.5cm}
\end{figure*}

Before moving to our NLO predictions for the cross sections, we find it useful to make a short digression on the gluon luminosities. Indeed, we would not want  the reader to be confused by some unnatural cross-section behaviours as a function of $\sqrt{s}$ or $y$ which we will show and to attribute these behaviours to QCD corrections to the hard scattering.  Contrary to $K^{\rm NLO}$ where the PDF impact is indirect\footnote{Yet, we have observed a bump in $K^{\rm NLO}$ because of the changing shape of PDF4LHC15\_nlo\_30 for low scales.}  because of a large cancellation of their effects in the ratio, the $gg$ contribution to the hadronic cross section will essentially be proportional to the square of gluon PDFs at low scales. As we have seen when discussing \cf{fig:PDF-comparison}, the conventional PDF sets typically exhibit a local minimum below 0.001 at scales below 2 GeV. In this region, however, the gluon PDFs are only poorly constrained by scarce data sensitive to gluons at smaller $x$ and larger scales.

To illustrate the typical effects that such a shape can induce on both the $\sqrt{s}$ or $y$ distribution on a low-scale process, we have plotted on \cf{fig:dLdtau_etac_Uncertainty} and \cf{fig:dLdtau_etab_Uncertainty} the corresponding  differential gluon luminosity which would normally multiply a simple $gg$ fusion process at LO. We have done so for our 3 chosen PDF sets, for 2 masses $M$  (3 and 9.5 GeV) and  3 corresponding scales $\mu_F$ ($0.5 M$, $M$ and $2 M$) as a function of $\sqrt{s}$ and $y$ in the ranges which will correspond to the NLO cross section plots which we will show in the next sections. 

It clearly appears that for PDF4LHC15\_nlo\_30, which is representative of usual PDF sets, and to a lesser extent for NNPDF31sx\_nlonllx\_as\_0118, both the $\sqrt{s}$ dependence at $y=0$ and the $y$ dependence at $\sqrt{s}=14$~TeV will strongly be distorted for $\mu_F=1.5$~GeV. Not only one observes a strong scale sensitivity in the luminosity magnitude\footnote{In a sense, this is acceptable since $\mu_F/\mu_{F,0}$ (with $\mu_{F,0}$ being where the evolution starts)  significantly varies between the $\mu_F$ values we took. Hence, the evolution can generate significantly more gluons at our larger scale choice.}, but the distributions are very different. More importantly, it is very improbable that any measured differential cross sections, even at low scales, would follow the trend of \cf{fig:NNPDF31sxNLONLL_dLdy_etac_Rap14TeV_Uncertainty} (and \cf{fig:PDF4LHCplots/PDF4LHC_dLdy_etac_Rap14TeV_Uncertainty}) with a yield showing a {\it global} maximum around $y=5$ at the LHC or a differential cross section at $y=0$ essentially constant between the Tevatron and the top LHC energy like on~\cf{fig:PDF4LHC_dLdtau_etac_EnEvo_Uncertainty}. As we will see, the corresponding NLO cross sections will be driven by this behaviour of the gluon luminosity.

\subsection{A digression on the $\eta_b$ detectability}
Having now at our disposal reliable NLO predictions for $\eta_b$ cross sections at hadron colliders, let us
address the question of the feasibility of such studies and in particular of the extraction of cross sections. 
Whereas prompt $\eta_c$ production at the LHC has now been the object of two experimental studies~\cite{Aaij:2014bga,Aaij:2019gsn} by the LHCb collaboration, the prospects for $\eta_b$ production studies are however less clear.

Compared to the $\eta_c$, we do not know much on the $\eta_b$ properties which was only discovered in 2008 by {\sc BaBar}~\cite{Aubert:2008ba} and whose mass is 9.4~GeV, while its $2S$ excitation, the $\eta_b(2S)$, was discovered in 2012 by Belle~\cite{Mizuk:2012pb} with a mass of 10.0~GeV.
The $\eta_b$ in fact has so far been observed only in $e^+e^-$ annihilations, by {\sc BaBar}~\cite{Aubert:2008ba,Aubert:2009as}, CLEO~\cite{Bonvicini:2009hs} and Belle~\cite{Mizuk:2012pb,Tamponi:2015xzb}. Most likely, future measurements will be carried out by Belle II~\cite{Kou:2018nap}. Its width has been measured to be on the order of 10 MeV. We note the good agreement with the LO estimate 
for $\Gamma(\eta_b \to gg) ={8\alphaS^2 |R_0|^2}/{3M_{\eta_b}^2}$ assuming
$\Gamma(\eta_b) \sim \Gamma(\eta_b \to gg)$. The agreement is confirmed up to NNLO~\cite{Feng:2017hlu}.
So far no measurement of any branching fractions are reported in the PDG~\cite{Zyla:2020zbs} and it is thus not clear in which decay channel it could be measured in hadroproduction. In addition, one needs to know the branching value with an acceptable uncertainty to derive a cross section to test state-of-the-art computations which do not address the decay but only the production.

Different theoretical ideas  have been pushed forward about the usable decay channels at the Tevatron and the LHC. For a long time, the decay into a pair of $J/\psi$ potentially clearly signalled by 4-muon events was considered to be a discovery channel in the busy environment of $pp$ collisions. Even though, from the beginning, physicists were aware that the branching fraction into this channel could be very small, we should stress here that the production cross sections  for $\eta_b$ at colliders are not small at all, as they are comparable to those for $\Upsilon(nS)$ which are routinely studied at the LHC. As such, small branching fractions could still yield observable rates. First estimations~\cite{Braaten:2000cm} reported\footnote{Note that this should then be multiplied by the square of $\Br(J/\psi\to \ell^+ \ell^-)$, \ie\ 6 \%.} $\Br(\eta_b \to J/\psi+J/\psi) = 7 \times 10^{-4\pm1}$,  using mass-rescaling arguments applied to $\eta_c \to \phi \phi$. However, this estimate was then questioned and searches via the detection of 2 charmed mesons were suggested~\cite{Maltoni:2004hv}. Actual computations based on NRQCD~\cite{Hao:2006nf,Gong:2008ue} later yielded a much smaller $\Br(\eta_b \to J/\psi+J/\psi)$, as low as $5 \times 10^{-8}$. It was however suggested that final-state interactions, beyond the effects included in NRQCD computations, could enhance the di-$J/\psi$ decay width by up 2 orders of magnitude~\cite{Santorelli:2007xg}. It is thus clear that until $\Br(\eta_b \to J/\psi+J/\psi)$  is actually measured elsewhere, it could not provide a way to derive cross-section measurements. Yet, given the current intense activity in $J/\psi+J/\psi$ studies with the observation~\cite{Aaij:2020fnh} of a di-$J/\psi$ resonance, a search for this decay channel at the LHC could still be an option as it may also be looked for at Belle II. Another channel subject of debates is that into 2 charmed mesons as the first branching-fraction estimate~\cite{Maltoni:2004hv}, on the order of a few per cent, was then also questioned~\cite{Jia:2006rx}. Finally, let us mention the inclusive channel $\eta_b \to J/\psi X$ which however suffers from large uncertainties owing to possible CO contributions~\cite{Hao:2007rb}. Other ideas could be found by looking at the many $\chi_b$ decay channels which have been analyses so far, see~\cite{Zyla:2020zbs}.

An alternative to these hadronic decays is the exclusive radiative decay, $\eta_b \to J/\psi \gamma$, whose branching is computed to be on the order of $2\times 10^{-7}$~\cite{Hao:2006nf}, which remains admittedly small. In principle, it suffers from smaller theory uncertainties compared to those above.  An even simpler branching to predict is that  of $\eta_b \to \gamma \gamma$, whose partial width is in fact known up to two loops~\cite{Czarnecki:2001zc,Feng:2015uha} in NRQCD\footnote{Other approaches may sometimes give different results, but not by factors differing by more than 2~\cite{Lansberg:2006sy}.}. In addition, some theoretical uncertainties cancel with those of the total width computed in NRQCD by assuming the dominance of the decay into two gluons, which is also known up to NNLO~\cite{Feng:2017hlu} (see above), yielding
\eqs{
\Br(\eta_b  \to \gamma \gamma ) = (4.8 \pm 0.7) \times 10^{-5}.
}
We thus believe that this channel should seriously be considered for a first extraction of the $\eta_b$ hadroproduction cross section
even though channels involving $J/\psi$ are probably easier to deal with as what regards the combinatoric background. Tools like {\sc EtabFDC}~\cite{Qiao:2008mw} should also definitely be helpful for experimental prospective studies.

As what regards the experimental setups where $\eta_b$ hadroproduction cross sections could be measured in $pp$ collisions, let us cite the LHC in the collider and fixed-target modes, in particular with the LHCb detector. The latter mode has been studied in details in~\cite{Brodsky:2012vg,Lansberg:2012kf,Massacrier:2015qba,Hadjidakis:2018ifr} as what regards quarkonium production. Its nominal $\sqrt{s}$ for 7~TeV proton beams reaches 114.6~GeV. Another possibility is the SPD detector at the NICA facility up to
$\sqrt{s}=27$~GeV~\cite{arbuzov2020physics}. We will provide predictions for these 3 setups.

\begin{figure*}[hbt!]
\centering
\subfloat[]{\includegraphics[width=0.66\columnwidth]{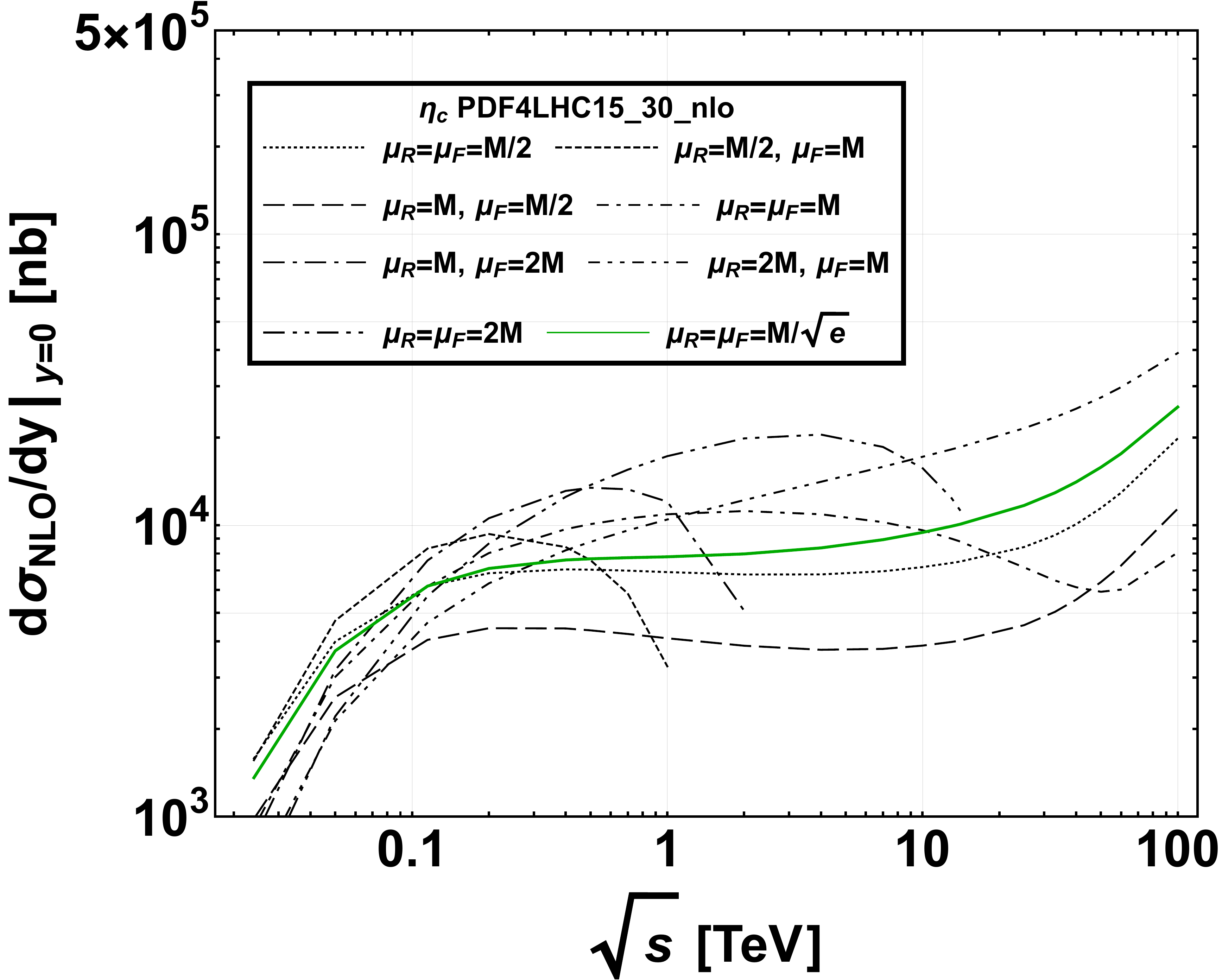}\label{fig:PDF4LHC_NLO_etac_EnEvo}}
\subfloat[]{\includegraphics[width=0.66\columnwidth]{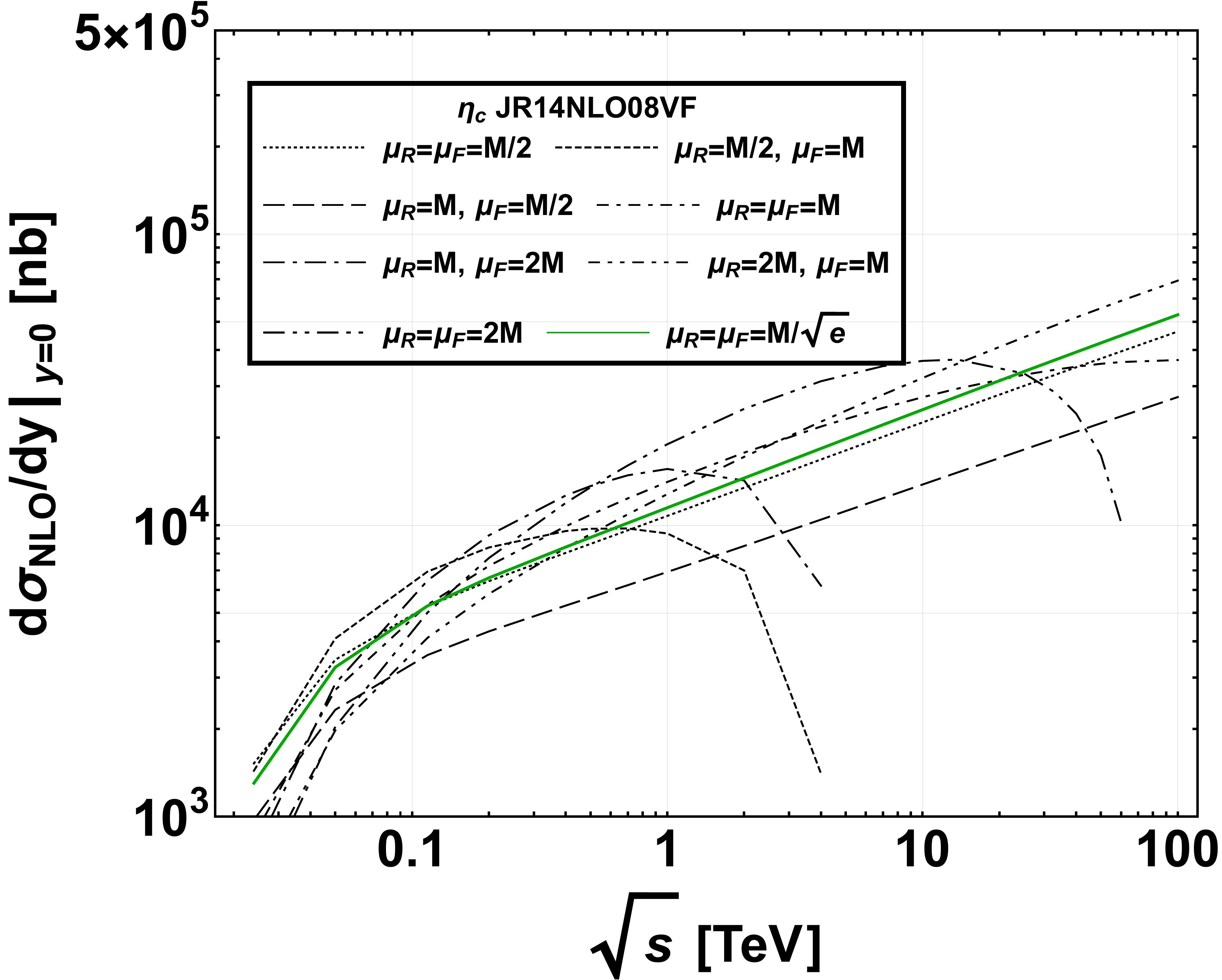}
\label{fig:JR14NLO08VF_NLO_etac_EnEvo}}
\subfloat[]{\includegraphics[width=0.66\columnwidth]{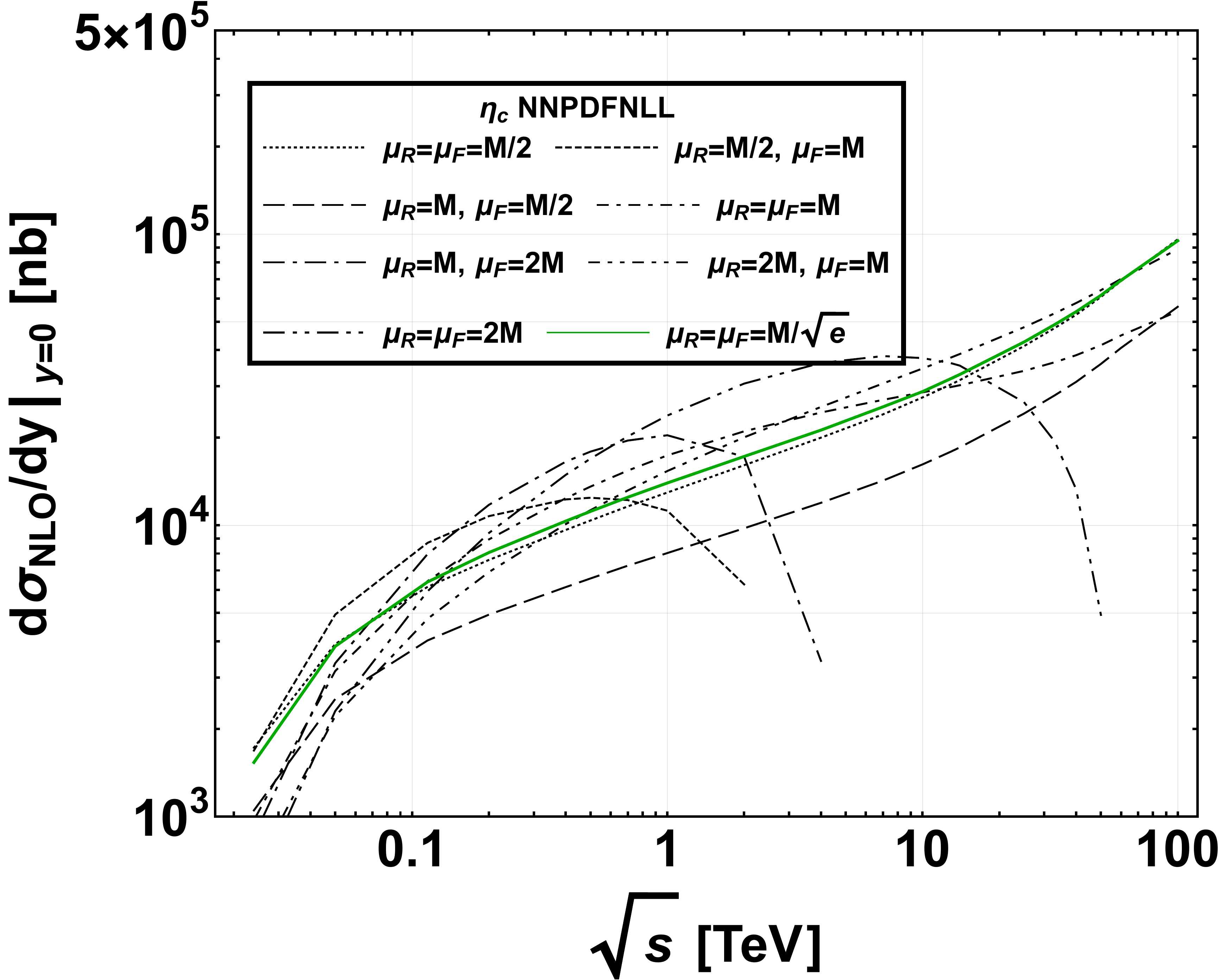}
\label{fig:NNPDF31sxNLONLL_NLO_etac_EnEvo}}\vspace*{-0.5cm}
\\
\subfloat[]{\includegraphics[width=0.66\columnwidth]{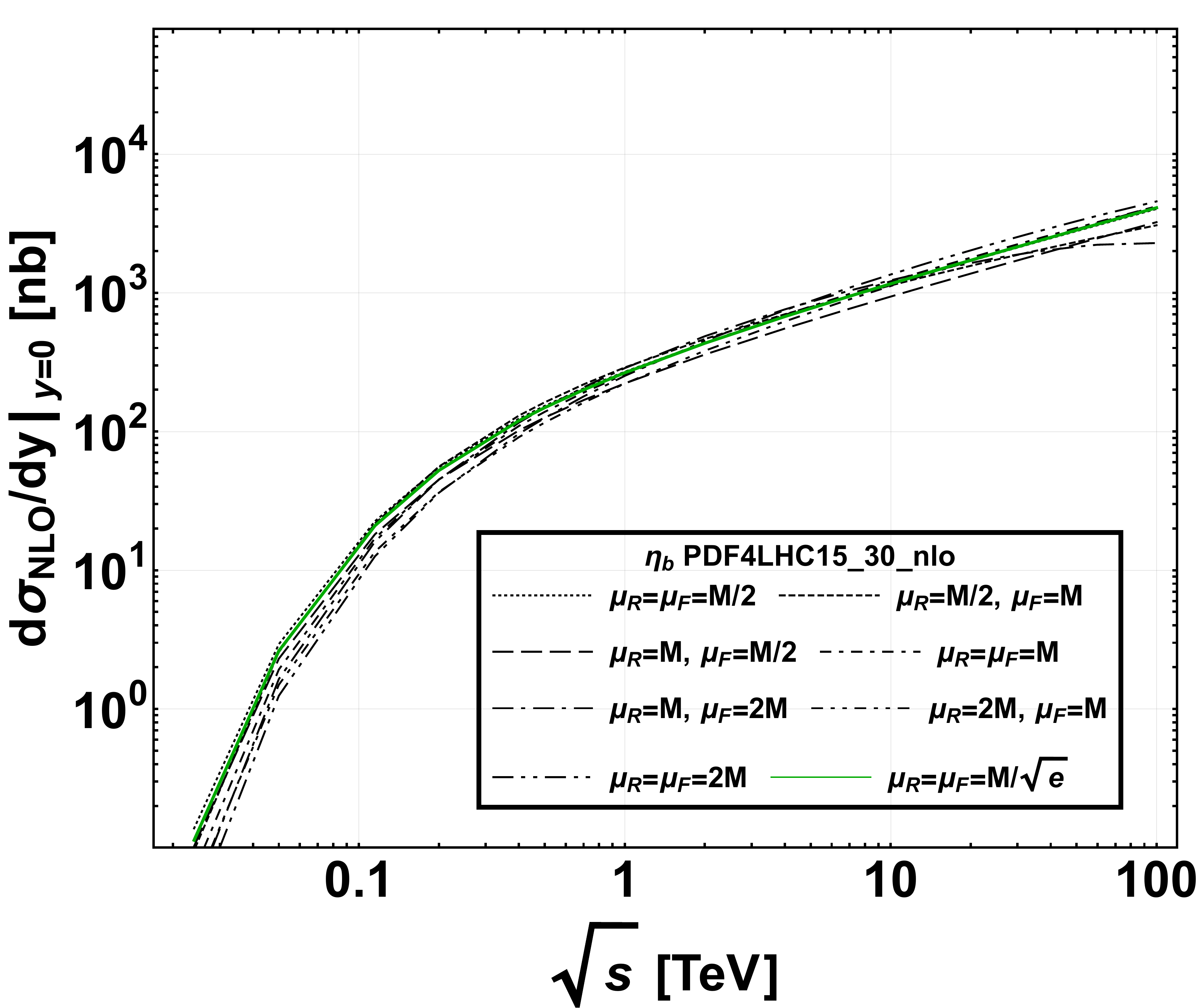}\label{fig:PDF4LHC_NLO_etab_EnEvo}}
\subfloat[]{\includegraphics[width=0.66\columnwidth]{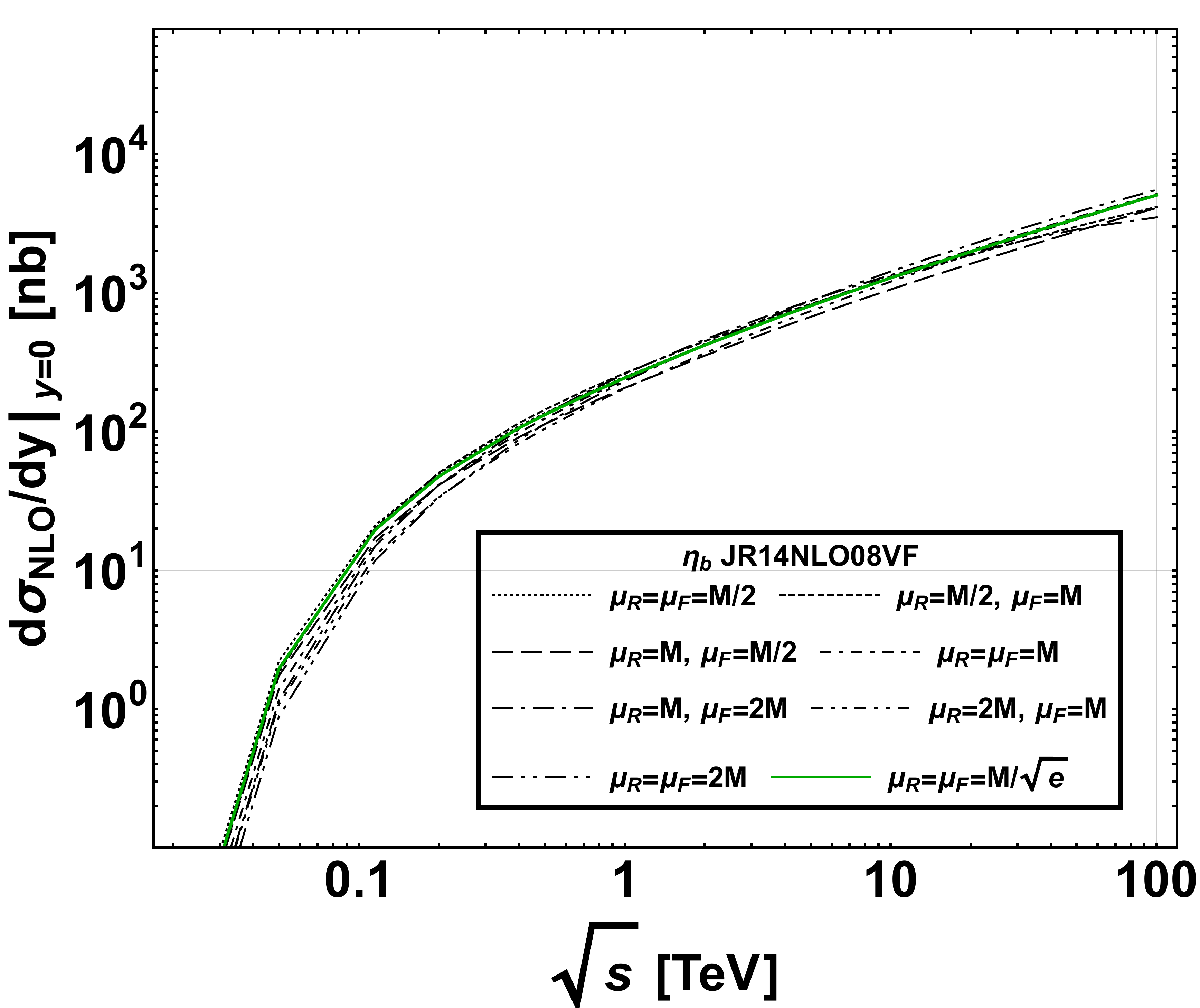}\label{figJR14NLO08VF_NLO_etab_EnEvo}}
\subfloat[]{\includegraphics[width=0.66\columnwidth]{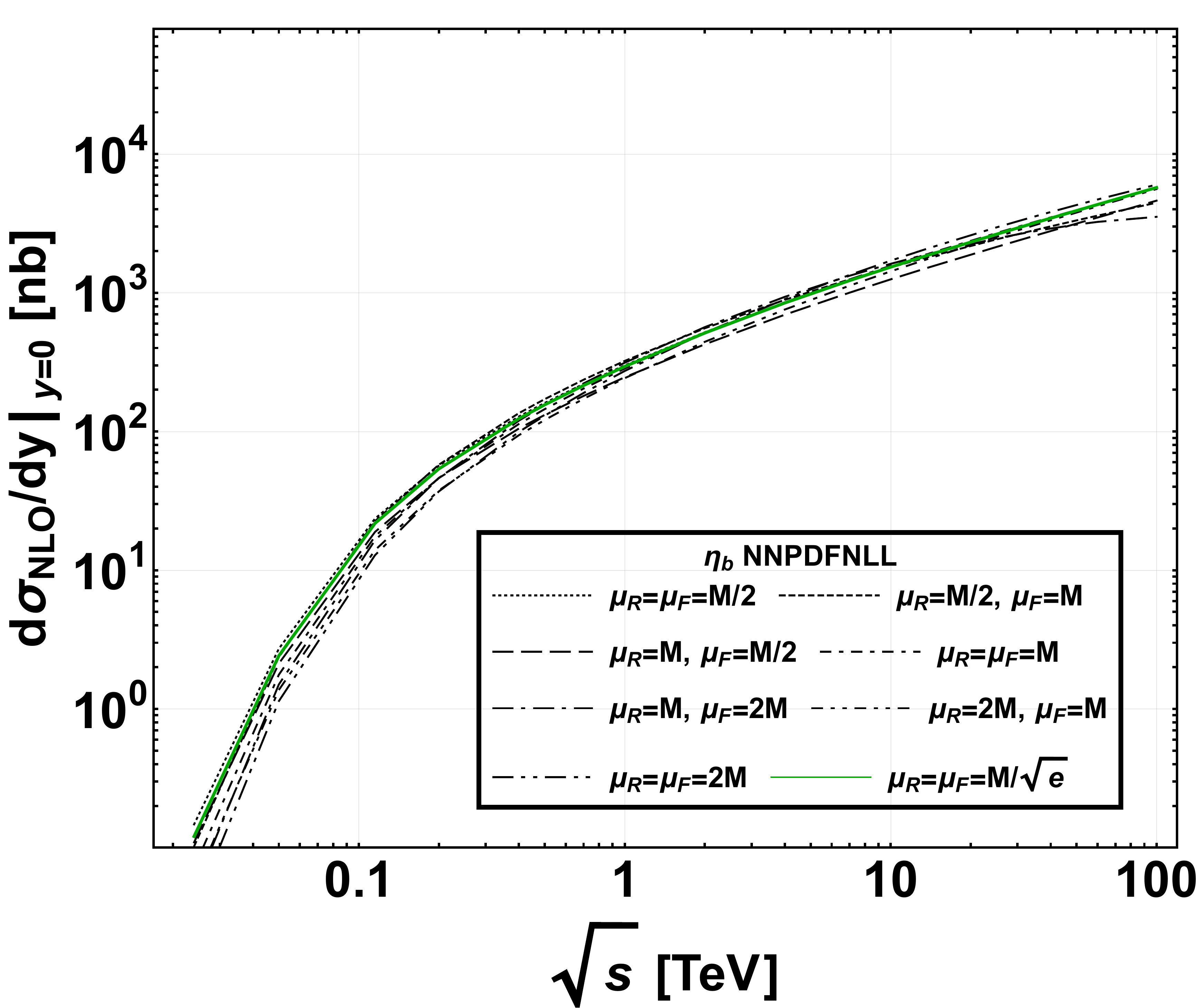}\label{fig:NNPDF31sxNLONLL_NLO_etab_EnEvob}}
\caption{$\frac{d\sigma^{\rm NLO}}{dy}|_{y=0}$ for $\eta_c$ (top) and  $\eta_b$ (bottom) (for PDF4LHC (left), JR14NLO08VF (middle),  NNPDFsxNLONLL (right)) as a function of $\sqrt{s}$ for the usual 7-point scale choices and our $\hat\mu_f$ scale with $\mu_R=\mu_F$.}\label{fig:sig-etaQ-scales}\vspace*{-0.5cm}
\end{figure*}

\begin{figure*}[hbt!]
\centering
\subfloat[]{\includegraphics[width=0.8\columnwidth]{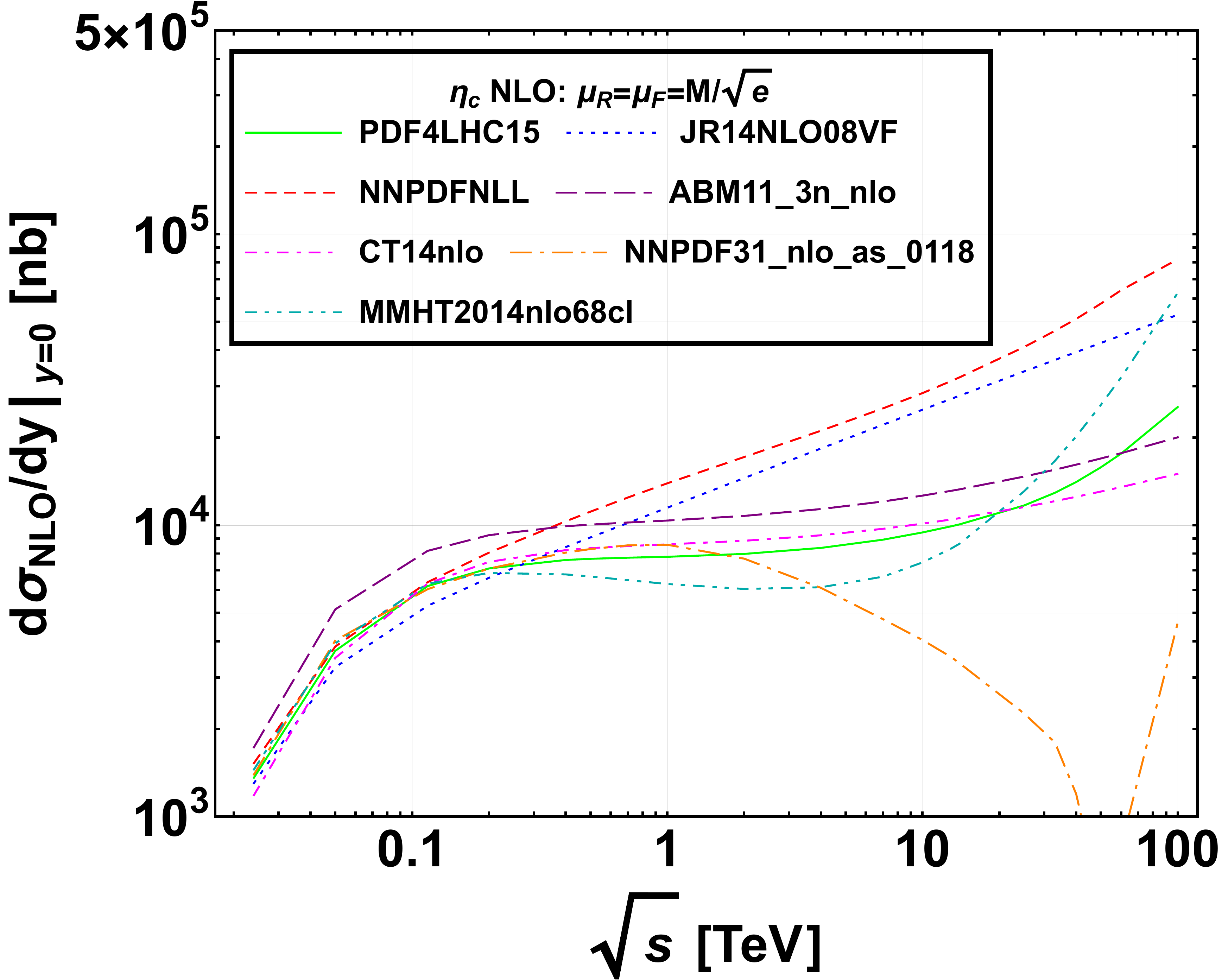}\label{fig:NLOcentralplotetacCol}}
\subfloat[]{\includegraphics[width=0.8\columnwidth]{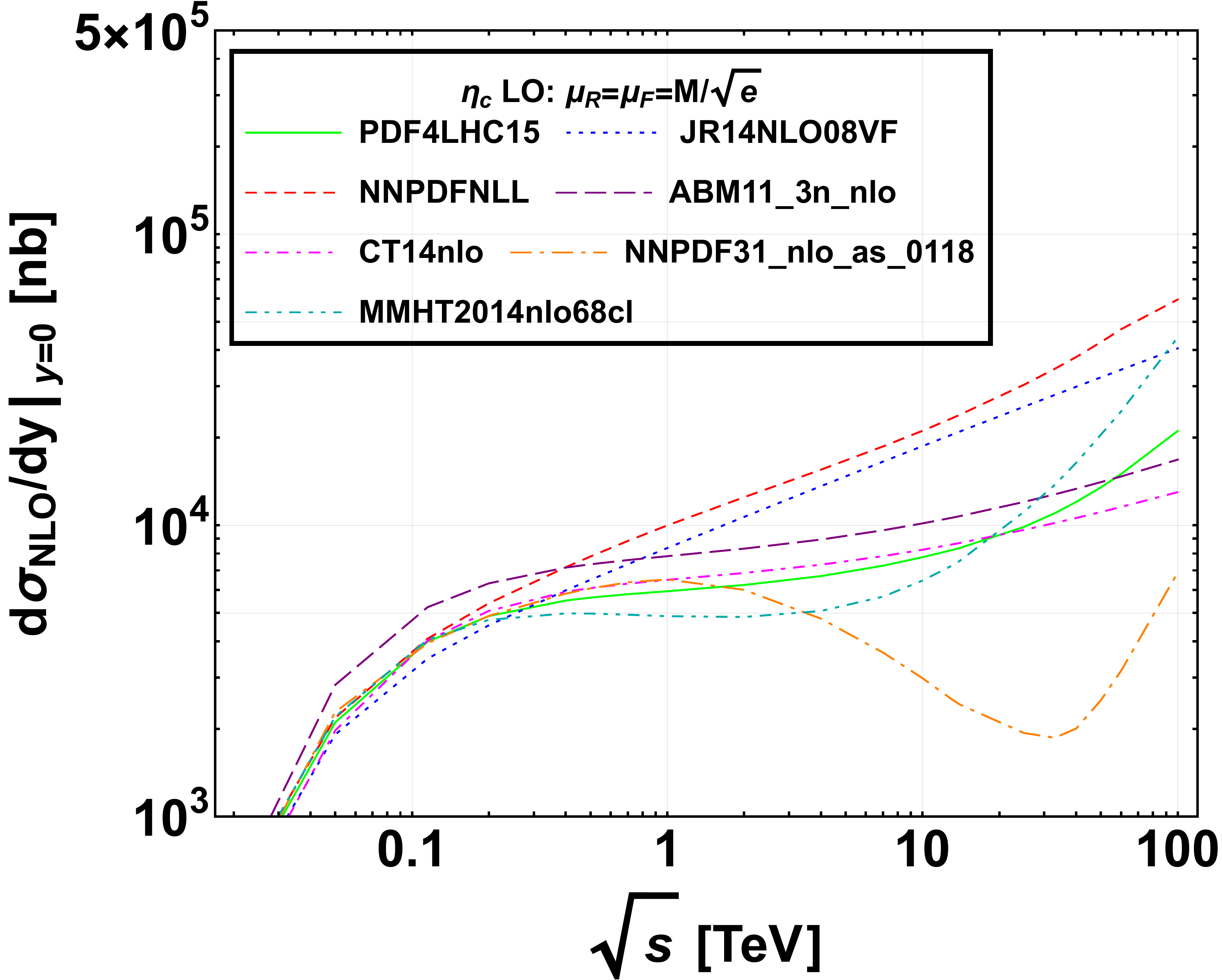}\label{fig:LOcentralplotetacCol}}
\caption{$\frac{d\sigma^{\rm NLO}}{dy}|_{y=0}$ for $\eta_c$ (a) at NLO and  (b) LO for various NLO PDF choice as a function of $\sqrt{s}$ for $\hat\mu_f$ scale with $\mu_R=\mu_F$.}\label{fig:centralplotetaQ}\vspace*{-0.5cm}
\end{figure*}

\begin{figure*}[hbt!]
\centering
\subfloat[]{\includegraphics[width=0.66\columnwidth]{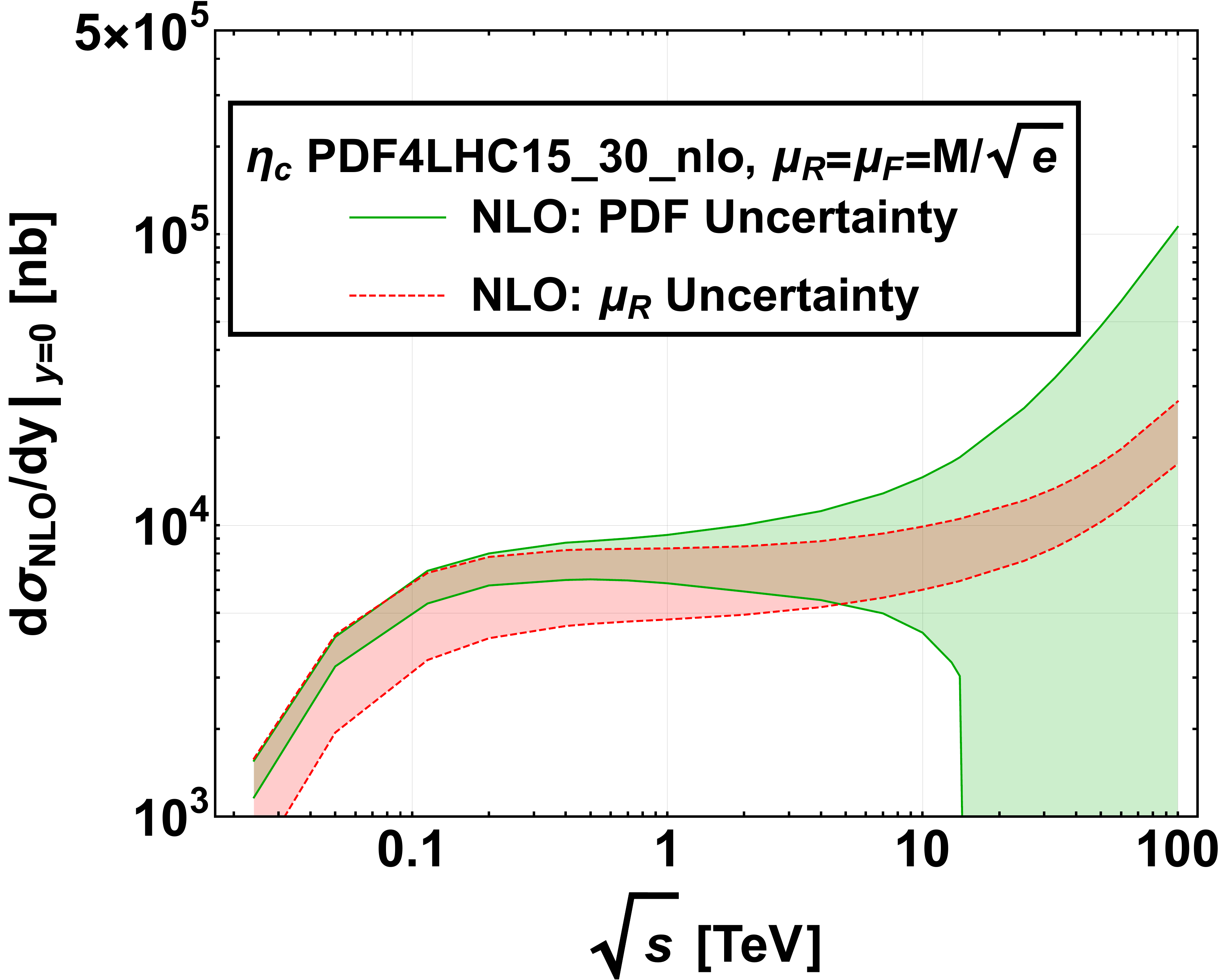}\label{fig:PDF4LHC_NLO_etac_EnEvo_Spec}}
\subfloat[]{\includegraphics[width=0.66\columnwidth]{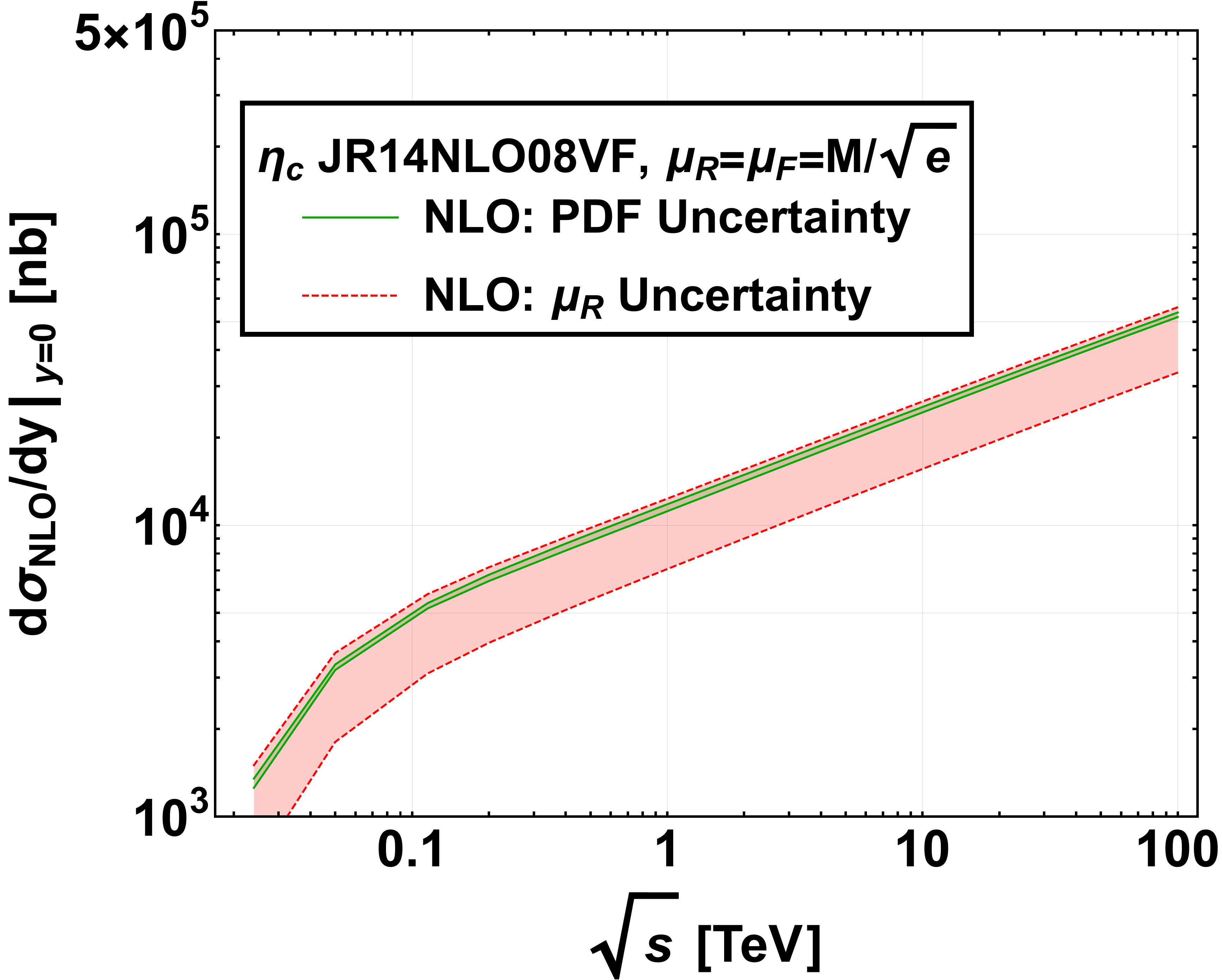}\label{fig:JR14NLO08VF_NLO_etac_EnEvo_Spec}}
\subfloat[]{\includegraphics[width=0.66\columnwidth]{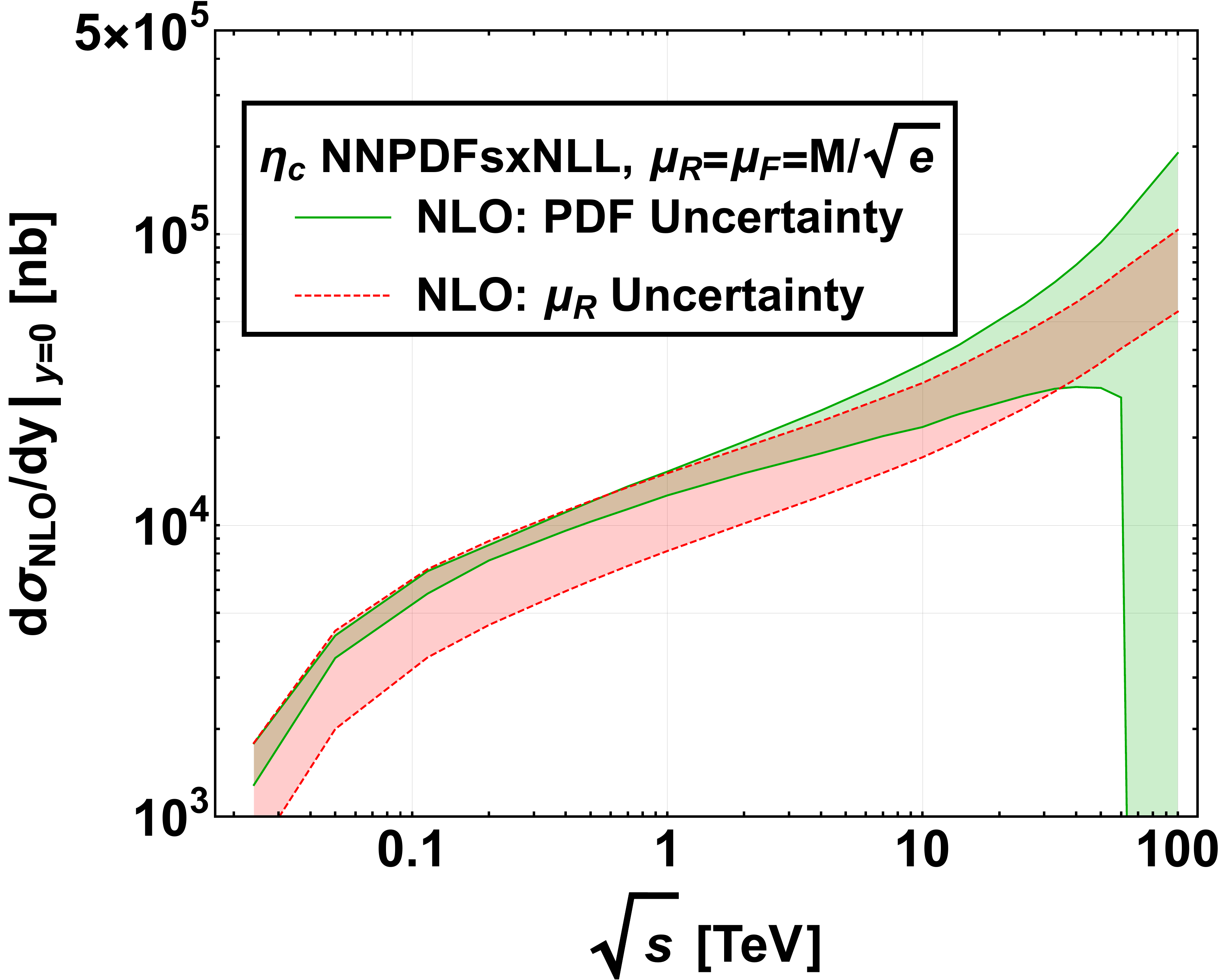}\label{fig:NNPDF31sxNLONLL_NLO_etac_EnEvo_Spec}}\vspace*{-0.5cm}
\\
\subfloat[]{\includegraphics[width=0.66\columnwidth]{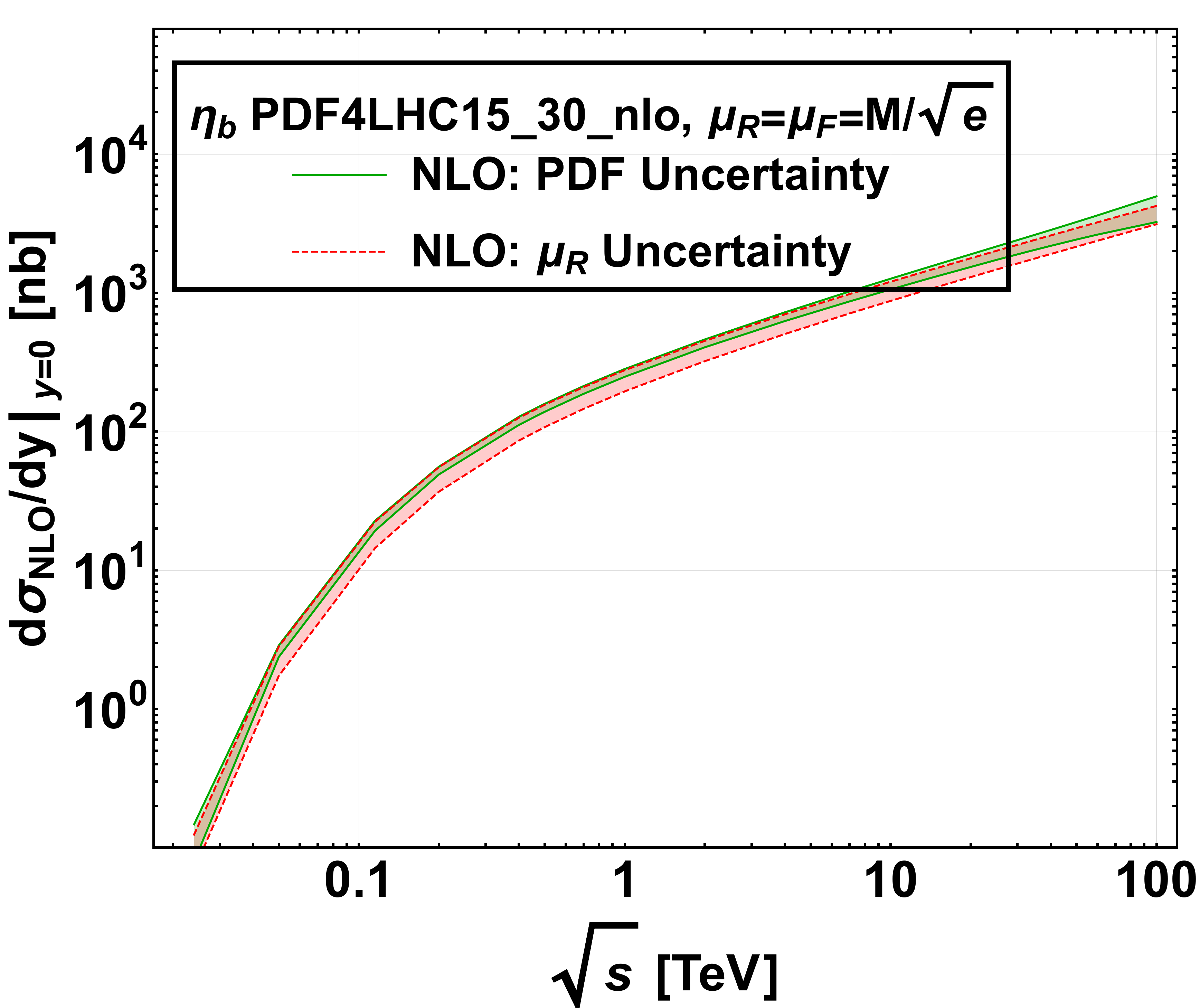}\label{fig:PDF4LHC_NLO_etab_EnEvo_Spec}}
\subfloat[]{\includegraphics[width=0.66\columnwidth]{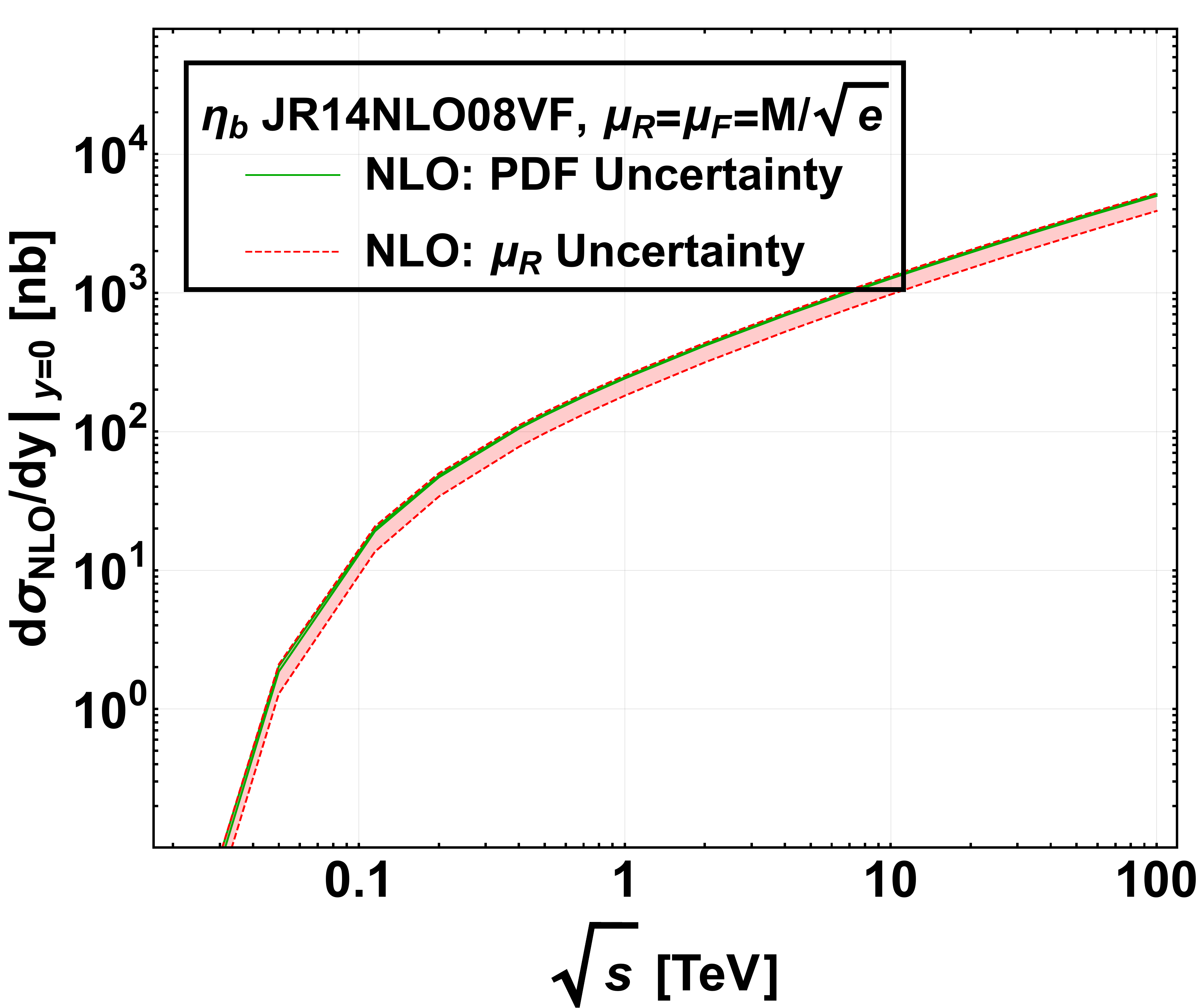}\label{fig:JR14NLO08VF_NLO_etab_EnEvo_Spec}}
\subfloat[]{\includegraphics[width=0.66\columnwidth]{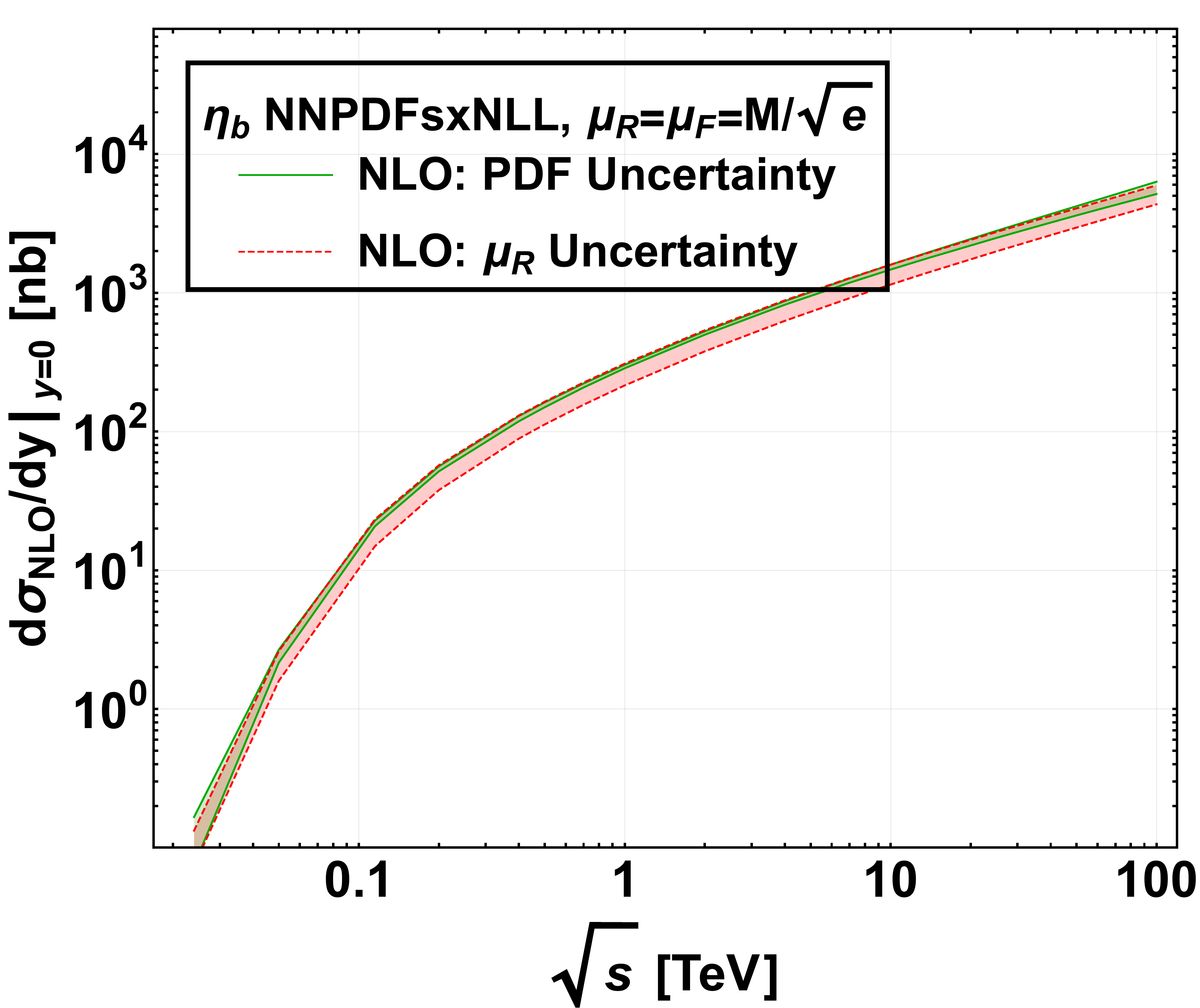}\label{fig:NNPDF31sxNLONLL_NLO_etab_EnEvo_Spec}}
\caption{
$\frac{d\sigma^{\rm NLO}}{dy}|_{y=0}$ for $\eta_c$ (top) and  $\eta_b$ (bottom) (for PDF4LHC (left), JR14NLO08VF (middle),  NNPDFsxNLONLL (right)) as a function of $\sqrt{s}$ for our $\hat\mu_f$ scale.
The green bands indicate the PDF uncertainty (for $\mu_R=\mu_F$) and the red band, the $\mu_R$ uncertainty
(for $\mu_R \in [M/2:2M]$).}\label{fig:sig-etaQ-PDFs}
\vspace*{-0.5cm}
\end{figure*}

\begin{figure*}[hbt!]
\centering
\subfloat[]{\includegraphics[width=0.66\columnwidth]{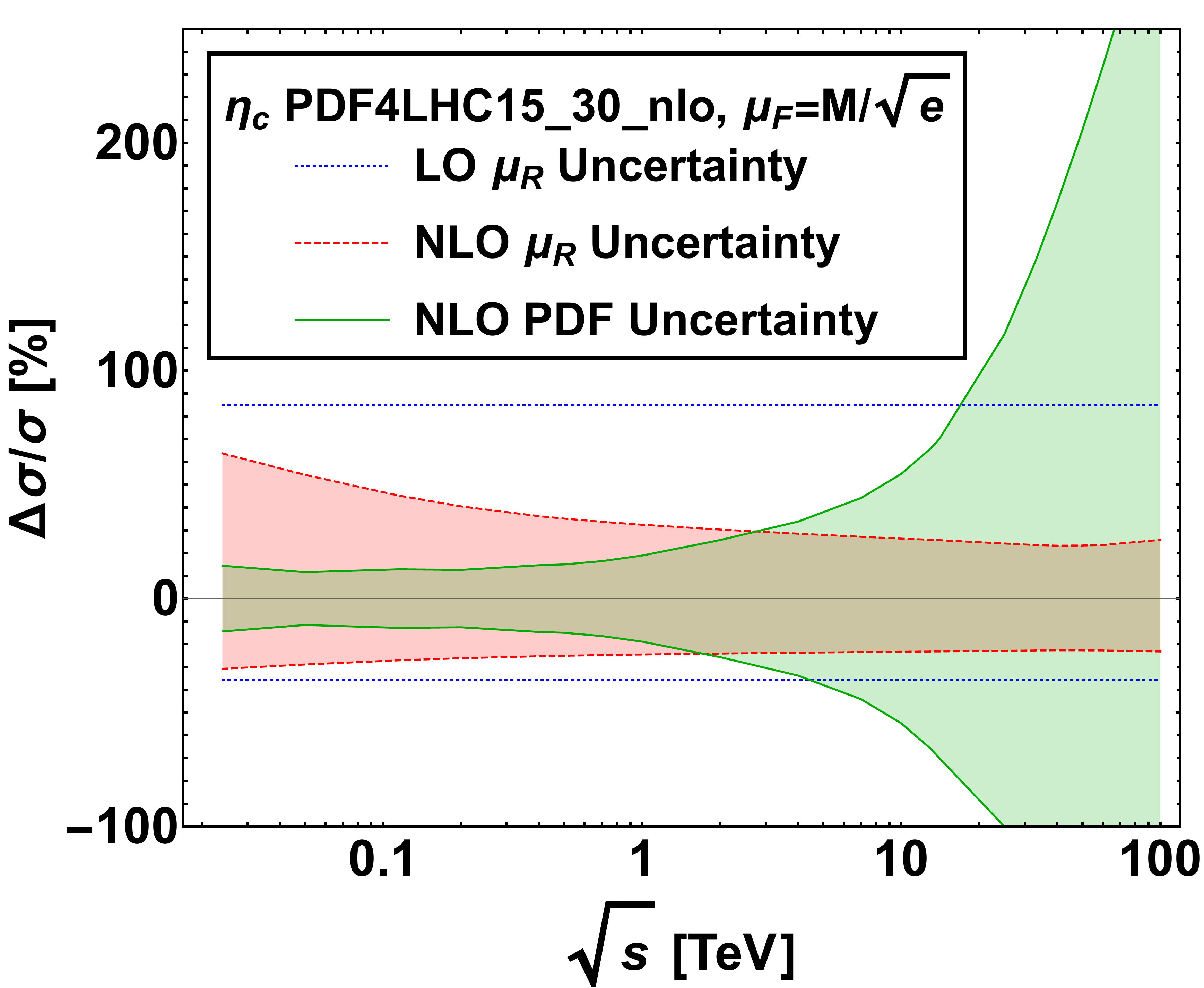}\label{fig:PDF4LHC_NLO_etac_EnEvo_SpecDelta}}
\subfloat[]{\includegraphics[width=0.66\columnwidth]{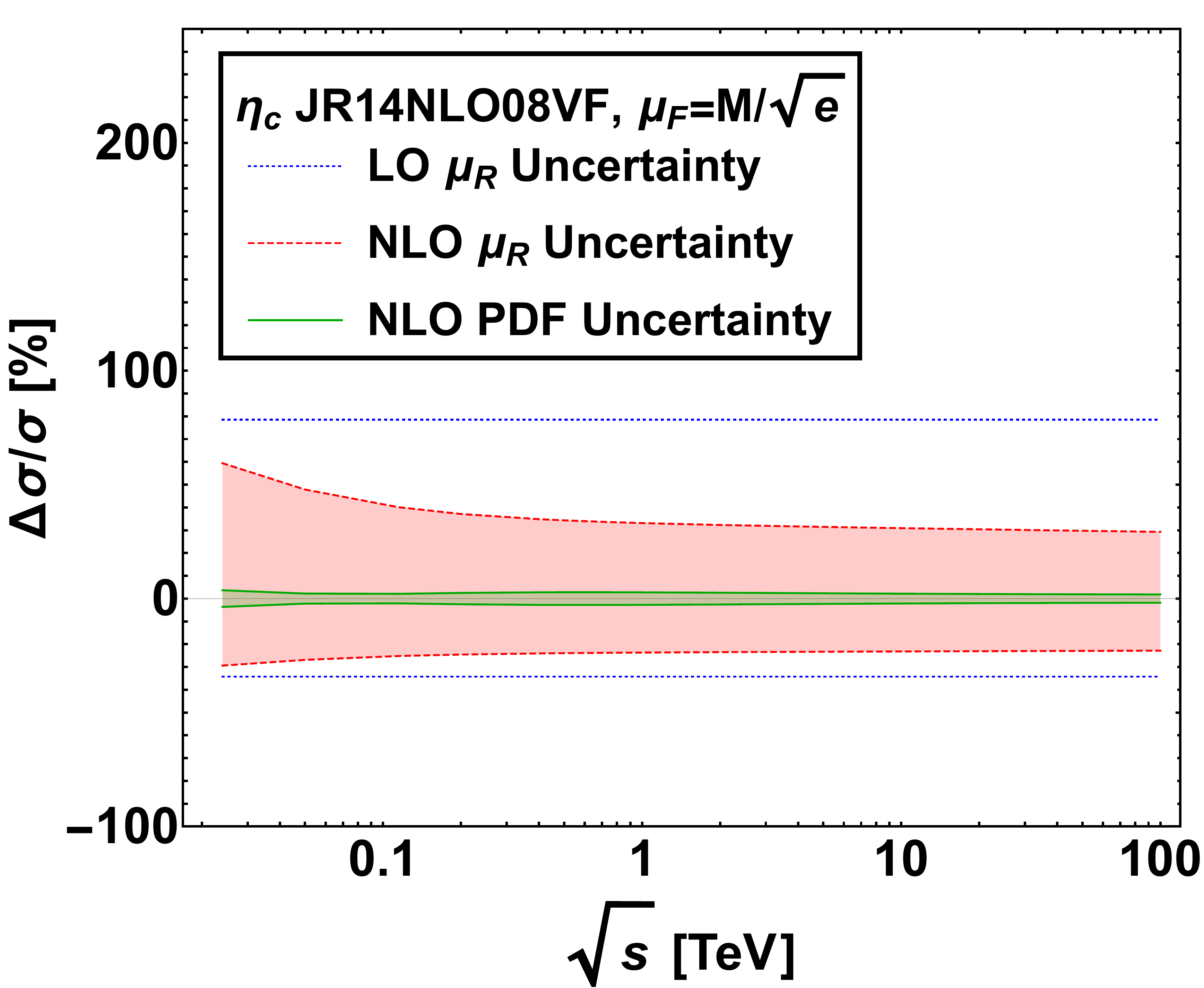}\label{fig:JR14NLO08VF_NLO_etac_EnEvo_SpecDelta}}
\subfloat[]{\includegraphics[width=0.66\columnwidth]{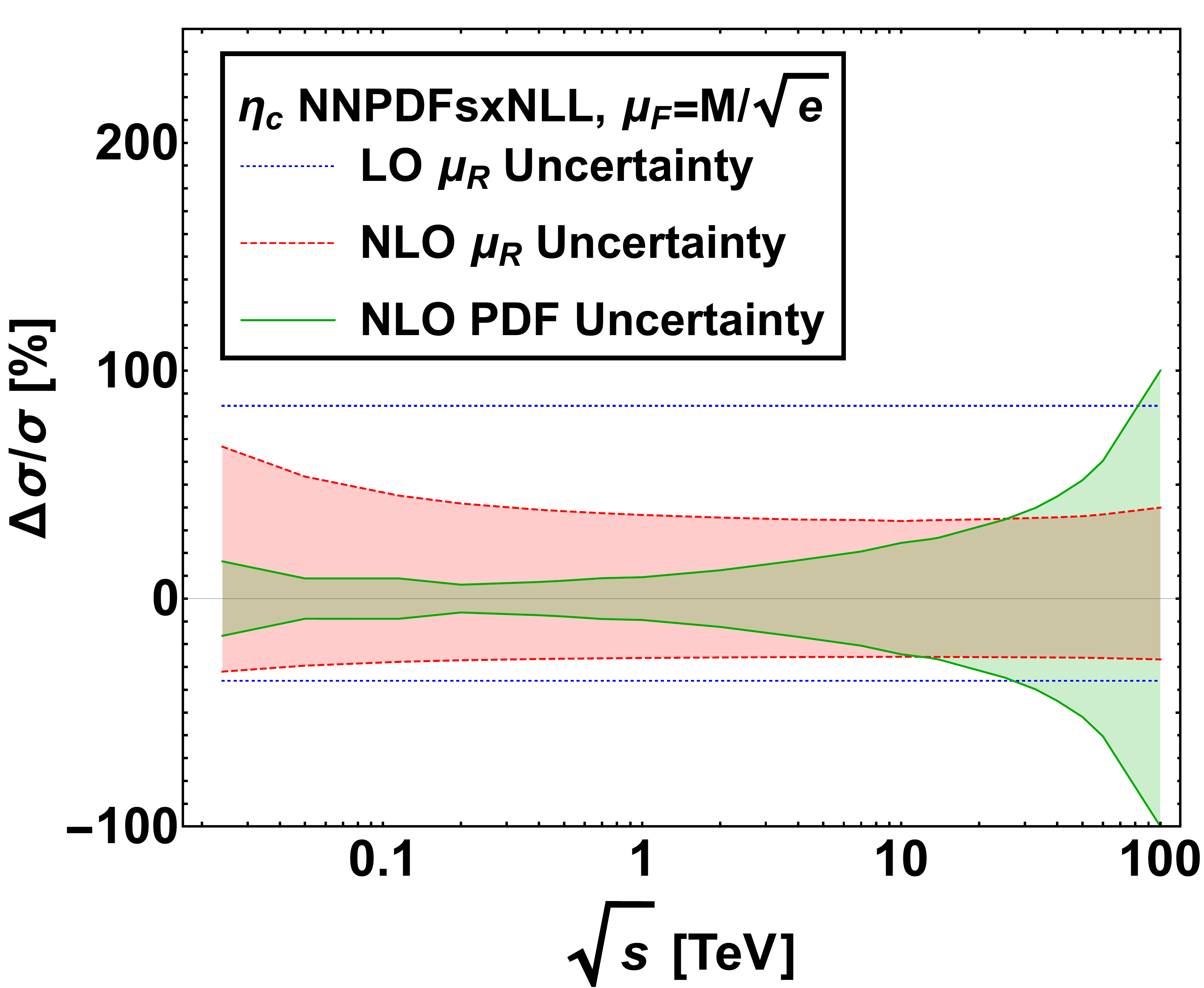}\label{fig:NNPDF31sxNLONLL_NLO_etac_EnEvo_SpecDelta}}\vspace*{-0.5cm}
\\
\subfloat[]{\includegraphics[width=0.66\columnwidth]{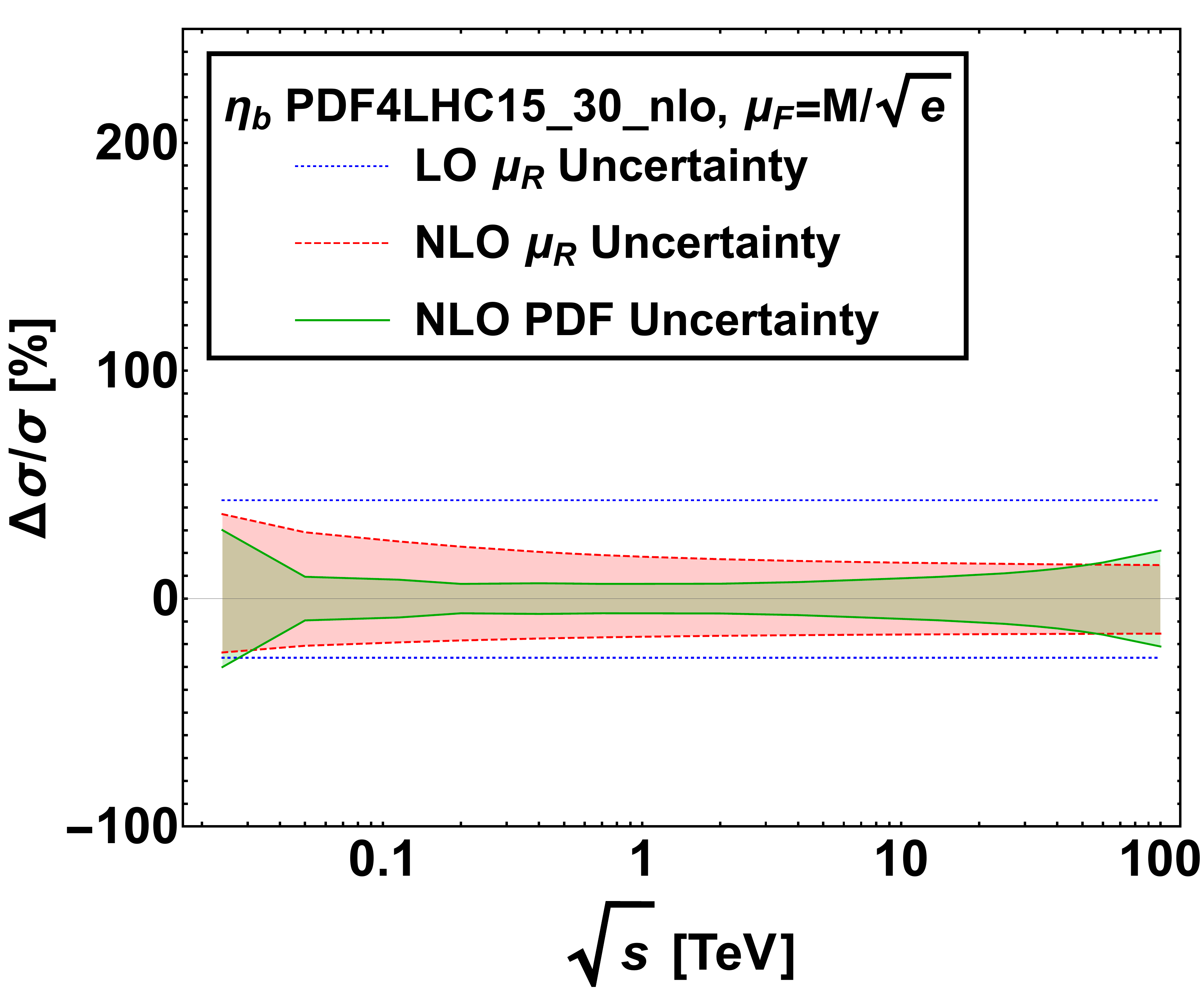}\label{fig:PDF4LHC_NLO_etab_EnEvo_SpecDelta}}
\subfloat[]{\includegraphics[width=0.66\columnwidth]{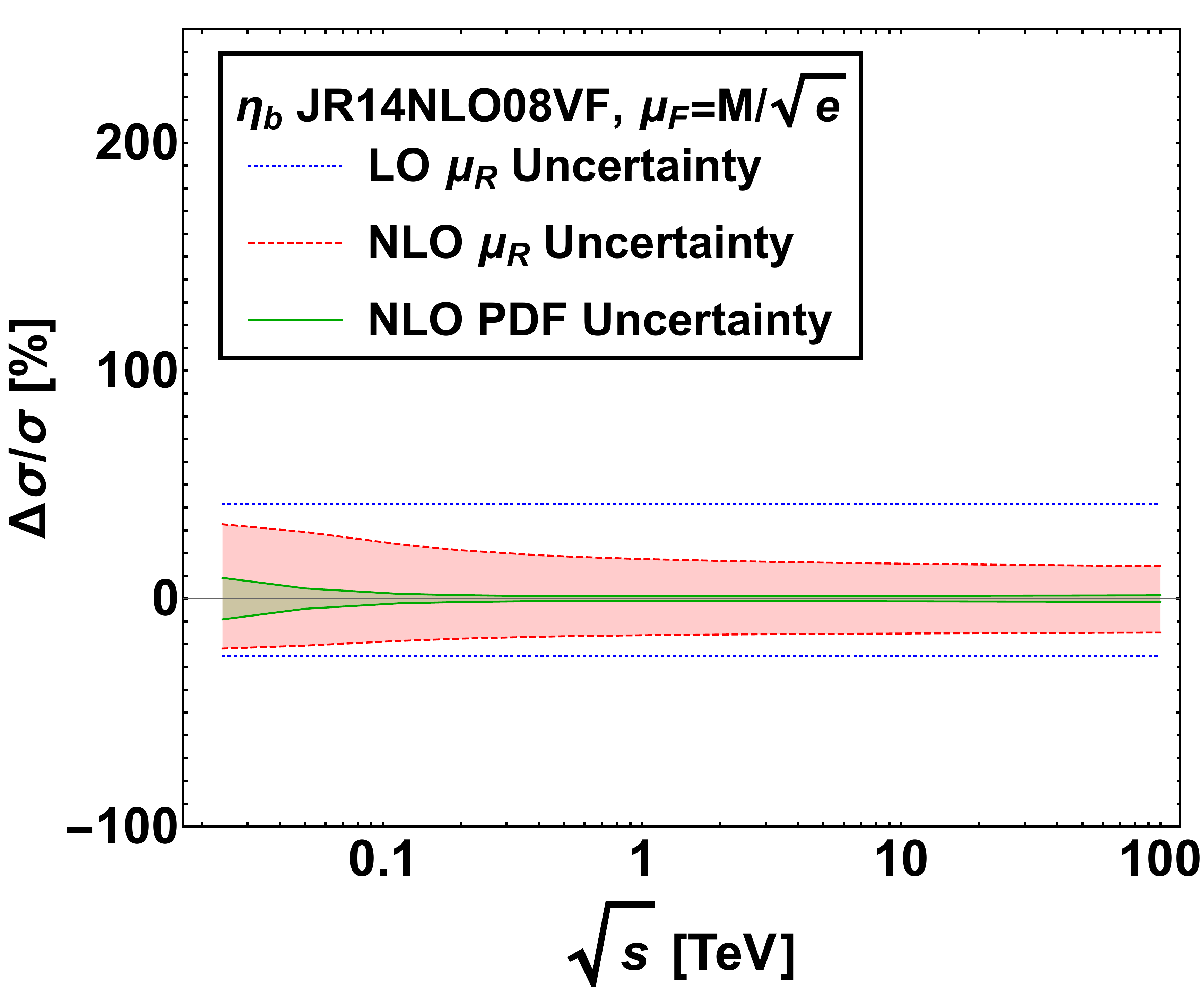}\label{fig:JR14NLO08VF_NLO_etab_EnEvo_SpecDelta}}
\subfloat[]{\includegraphics[width=0.66\columnwidth]{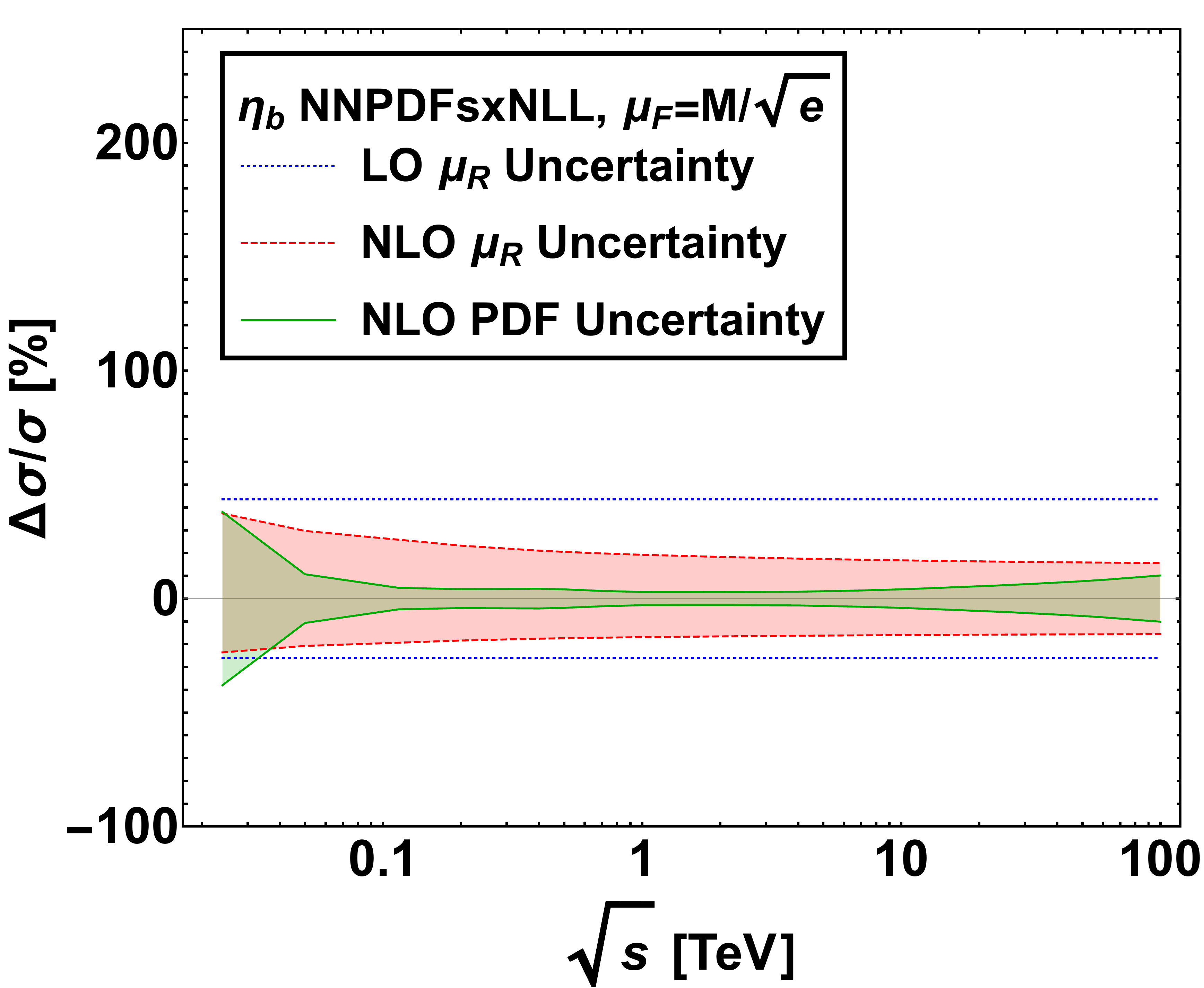}\label{fig:NNPDF31sxNLONLL_NLO_etab_EnEvo_SpecDelta}}
\caption{Relative uncertainties  $\Delta \sigma/\sigma$ from $\mu_R$ (for $\mu_R \in [M/2:2M]$; red band) and PDF (for $\mu_R=\mu_F$; green band) at NLO for $\eta_c$ (top) and  $\eta_b$ (bottom) as a function of $\sqrt{s}$ compared to the $\mu_R$ uncertainty at LO for PDF4LHC (left), JR14NLO08VF (middle) and NNPDFsxNLONLL (right).}\label{fig:deltasigma}
\vspace*{-0.5cm}
\end{figure*}

\begin{figure*}[hbt!] 
\centering
\subfloat[]{\includegraphics[width=0.66\columnwidth]{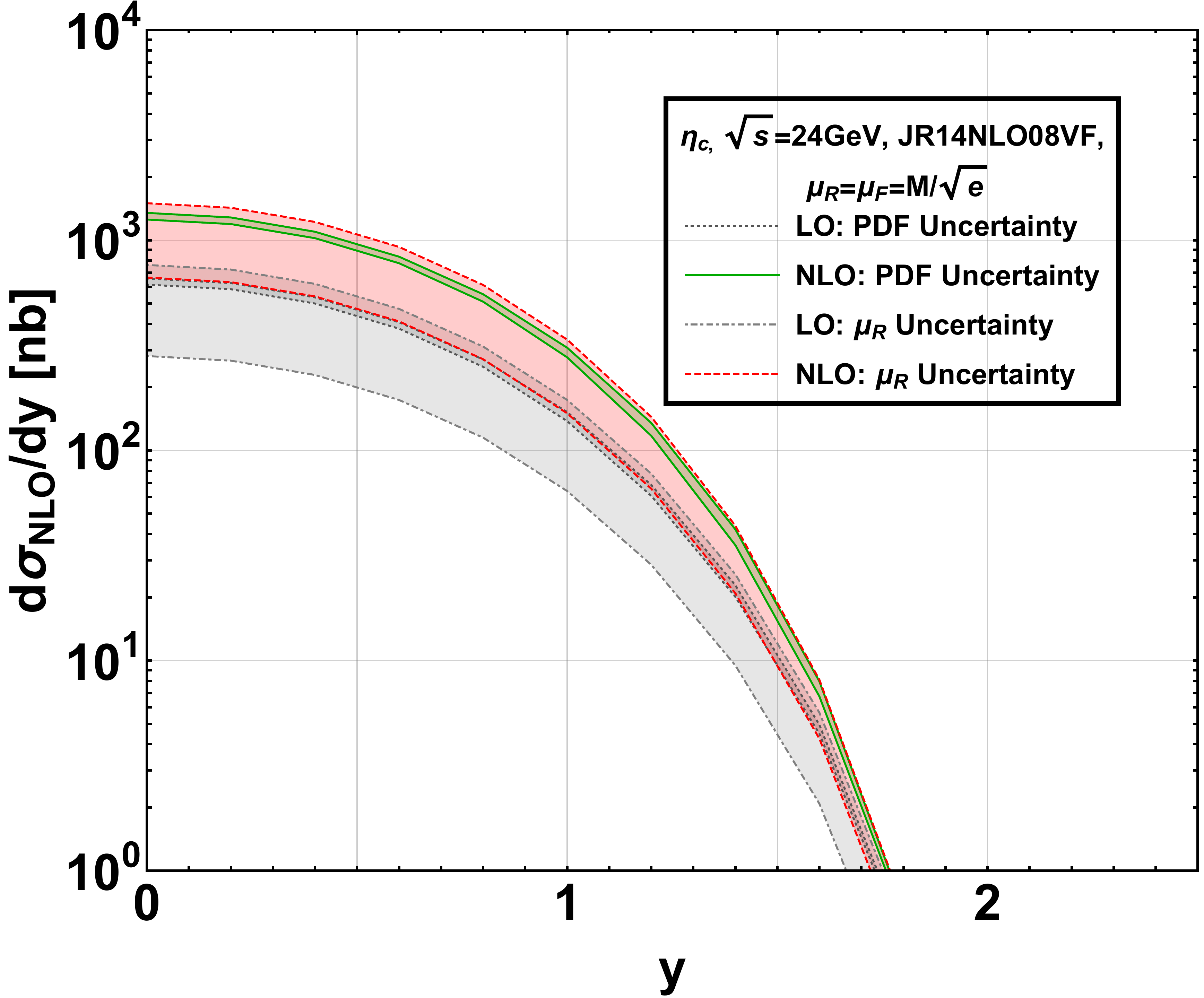}\label{fig:JR14NLO08VF_NLO_etac_Rap24GeV_Log_Spec}}
\subfloat[]{\includegraphics[width=0.66\columnwidth]{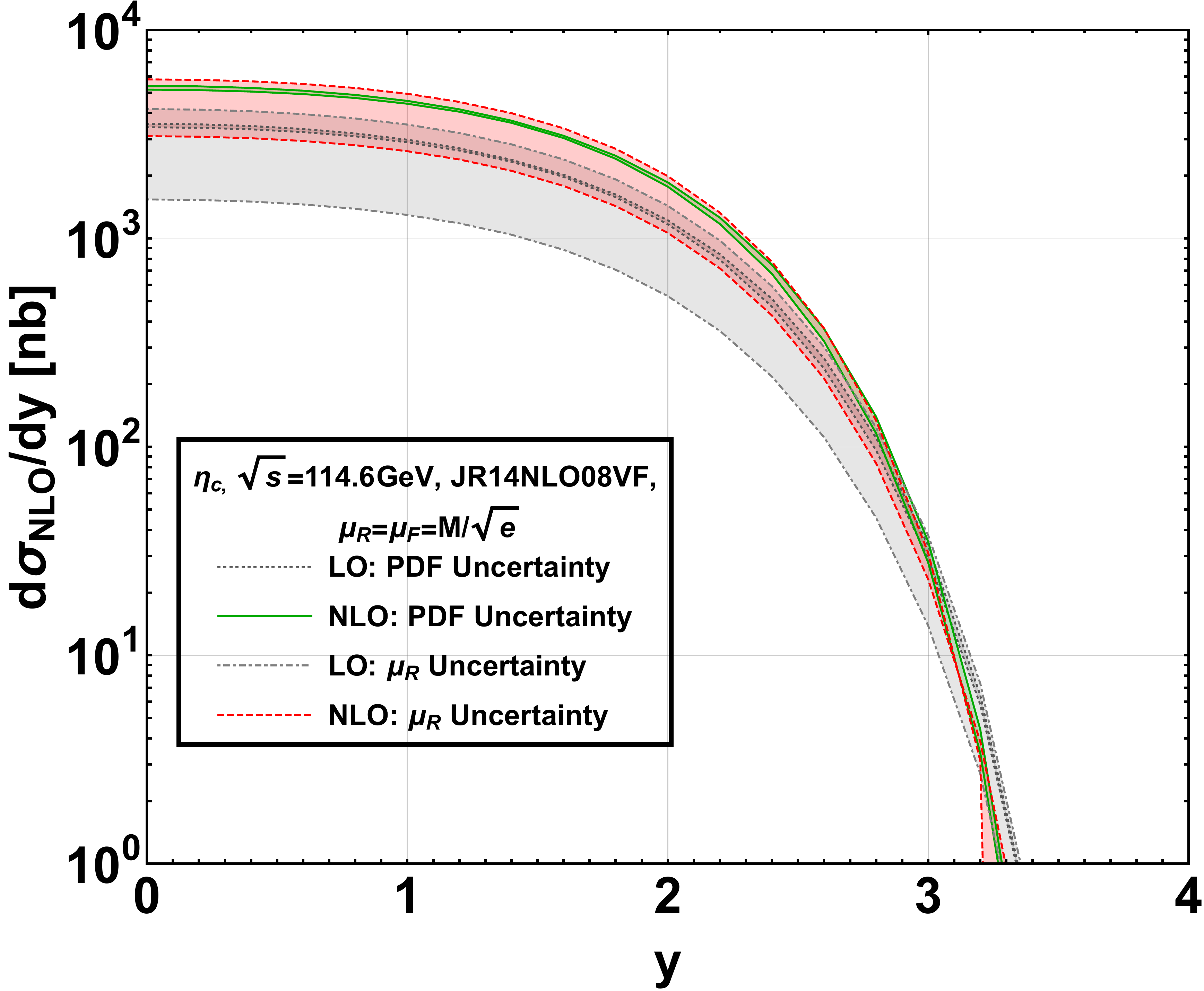}\label{fig:JR14NLO08VF_NLO_etac_Rap114p6GeV_Log_Spec}}
\subfloat[]{\includegraphics[width=0.66\columnwidth]{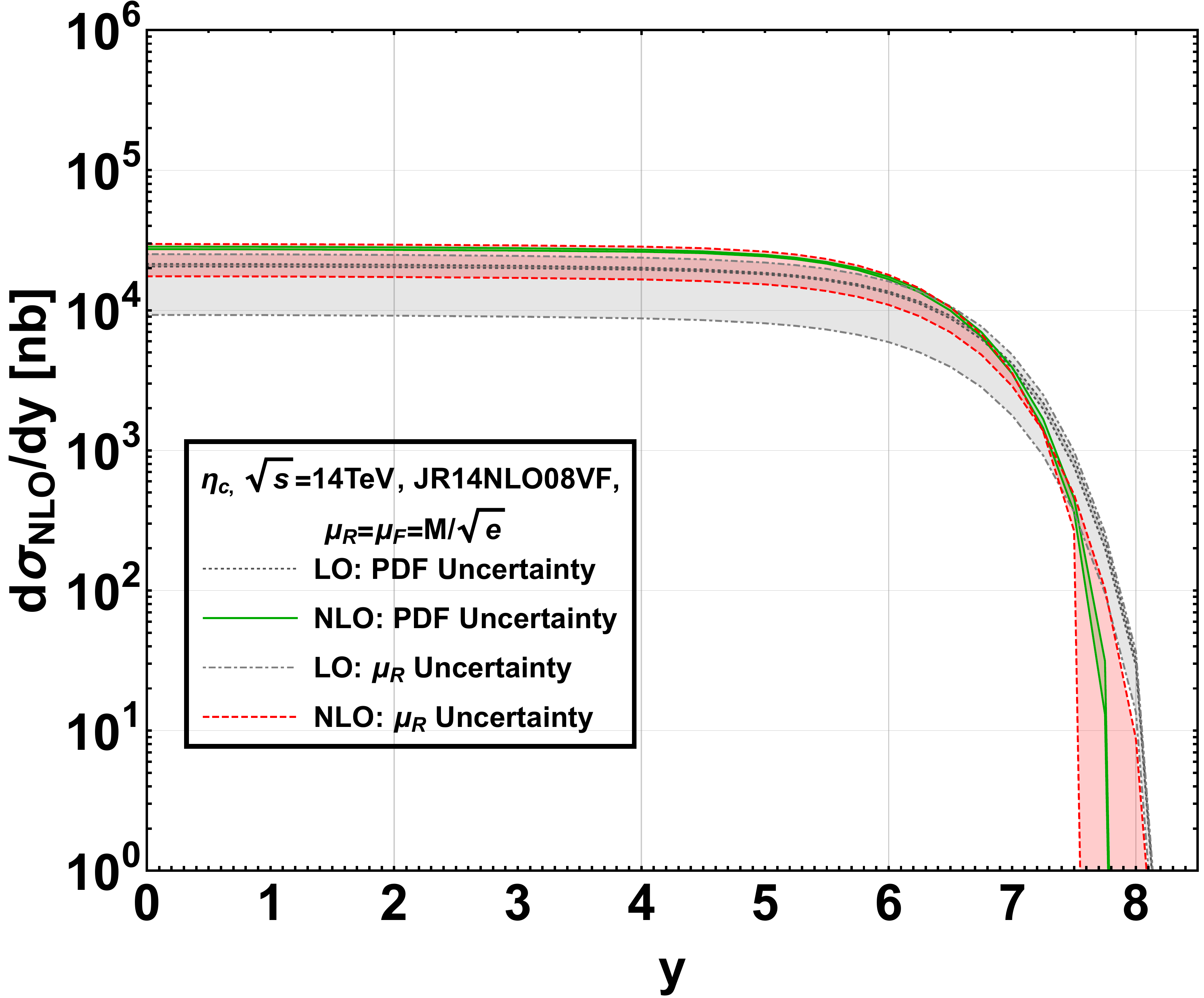}\label{fig:JR14NLO08VF_NLO_etac_Rap14TeV_Log_Spec}}\vspace*{-0.5cm}
\\
\subfloat[]{\includegraphics[width=0.66\columnwidth]{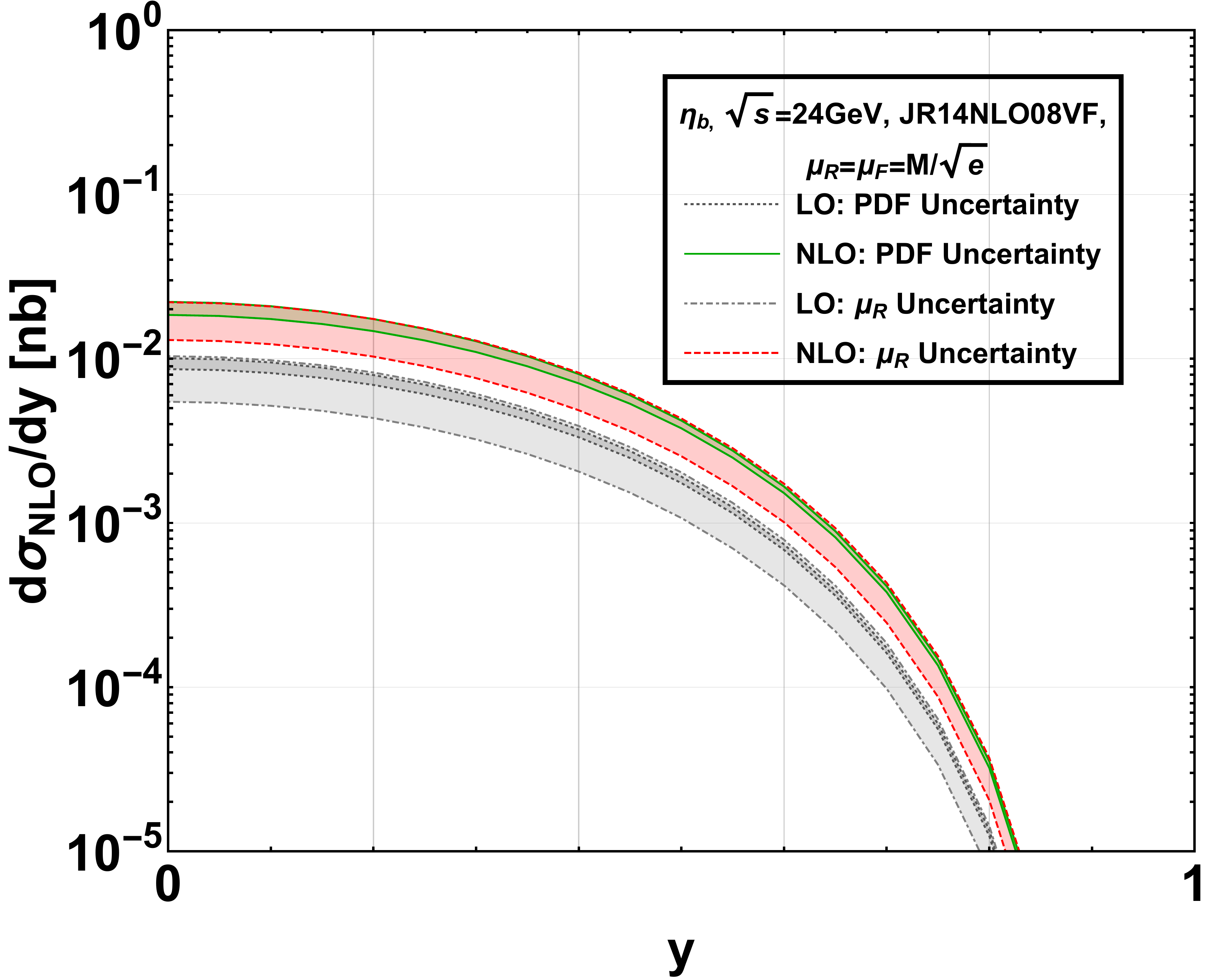}\label{fig:JR14NLO08VF_NLO_etab_Rap24GeV_Log_Spec}}
\subfloat[]{\includegraphics[width=0.66\columnwidth]{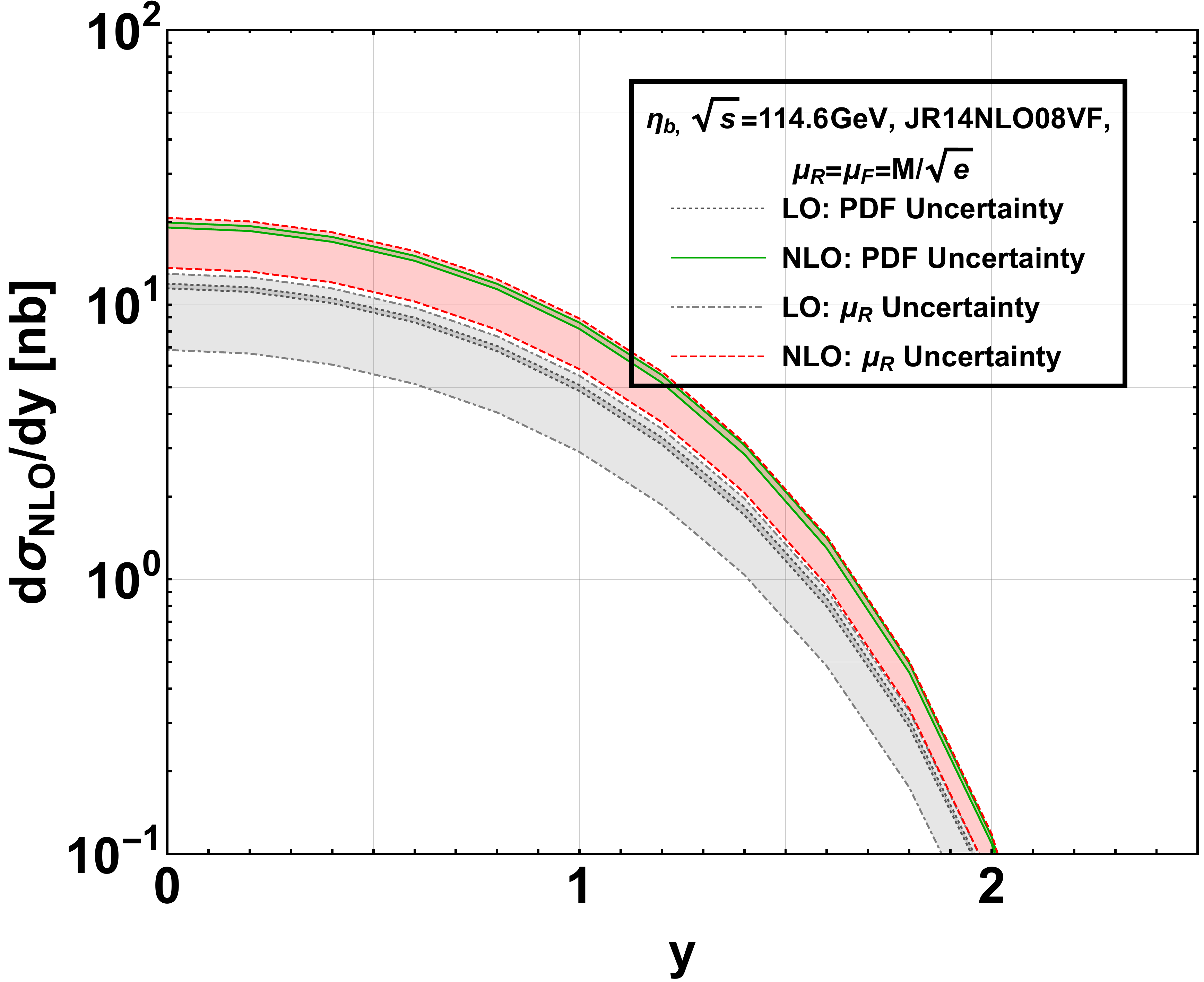}\label{fig:JR14NLO08VF_NLO_etab_Rap114p6GeV_Log_Spec}}
\subfloat[]{\includegraphics[width=0.66\columnwidth]{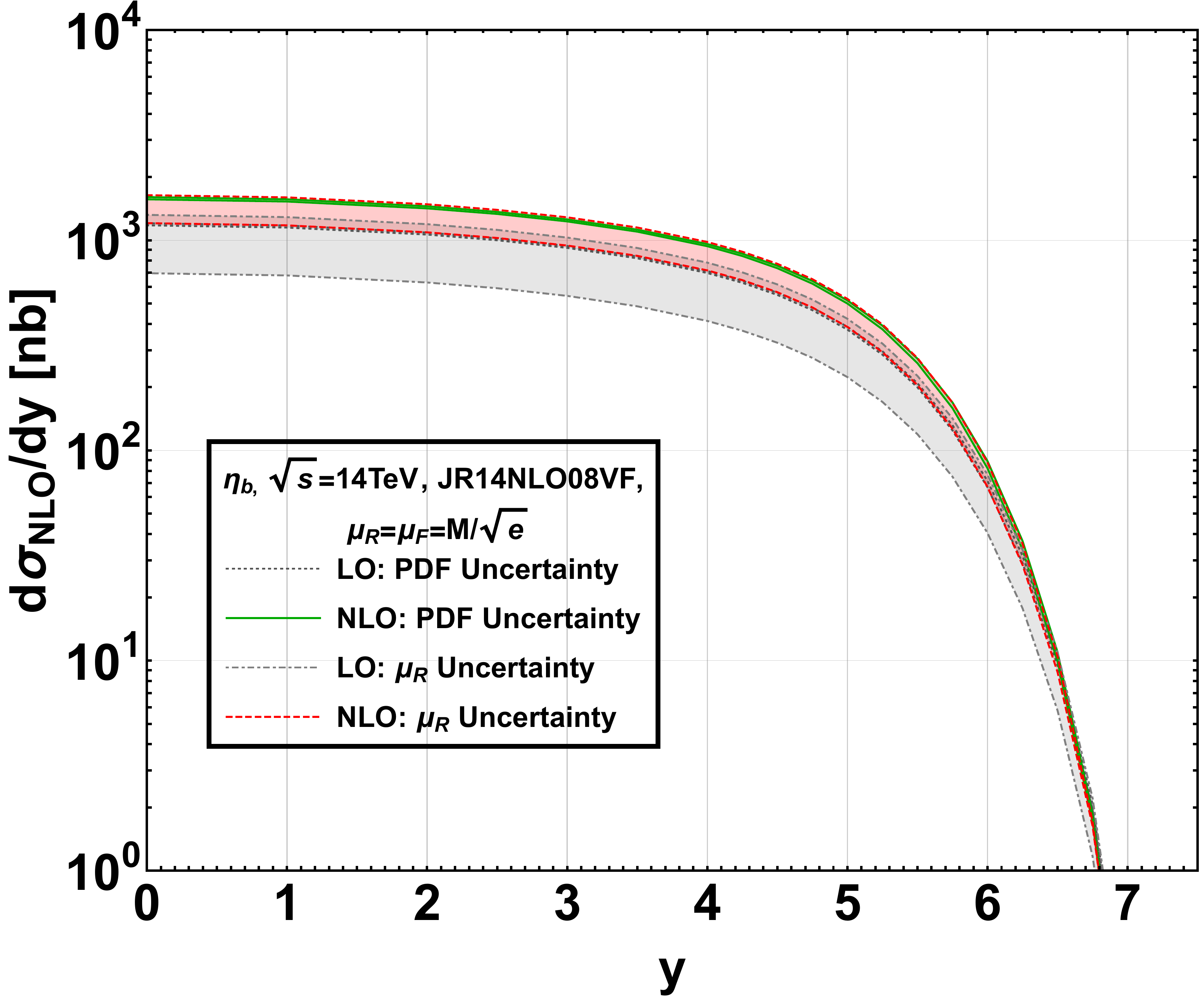}\label{fig:JR14NLO08VF_NLO_etab_Rap14TeV_Log_Spec}}
\caption{
$\frac{d\sigma^{\rm NLO}}{dy}$ for $\eta_c$ (top) and  $\eta_b$ (bottom) as a function of $y$ at $\sqrt{s}=$~24 GeV(left), 114.6 GeV (middle), 14~TeV (right), for $\mu_F=\hat\mu_F=\mu_R$ for JR14NLO08VF at LO and NLO. The green (gray) bands indicate the PDF uncertainty (for $\mu_R=\mu_F$)  and the red (beige) band, the $\mu_R$ uncertainty
(for $\mu_R \in [M/2:2M]$) at NLO (LO).}\label{fig:dsigdy-etaQ}
\vspace*{-0.5cm}
\end{figure*}

\subsection{Cross section predictions}

We are now finally in position to show our results for the cross sections as a function of $\sqrt{s}$ and of $y$ for selected values of $\sqrt{s}$ which correspond to experimental setups where we believe the challenging measurement of $P_T$-integrated  $\eta_Q$ yield could be performed in the future. These are the LHC at 14 TeV in the collider mode and in the fixed-target mode at 114.6~GeV, having in mind in particular the LHCb detector, and to the SPD experiment at NICA which could run up to 27 GeV.

As our study mainly addresses the interplay between the size of the NLO corrections and the choice of scales, mainly that of $\mu_F$, we start by showing,  $d\sigma^{\rm NLO}_{\eta_c}/dy|_{y=0}$ as a function of $\sqrt{s}$, for the 7-point scale choice (in black) and $\hat{\mu}_F$ (in green) for the central eigensets of our 3 PDF sets. These plots are the exact counterpart of the $K^{\rm NLO}|_{y=0}$ plots of \cf{fig:K-etaQ} which allowed us to assess the much better convergence of the NLO results at high energies when taking $\mu_F=\hat{\mu}_F$. The essential difference here is that the PDF effects do not cancel. 

As what regards the $\eta_c$ results shown on the first row of \cf{fig:sig-etaQ-scales}, we first note that 3 out of the 7 scale choices leads to negative cross section (see the incomplete curves) irrespective of the PDF choice. No matter what the PDF shape is, too ``large'' a value of $\mu_F$ inevitably leads to unacceptable results. This observation is obviously in line with previous studies~\cite{Feng:2015cba,Mangano:1996kg,Schuler:1994hy}. We now know that it is related to large negative contributions away from threshold due to a process-dependent oversubtraction of the collinear divergences which cannot be compensated by the PDF evolution, in particular at low scales where they are not much evolved.  Now, as anticipated during the discussion of $K^{\rm NLO}|_{y=0}$ plots, choosing $\mu_F=\hat{\mu}_F$ provides much more sound results. The results are particularly good up to the FCC energies with the JR14NLO08VF PDF as can be seen on \cf{fig:JR14NLO08VF_NLO_etac_EnEvo}.

Yet, as we discussed in section~\ref{sec:Gluon-luminosities}, most of the conventional PDFs exhibit, at low scales, a local minimum for $x$ around or below 0.001. This results in gluon luminosities --which correspond the expected kinematical dependencies for a simple LO $gg$ fusion--  essentially constant in the TeV range. Without any surprise, this what we observe for the $\hat{\mu}_F$ curve using PDF4LHC15\_nlo\_30 (\cf{fig:PDF4LHC_NLO_etac_EnEvo}) and to a lesser extent for NNPDF31sx\_nlonllx\_as\_0118 (\cf{fig:NNPDF31sxNLONLL_NLO_etac_EnEvo}). As expected, we observe the same with MMHT14nlo, CT14nlo and NNPDF31\_nlo\_as\_0118 on \cf{fig:centralplotetaQ} both at LO and NLO.  

The $\eta_c$ NLO energy dependence  admittedly does not make sense when it remains a constant  between $\sqrt{s}=10$~GeV and 10~TeV ! We urge the global fitters to examine whether global NLO fits cannot in fact be slightly amended in order to yield monotonous gluon PDFs at scale below 2~GeV. It is very important to realise that such a distorted shape is not at all due to the NLO corrections, but entirely due to the PDFs, as the JR14NLO08VF with monotonous gluon PDF case shows (\cf{fig:JR14NLO08VF_NLO_etac_EnEvo} and \cf{fig:NLOcentralplotetacCol}). 

Let us now turn to the $\eta_b$ case for which we know that the issue of over-subtraction is less problematic. Indeed, one only sees, on the second row of \cf{fig:sig-etaQ-scales}, a slight deviation at the FCC energies for the curves for $\mu_F=2M$ and  $\mu_F=0.5M$ which admittedly is the most critical one according to our $K^{\rm NLO}|_{y=0}$ analysis (see \cf{fig:K-etaQ}). 
Even though the $K_{\eta_b}^{\rm NLO}|_{y=0}$ are significantly $\mu_F$-dependent, the $d\sigma^{\rm NLO}_{\eta_b}/dy|_{y=0}$ 
are much less $\mu_F$-dependent as the difference in the --process-independent--  gluon evolution induced by the different chosen $\mu_F$ values  efficiently compensates the explicit $\mu_F$ dependence of the --process-dependent-- hard scattering for $\eta_b$. In such a case, the conventional scale choices would range between 5 and 20~GeV. Clearly, for $\eta_c$, such a compensation does not work at NLO and the only solution to the issue remains our scale choice $\mu_F=\hat{\mu}_F$.

This is why for the following plots we stick to this scale choice which we consider to give, at the moment, the best possible NLO predictions for $\eta_c$ production in the TeV range. On \cf{fig:sig-etaQ-PDFs}, we show again $d\sigma^{\rm NLO}_{\eta_Q}/dy|_{y=0}$ but with the PDF uncertainty associated with our PDF set choices for $\mu_R=\hat{\mu}_F$ (green band) 
and the $\mu_R$ uncertainty for $\mu_R \in [M/2:2M]$ (red band). The same observation as above can be made. Since $\hat{\mu}_F=1.82$~GeV for $\eta_c$, PDF4LHC15\_nlo\_30 results show a large distortion due the PDF eigenset shapes and, clearly, nobody would expect to see it in any experimental data in the future\footnote{We stress that $d\sigma^{\rm NLO}_{\eta_c}/dy|_{y=0}$ for $\mu_F=\hat{\mu}_F$  only get negative on \cf{fig:PDF4LHC_NLO_etac_EnEvo_Spec} and \cf{fig:NNPDF31sxNLONLL_NLO_etac_EnEvo_Spec} because some gluon NLO PDF eigensets get negative. Negative cross sections would disappear at LO since the gluon PDF are squared at $y=0$.}. The situation is better for NNPDF31sx\_nlonllx\_as\_0118 and remains very good for JR14NLO08VF. For $\eta_b$, the 3 PDF sets yield similar results. The smaller uncertainty band for JR14NLO08VF (\cf{fig:JR14NLO08VF_NLO_etab_EnEvo_Spec}) simply comes from the very small PDF uncertainty of this set. It is admittedly much smaller than the conventional ones. Possible experimental data should be able to test such PDFs rather straightforwardly.

To go further in the analysis of our improved NLO results with $\mu_F=\hat{\mu}_F$, 
we have plotted on \cf{fig:deltasigma} the relative uncertainties, dubbed as $\Delta \sigma/\sigma$, from the PDF uncertainties and from $\mu_R$ variations, obtained  by normalising the upper and lower values of $d\sigma_{\eta_Q}/dy|_{y=0}$ by the respective central ones, \ie\ from the central eigenset for the PDF uncertainty and from $\mu_R=M_\Q$ for the $\mu_R$ uncertainty.
The resulting  (green and red) bands   are compared to the LO $\mu_R$ uncertainty (horizontal dotted blue line)--which is obviously a constant. Our first observation is that the renormalisation scale uncertainty is clearly reduced at NLO, which is a good sign of the $\alphaS$ convergence even at these low scales. Second we note that the scale uncertainty for $\eta_c$ is smaller, for $\sqrt{s}>2$~TeV, than that from PDF4LHC15\_nlo\_30, on the order of 30 \% and then steadily growing; thich is representative of what NLO global fits would give. This means that forthcoming $\eta_c$ data at the LHC with a precision of 10 \%, or lower, should already be enough to improve PDF fits, even taking into account the $\mu_R$ uncertainty. For $\eta_b$, both are of similar sizes and one should probably look at $d\sigma^{\rm NLO}_{\eta_Q}/dy$ at fixed $\sqrt{s}$ or at various $\sqrt{s}$ if available to get more discriminating power on the PDF along the lines of~\cite{Zenaiev:2015rfa,Zenaiev:2019ktw,Gauld:2016kpd}. The same holds for $\eta_c$ at lower energies, which should then be differential in $y$. At this point, we wish to stress that we have decided not to vary the charm and beauty quark masses\footnote{Typical variations for quarkonium production studies are $m_c\in[1.4:1.6]$~GeV and $m_b\in[4.5:5.0]$~GeV.} which are also usually considered to yield additional theoretical uncertainties. We however note that the induced variations are highly correlated in $y$ (and $\sqrt{s}$) and such correlations can expediently be used to alleviate their effect in a possible PDF fit.

In view of the above observations, we have decided to show $d\sigma^{\rm NLO}_{\eta_Q}/dy$ at fixed $\sqrt{s}$ on \cf{fig:dsigdy-etaQ} only for our $\mu_F$ scale choice, only using JR14NLO08VF and its uncertainty (green band), along with the $\mu_R$ uncertainty (red band) compared to the LO results (gray and beige bands). Our objective with these plots is to provide NLO predictions to motivate prospects for measurements and then the extraction of constraints on PDFs, rather than to test quarkonium-production models.  From a kinematical viewpoint, $d\sigma^{\rm NLO}_{\eta_c}/dy$ measurement should offer reliable constraints on the gluon PDF $x$ dependence in the approximate range $[5\times 10^{-2}:1]$ for SPD, $[10^{-2}:1]$ for AFTER@LHCb and $[10^{-6}:5 \times 10^{-2}]$ for LHCb (at 14 TeV). At SPD, the $\eta_c$ production cross section is expected to be on the order of 1 pb. At AFTER@LHC, it would grow to 10 pb to reach 200 pb at the LHC. Of course, to these, one should apply rapidity-acceptance cuts, besides the appropriate branching fractions. As for the $\eta_b$, the $y$-integrated cross section, $\sigma^{\rm NLO}_{\eta_b}(s)$ is respectively expected to be 0.5~nb,  60~nb and 10~pb.

\section{Conclusion and outlook}
In this work, we have addressed the unphysical predictions of the collinear and NRQCD factorisations for the $P_T$-integrated quarkonium production, whereby negative cross sections are obtained for most of the conventional factorisation and renormalisation scale choices at LHC energies down to RHIC energies in some cases. In particular, we have focused on the pseudoscalar case, which is by far the simplest to tackle with analytical results available for the total cross section available since the mid nineties. On the way, we have provided analytical results for the rapidity differential cross sections, which were not available elsewhere. 

 We have shown that this unphysical behaviour can be explained through the high-energy limit of the partonic cross section, 
which is negative unless $\mu_F$ is chosen to be relatively small compared to the quarkonium mass. This negative limit can be ultimately traced back to an over-subtraction of the Altarelli-Parisi counterterm in the $\msbar$ scheme to absorb the collinear divergences inside the PDFs. This over-subtraction should usually be returned via the DGLAP evolution with steeper PDFs. However, the high-energy-limit values ${A}_{a}$ are \textit{process-dependent} while the DGLAP evolution is clearly \textit{process-independent} at fixed scales. This over-subtraction cannot be returned in a global manner by the PDFs and this mismatch badly affects the charmonium phenomenology as $\alphaS$ is not very small and the PDF evolution is limited which results in flat PDF shapes in the mid and low-$x$ region. 

Our solution to this issue is to propose a new scale setting, $\hat{\mu}_F$, which is based on the simple criterion that the partonic cross-section vanish at large $\hat{s}$ (or small $z$). This is to be understood as that the real-emission contributions coming from the initial partons are entirely absorbed in the PDF. Although less ambitious, this is somewhat equivalent to a resummation picture, yet much simpler to implement. 

We have demonstrated the efficiency of this new scale setting for a wide class of rapidity and energy shapes. We have applied this scale setting to $\eta_{c,b}$ production, a fictitious light elementary scalar boson $\tilde{H}^0$ and also for the real BEH boson $H^0$ with different mass values of a fictitious heavy-quark active in the loop. The success of this scale demonstrates that the issue we tackle is not in principle limited to quarkonium but rather to processes occurring at low scales, in particular when some unfavourable effects add up.

Having cured the NLO $\eta_{c,b}$ phenomenology from these negative cross sections, we have then provided predictions for SPD, AFTER@LHC, LHC up to FCC energies. Naturally, our NLO $\eta_{c}$ predictions bore on PDFs at low scales. Indeed, our scale choice for $\eta_{c}$ amounts to $2m_c/\sqrt{e}$, thus 1.82 GeV which is well within the usual scale range for a produced system whose mass is 3 GeV. However, at such a scale, conventional PDFs such as PDF4LHC15\_nlo\_30 which we used exhibit a local minimum for 
$x$ close or below 0.001. This translates into a distortion of the energy and rapidity dependence of the  cross-section in the TeV range that nobody would expect to see in experimental data. 

We thus encourage the PDF community to see what would be the outcome of a global NLO PDF fit preventing a local minimum in $x g(x)$, which is by the way absent in NNPDF31sx\_nlonllx\_as\_0118, where resummation effects have been taken into account in the PDF fit, and in JR14NLO08, where gluons are evolved from a lower scale. This local minimum also seems to contradict the recent analysis of Flett~\etal~\cite{Flett:2020duk}.

Similarly, we encourage the LHC experimental community to study such $P_T$-integrated quarkonium cross section, despite the likely challenging decay channels which should be used. $\eta_c$ has already been studied by LHCb at finite $P_T$ ($P_T>6$~GeV), they can definitely push down to 0 with a targeted effort, hopefully motivated by our study. As for $\eta_b$, which remains unobserved in hadroproduction, we have gathered some suggestions on how to extract its production cross section at the LHC. Indeed, these cross sections are definitely very large. Once they are available, we believe them to be ideal to better constrain the gluon PDF at low scales, and thus the gluon content of the proton in general, despite the remaining theoretical uncertainties inherent to quarkonium production. Extension to nuclei could then be considered along the lines of \cite{Kusina:2017gkz,Lansberg:2016deg}.

In the near future, it remains to be investigated the effect of such a scale setting for $\chi_c$ and $J/\psi$ production. The situation for the $\chi_c$ and $J/\psi$ is much worse~\cite{Feng:2015cba,Brodsky:2009cf} than that for $\eta_c$ as one already encounters negative yields in hadroproduction at $\sqrt{s}$ as low as a couple of hundreds of GeV. With our scale setting, we expect that both $\chi_c$ and $J/\psi$ NLO cross section will stabilise and give physical cross-section results which  can then be used in NLO PDF fits.

\vspace*{-.5cm}
\section*{Acknowledgements}
\vspace*{-.25cm}

We are indebted to Y. Feng for cross checks of our codes with FDC. We thank S.~Abreu, N.~Armesto, S.~Barsuk, M.~Becchetti, V.~Bertone, D.~Boer, M.~Bonvini, C.~Duhr, M.G. Echevarria, C.~Flett, C.~Flore, M.~Garzelli, B. Gong, R.~Harlander, L.~Harland-Lang, T.~Kasemets, A.~Kusina, M.~Mangano, S.~Marzani, M.~Nefedov, C.~Pisano, J.W.~Qiu, H.~Sazdjian, I.~Schienbein, H.S.~Shao, A.~Signori, M.~Spira, L.~Szymanowski, A.~Usachov, S.~Wallon, J.X.~Wang, for their help, insightful comments and useful suggestions on our study.

This project has received funding from the European Union’s Horizon 2020 research and innovation programme under the grant agreement No.824093 in order to contribute to the EU Virtual AccessNLOAccess. This work was also partly supported by the French CNRS via the IN2P3 project GLUE@NLO and via the Franco-Chinese LIA FCPPL (Quarkonium4AFTER), by the Paris-Saclay U. via the P2I Department and by the P2IO Labex via the Gluodynamics project. M.A.O.'s work was partly supported by the ERC grant 637019 ``MathAm''. 
\vspace*{-.5cm}

\bibliographystyle{utphys}

\bibliography{CSM-NLO_Scale}

\providecommand{\href}[2]{#2}\begingroup\raggedright\begin{thebibliography}{100}

\bibitem{Kramer:2001hh}
M.~Kraemer, ``{Quarkonium production at high-energy colliders},''
  \href{http://dx.doi.org/10.1016/S0146-6410(01)00154-5}{{\em Prog. Part. Nucl.
  Phys.} {\bfseries 47} (2001) 141--201},
\href{http://arxiv.org/abs/hep-ph/0106120}{{\ttfamily arXiv:hep-ph/0106120
  [hep-ph]}}.

\bibitem{Brambilla:2004wf}
{\bfseries Quarkonium Working Group} Collaboration, N.~Brambilla {\em et~al.},
  ``{Heavy quarkonium physics},''
\href{http://arxiv.org/abs/hep-ph/0412158}{{\ttfamily arXiv:hep-ph/0412158
  [hep-ph]}}.

\bibitem{Lansberg:2006dh}
J.~P. Lansberg, ``{$J/\psi$, $\psi'$ and $\Upsilon$ production at hadron
  colliders: A Review},''
  \href{http://dx.doi.org/10.1142/S0217751X06033180}{{\em Int. J. Mod. Phys.}
  {\bfseries A21} (2006) 3857--3916},
\href{http://arxiv.org/abs/hep-ph/0602091}{{\ttfamily arXiv:hep-ph/0602091
  [hep-ph]}}.

\bibitem{Brambilla:2010cs}
N.~Brambilla {\em et~al.}, ``{Heavy Quarkonium: Progress, Puzzles, and
  Opportunities},''
  \href{http://dx.doi.org/10.1140/epjc/s10052-010-1534-9}{{\em Eur. Phys. J.}
  {\bfseries C71} (2011) 1534},
\href{http://arxiv.org/abs/1010.5827}{{\ttfamily arXiv:1010.5827 [hep-ph]}}.

\bibitem{Andronic:2015wma}
A.~Andronic {\em et~al.}, ``{Heavy-flavour and quarkonium production in the LHC
  era: from proton–proton to heavy-ion collisions},''
  \href{http://dx.doi.org/10.1140/epjc/s10052-015-3819-5}{{\em Eur. Phys. J.}
  {\bfseries C76} no.~3, (2016) 107},
\href{http://arxiv.org/abs/1506.03981}{{\ttfamily arXiv:1506.03981 [nucl-ex]}}.

\bibitem{Lansberg:2019adr}
J.-P. Lansberg, ``{New Observables in Inclusive Production of Quarkonia},''
  \href{http://dx.doi.org/10.1016/j.physrep.2020.08.007}{{\em Phys. Rept.}
  {\bfseries 889} (2020) 1--106},
\href{http://arxiv.org/abs/1903.09185}{{\ttfamily arXiv:1903.09185 [hep-ph]}}.

\bibitem{Halzen:1984rq}
F.~Halzen, F.~Herzog, E.~W.~N. Glover, and A.~D. Martin, ``{The $J/\psi$ as a
  Trigger in $\bar{p} p$ Collisions},''
\href{http://dx.doi.org/10.1103/PhysRevD.30.700}{{\em Phys. Rev.} {\bfseries
  D30} (1984) 700}.

\bibitem{Martin:1987ww}
A.~D. Martin, C.~K. Ng, and W.~J. Stirling, ``{Inelastic Leptoproduction of
  $J/\psi$ as a Probe of the Small X Behavior of the Gluon Structure
  Function},''
\href{http://dx.doi.org/10.1016/0370-2693(87)91347-5}{{\em Phys. Lett.}
  {\bfseries B191} (1987) 200}.

\bibitem{Martin:1987vw}
A.~D. Martin, R.~G. Roberts, and W.~J. Stirling, ``{Structure Function Analysis
  and psi, Jet, W, Z Production: Pinning Down the Gluon},''
\href{http://dx.doi.org/10.1103/PhysRevD.37.1161}{{\em Phys. Rev.} {\bfseries
  D37} (1988) 1161}.

\bibitem{Jung:1992uj}
H.~Jung, G.~A. Schuler, and J.~Terron, ``{J / psi production mechanisms and
  determination of the gluon density at HERA},''
\href{http://dx.doi.org/10.1142/S0217751X92003604}{{\em Int. J. Mod. Phys.}
  {\bfseries A7} (1992) 7955--7988}.

\bibitem{Boer:2012bt}
D.~Boer and C.~Pisano, ``{Polarized gluon studies with charmonium and
  bottomonium at LHCb and AFTER},''
  \href{http://dx.doi.org/10.1103/PhysRevD.86.094007}{{\em Phys. Rev.}
  {\bfseries D86} (2012) 094007},
\href{http://arxiv.org/abs/1208.3642}{{\ttfamily arXiv:1208.3642 [hep-ph]}}.

\bibitem{Dunnen:2014eta}
W.~J. den Dunnen, J.~P. Lansberg, C.~Pisano, and M.~Schlegel, ``{Accessing the
  Transverse Dynamics and Polarization of Gluons inside the Proton at the
  LHC},'' \href{http://dx.doi.org/10.1103/PhysRevLett.112.212001}{{\em Phys.
  Rev. Lett.} {\bfseries 112} (2014) 212001},
\href{http://arxiv.org/abs/1401.7611}{{\ttfamily arXiv:1401.7611 [hep-ph]}}.

\bibitem{Boer:2016bfj}
D.~Boer, ``{Gluon TMDs in quarkonium production},''
  \href{http://dx.doi.org/10.1007/s00601-016-1198-6}{{\em Few Body Syst.}
  {\bfseries 58} no.~2, (2017) 32},
\href{http://arxiv.org/abs/1611.06089}{{\ttfamily arXiv:1611.06089 [hep-ph]}}.

\bibitem{Lansberg:2017dzg}
J.-P. Lansberg, C.~Pisano, F.~Scarpa, and M.~Schlegel, ``{Pinning down the
  linearly-polarised gluons inside unpolarised protons using quarkonium-pair
  production at the LHC},''
  \href{http://dx.doi.org/10.1016/j.physletb.2018.08.004,
  10.1016/j.physletb.2019.01.057}{{\em Phys. Lett.} {\bfseries B784} (2018)
  217--222}, \href{http://arxiv.org/abs/1710.01684}{{\ttfamily arXiv:1710.01684
  [hep-ph]}}.
[Erratum: Phys. Lett.B791,420(2019)].

\bibitem{Lansberg:2017tlc}
J.-P. Lansberg, C.~Pisano, and M.~Schlegel, ``{Associated production of a
  dilepton and a $\Upsilon(J/\psi)$ at the LHC as a probe of gluon transverse
  momentum dependent distributions},''
  \href{http://dx.doi.org/10.1016/j.nuclphysb.2017.04.011}{{\em Nucl. Phys.}
  {\bfseries B920} (2017) 192--210},
\href{http://arxiv.org/abs/1702.00305}{{\ttfamily arXiv:1702.00305 [hep-ph]}}.

\bibitem{Lansberg:2018fwx}
J.-P. Lansberg, C.~Pisano, F.~Scarpa, and M.~Schlegel, ``{Probing the gluon
  TMDs with quarkonia},'' \href{http://dx.doi.org/10.22323/1.316.0159}{{\em
  PoS} {\bfseries DIS2018} (2018) 159},
\href{http://arxiv.org/abs/1808.09866}{{\ttfamily arXiv:1808.09866 [hep-ph]}}.

\bibitem{Bacchetta:2018ivt}
A.~Bacchetta, D.~Boer, C.~Pisano, and P.~Taels, ``{Gluon TMDs and NRQCD matrix
  elements in $J/\psi$ production at an EIC},''
  \href{http://dx.doi.org/10.1140/epjc/s10052-020-7620-8}{{\em Eur. Phys. J.}
  {\bfseries C80} no.~1, (2020) 72},
\href{http://arxiv.org/abs/1809.02056}{{\ttfamily arXiv:1809.02056 [hep-ph]}}.

\bibitem{DAlesio:2019qpk}
U.~D'Alesio, F.~Murgia, C.~Pisano, and P.~Taels, ``{Azimuthal asymmetries in
  semi-inclusive $J/\psi\,+\,\mathrm{jet}$ production at an EIC},''
  \href{http://dx.doi.org/10.1103/PhysRevD.100.094016}{{\em Phys. Rev.}
  {\bfseries D100} no.~9, (2019) 094016},
\href{http://arxiv.org/abs/1908.00446}{{\ttfamily arXiv:1908.00446 [hep-ph]}}.

\bibitem{Kishore:2019fzb}
R.~Kishore, A.~Mukherjee, and S.~Rajesh, ``{Sivers asymmetry in the
  photoproduction of a $J/\psi$ and a jet at the EIC},''
  \href{http://dx.doi.org/10.1103/PhysRevD.101.054003}{{\em Phys. Rev.}
  {\bfseries D101} no.~5, (2020) 054003},
\href{http://arxiv.org/abs/1908.03698}{{\ttfamily arXiv:1908.03698 [hep-ph]}}.

\bibitem{Scarpa:2019fol}
F.~Scarpa, D.~Boer, M.~G. Echevarria, J.-P. Lansberg, C.~Pisano, and
  M.~Schlegel, ``{Studies of gluon TMDs and their evolution using
  quarkonium-pair production at the LHC},''
  \href{http://dx.doi.org/10.1140/epjc/s10052-020-7619-1}{{\em Eur. Phys. J.}
  {\bfseries C80} no.~2, (2020) 87},
\href{http://arxiv.org/abs/1909.05769}{{\ttfamily arXiv:1909.05769 [hep-ph]}}.

\bibitem{Boer:2020bbd}
D.~Boer, U.~D'Alesio, F.~Murgia, C.~Pisano, and P.~Taels, ``{$J/\psi$ meson
  production in SIDIS: matching high and low transverse momentum},''
  \href{http://dx.doi.org/10.1007/JHEP09(2020)040}{{\em JHEP} {\bfseries 09}
  (2020) 040},
\href{http://arxiv.org/abs/2004.06740}{{\ttfamily arXiv:2004.06740 [hep-ph]}}.

\bibitem{Kramer:1995nb}
M.~Kraemer, ``{QCD corrections to inelastic J / psi photoproduction},''
  \href{http://dx.doi.org/10.1016/0550-3213(95)00568-4}{{\em Nucl. Phys.}
  {\bfseries B459} (1996) 3--50},
\href{http://arxiv.org/abs/hep-ph/9508409}{{\ttfamily arXiv:hep-ph/9508409
  [hep-ph]}}.

\bibitem{Artoisenet:2008fc}
P.~Artoisenet, J.~M. Campbell, J.~Lansberg, F.~Maltoni, and F.~Tramontano,
  ``{$\Upsilon$ Production at Fermilab Tevatron and LHC Energies},''
  \href{http://dx.doi.org/10.1103/PhysRevLett.101.152001}{{\em Phys. Rev.
  Lett.} {\bfseries 101} (2008) 152001},
  \href{http://arxiv.org/abs/0806.3282}{{\ttfamily arXiv:0806.3282 [hep-ph]}}.

\bibitem{Lansberg:2008gk}
J.~Lansberg, ``{On the mechanisms of heavy-quarkonium hadroproduction},''
  \href{http://dx.doi.org/10.1140/epjc/s10052-008-0826-9}{{\em Eur. Phys. J. C}
  {\bfseries 61} (2009) 693--703},
  \href{http://arxiv.org/abs/0811.4005}{{\ttfamily arXiv:0811.4005 [hep-ph]}}.

\bibitem{Lansberg:2009db}
J.~Lansberg, ``{Real next-to-next-to-leading-order QCD corrections to $J/\psi$
  and Upsilon hadroproduction in association with a photon},''
  \href{http://dx.doi.org/10.1016/j.physletb.2009.07.067}{{\em Phys. Lett. B}
  {\bfseries 679} (2009) 340--346},
  \href{http://arxiv.org/abs/0901.4777}{{\ttfamily arXiv:0901.4777 [hep-ph]}}.

\bibitem{Gong:2012ah}
B.~Gong, J.-P. Lansberg, C.~Lorce, and J.~Wang, ``{Next-to-leading-order QCD
  corrections to the yields and polarisations of J/Psi and Upsilon directly
  produced in association with a Z boson at the LHC},''
  \href{http://dx.doi.org/10.1007/JHEP03(2013)115}{{\em JHEP} {\bfseries 03}
  (2013) 115},
\href{http://arxiv.org/abs/1210.2430}{{\ttfamily arXiv:1210.2430 [hep-ph]}}.

\bibitem{Lansberg:2013qka}
J.-P. Lansberg and H.-S. Shao, ``{Production of $J/\psi + \eta_{c}$ versus
  $J/\psi + J/\psi$ at the LHC: Importance of Real $\alpha^{5}_{s}$
  Corrections},'' \href{http://dx.doi.org/10.1103/PhysRevLett.111.122001}{{\em
  Phys. Rev. Lett.} {\bfseries 111} (2013) 122001},
\href{http://arxiv.org/abs/1308.0474}{{\ttfamily arXiv:1308.0474 [hep-ph]}}.

\bibitem{Lansberg:2014swa}
J.-P. Lansberg and H.-S. Shao, ``{J/psi -pair production at large momenta:
  Indications for double parton scatterings and large $\alpha_s^5$
  contributions},''
  \href{http://dx.doi.org/10.1016/j.physletb.2015.10.083}{{\em Phys. Lett.}
  {\bfseries B751} (2015) 479--486},
\href{http://arxiv.org/abs/1410.8822}{{\ttfamily arXiv:1410.8822 [hep-ph]}}.

\bibitem{Chang:1979nn}
C.-H. Chang, ``{Hadronic Production of $J/\psi$ Associated With a Gluon},''
\href{http://dx.doi.org/10.1016/0550-3213(80)90175-3}{{\em Nucl. Phys.}
  {\bfseries B172} (1980) 425--434}.

\bibitem{Berger:1980ni}
E.~L. Berger and D.~L. Jones, ``{Inelastic Photoproduction of J/psi and Upsilon
  by Gluons},''
\href{http://dx.doi.org/10.1103/PhysRevD.23.1521}{{\em Phys. Rev.} {\bfseries
  D23} (1981) 1521--1530}.

\bibitem{Baier:1983va}
R.~Baier and R.~Ruckl, ``{Hadronic Collisions: A Quarkonium Factory},''
\href{http://dx.doi.org/10.1007/BF01572254}{{\em Z. Phys.} {\bfseries C19}
  (1983) 251}.

\bibitem{Bodwin:1994jh}
G.~T. Bodwin, E.~Braaten, and G.~P. Lepage, ``{Rigorous QCD analysis of
  inclusive annihilation and production of heavy quarkonium},''
  \href{http://dx.doi.org/10.1103/PhysRevD.55.5853,
  10.1103/PhysRevD.51.1125}{{\em Phys. Rev.} {\bfseries D51} (1995)
  1125--1171}, \href{http://arxiv.org/abs/hep-ph/9407339}{{\ttfamily
  arXiv:hep-ph/9407339 [hep-ph]}}.
[Erratum: Phys. Rev.D55,5853(1997)].

\bibitem{Fritzsch:1977ay}
H.~Fritzsch, ``{Producing Heavy Quark Flavors in Hadronic Collisions: A Test of
  Quantum Chromodynamics},''
\href{http://dx.doi.org/10.1016/0370-2693(77)90108-3}{{\em Phys. Lett.}
  {\bfseries 67B} (1977) 217--221}.

\bibitem{Halzen:1977rs}
F.~Halzen, ``{Cvc for Gluons and Hadroproduction of Quark Flavors},''
\href{http://dx.doi.org/10.1016/0370-2693(77)90144-7}{{\em Phys. Lett.}
  {\bfseries 69B} (1977) 105--108}.

\bibitem{Lansberg:2016rcx}
J.-P. Lansberg and H.-S. Shao, ``{Associated production of a quarkonium and a Z
  boson at one loop in a quark-hadron-duality approach},''
  \href{http://dx.doi.org/10.1007/JHEP10(2016)153}{{\em JHEP} {\bfseries 10}
  (2016) 153},
\href{http://arxiv.org/abs/1608.03198}{{\ttfamily arXiv:1608.03198 [hep-ph]}}.

\bibitem{Lansberg:2020rft}
J.-P. Lansberg, H.-S. Shao, N.~Yamanaka, Y.-J. Zhang, and C.~No\^us,
  ``{Complete NLO QCD study of single- and double-quarkonium hadroproduction in
  the colour-evaporation model at the Tevatron and the LHC},''
  \href{http://dx.doi.org/10.1016/j.physletb.2020.135559}{{\em Phys. Lett.}
  {\bfseries B807} (2020) 135559},
\href{http://arxiv.org/abs/2004.14345}{{\ttfamily arXiv:2004.14345 [hep-ph]}}.

\bibitem{Feng:2015cba}
Y.~Feng, J.-P. Lansberg, and J.-X. Wang, ``{Energy dependence of
  direct-quarkonium production in $pp$ collisions from fixed-target to LHC
  energies: complete one-loop analysis},''
  \href{http://dx.doi.org/10.1140/epjc/s10052-015-3527-1}{{\em Eur. Phys. J.}
  {\bfseries C75} no.~7, (2015) 313},
\href{http://arxiv.org/abs/1504.00317}{{\ttfamily arXiv:1504.00317 [hep-ph]}}.

\bibitem{Maltoni:2006yp}
F.~Maltoni {\em et~al.}, ``{Analysis of charmonium production at fixed-target
  experiments in the NRQCD approach},''
  \href{http://dx.doi.org/10.1016/j.physletb.2006.05.010}{{\em Phys. Lett.}
  {\bfseries B638} (2006) 202--208},
\href{http://arxiv.org/abs/hep-ph/0601203}{{\ttfamily arXiv:hep-ph/0601203
  [hep-ph]}}.

\bibitem{Brodsky:2009cf}
S.~J. Brodsky and J.-P. Lansberg, ``{Heavy-Quarkonium Production in High Energy
  Proton-Proton Collisions at RHIC},''
  \href{http://dx.doi.org/10.1103/PhysRevD.81.051502}{{\em Phys. Rev. D}
  {\bfseries 81} (2010) 051502},
  \href{http://arxiv.org/abs/0908.0754}{{\ttfamily arXiv:0908.0754 [hep-ph]}}.

\bibitem{Schuler:1994hy}
G.~A. Schuler, ``{Quarkonium production and decays},'' 1994.

\bibitem{Mangano:1996kg}
M.~L. Mangano and A.~Petrelli, ``{NLO quarkonium production in hadronic
  collisions},'' \href{http://dx.doi.org/10.1142/S0217751X97002048}{{\em Int.
  J. Mod. Phys.} {\bfseries A12} (1997) 3887--3897},
\href{http://arxiv.org/abs/hep-ph/9610364}{{\ttfamily arXiv:hep-ph/9610364
  [hep-ph]}}.

\bibitem{Ozcelik:2019qze}
M.~A. Ozcelik, ``{Constraining gluon PDFs with quarkonium production},''
  \href{http://dx.doi.org/10.22323/1.352.0159}{{\em PoS} {\bfseries DIS2019}
  (2019) 159},
\href{http://arxiv.org/abs/1907.01400}{{\ttfamily arXiv:1907.01400 [hep-ph]}}.

\bibitem{Brock:1993sz}
{\bfseries CTEQ} Collaboration, R.~Brock {\em et~al.}, ``{Handbook of
  perturbative QCD: Version 1.0},''
\href{http://dx.doi.org/10.1103/RevModPhys.67.157}{{\em Rev. Mod. Phys.}
  {\bfseries 67} (1995) 157--248}.

\bibitem{tHooft:1972tcz}
G.~'t~Hooft and M.~Veltman, ``{Regularization and Renormalization of Gauge
  Fields},'' \href{http://dx.doi.org/10.1016/0550-3213(72)90279-9}{{\em Nucl.
  Phys. B} {\bfseries 44} (1972) 189--213}.

\bibitem{Kuhn:1992qw}
J.~H. Kuhn and E.~Mirkes, ``{QCD corrections to toponium production at hadron
  colliders},'' \href{http://dx.doi.org/10.1103/PhysRevD.48.179}{{\em Phys.
  Rev. D} {\bfseries 48} (1993) 179--189},
  \href{http://arxiv.org/abs/hep-ph/9301204}{{\ttfamily arXiv:hep-ph/9301204}}.

\bibitem{Petrelli:1997ge}
A.~Petrelli, M.~Cacciari, M.~Greco, F.~Maltoni, and M.~L. Mangano, ``{NLO
  production and decay of quarkonium},''
  \href{http://dx.doi.org/10.1016/S0550-3213(97)00801-8}{{\em Nucl. Phys. B}
  {\bfseries 514} (1998) 245--309},
  \href{http://arxiv.org/abs/hep-ph/9707223}{{\ttfamily arXiv:hep-ph/9707223}}.

\bibitem{Klasen:2004tz}
M.~Klasen, B.~Kniehl, L.~Mihaila, and M.~Steinhauser, ``{$J/\psi$ plus jet
  associated production in two-photon collisions at next-to-leading order},''
  \href{http://dx.doi.org/10.1016/j.nuclphysb.2005.02.009}{{\em Nucl. Phys. B}
  {\bfseries 713} (2005) 487--521},
  \href{http://arxiv.org/abs/hep-ph/0407014}{{\ttfamily arXiv:hep-ph/0407014}}.

\bibitem{Altarelli:1977zs}
G.~Altarelli and G.~Parisi, ``{Asymptotic Freedom in Parton Language},''
  \href{http://dx.doi.org/10.1016/0550-3213(77)90384-4}{{\em Nucl. Phys. B}
  {\bfseries 126} (1977) 298--318}.

\bibitem{Wang:2004du}
J.-X. Wang, ``{Progress in FDC project},''
  \href{http://dx.doi.org/10.1016/j.nima.2004.07.094}{{\em Nucl. Instrum. Meth.
  A} {\bfseries 534} (2004) 241--245},
  \href{http://arxiv.org/abs/hep-ph/0407058}{{\ttfamily arXiv:hep-ph/0407058}}.

\bibitem{Harlander:2009my}
R.~V. Harlander, H.~Mantler, S.~Marzani, and K.~J. Ozeren, ``{Higgs production
  in gluon fusion at next-to-next-to-leading order QCD for finite top mass},''
  \href{http://dx.doi.org/10.1140/epjc/s10052-010-1258-x}{{\em Eur. Phys. J. C}
  {\bfseries 66} (2010) 359--372},
  \href{http://arxiv.org/abs/0912.2104}{{\ttfamily arXiv:0912.2104 [hep-ph]}}.

\bibitem{Dawson:1990zj}
S.~Dawson, ``{Radiative corrections to Higgs boson production},''
  \href{http://dx.doi.org/10.1016/0550-3213(91)90061-2}{{\em Nucl. Phys. B}
  {\bfseries 359} (1991) 283--300}.

\bibitem{Graudenz:1992pv}
D.~Graudenz, M.~Spira, and P.~Zerwas, ``{QCD corrections to Higgs boson
  production at proton proton colliders},''
  \href{http://dx.doi.org/10.1103/PhysRevLett.70.1372}{{\em Phys. Rev. Lett.}
  {\bfseries 70} (1993) 1372--1375}.

\bibitem{Spira:1995rr}
M.~Spira, A.~Djouadi, D.~Graudenz, and P.~Zerwas, ``{Higgs boson production at
  the LHC},'' \href{http://dx.doi.org/10.1016/0550-3213(95)00379-7}{{\em Nucl.
  Phys. B} {\bfseries 453} (1995) 17--82},
  \href{http://arxiv.org/abs/hep-ph/9504378}{{\ttfamily arXiv:hep-ph/9504378}}.

\bibitem{Candido:2020yat}
A.~Candido, S.~Forte, and F.~Hekhorn, ``{Can $\overline{\rm { MS}}$ parton
  distributions be negative?},''
\href{http://arxiv.org/abs/2006.07377}{{\ttfamily arXiv:2006.07377 [hep-ph]}}.

\bibitem{Maltoni:2007tc}
F.~Maltoni, T.~McElmurry, R.~Putman, and S.~Willenbrock, ``{Choosing the
  Factorization Scale in Perturbative QCD},''
\href{http://arxiv.org/abs/hep-ph/0703156}{{\ttfamily arXiv:hep-ph/0703156
  [HEP-PH]}}.

\bibitem{Ball:2017otu}
R.~D. Ball, V.~Bertone, M.~Bonvini, S.~Marzani, J.~Rojo, and L.~Rottoli,
  ``{Parton distributions with small-x resummation: evidence for BFKL dynamics
  in HERA data},'' \href{http://dx.doi.org/10.1140/epjc/s10052-018-5774-4}{{\em
  Eur. Phys. J. C} {\bfseries 78} no.~4, (2018) 321},
  \href{http://arxiv.org/abs/1710.05935}{{\ttfamily arXiv:1710.05935
  [hep-ph]}}.

\bibitem{Abdolmaleki:2018jln}
{\bfseries xFitter Developers' Team} Collaboration, H.~Abdolmaleki {\em
  et~al.}, ``{Impact of low-$x$ resummation on QCD analysis of HERA data},''
  \href{http://dx.doi.org/10.1140/epjc/s10052-018-6090-8}{{\em Eur. Phys. J.}
  {\bfseries C78} no.~8, (2018) 621},
\href{http://arxiv.org/abs/1802.00064}{{\ttfamily arXiv:1802.00064 [hep-ph]}}.

\bibitem{Buckley:2014ana}
A.~Buckley, J.~Ferrando, S.~Lloyd, K.~Nordström, B.~Page, M.~Rüfenacht,
  M.~Schönherr, and G.~Watt, ``{LHAPDF6: parton density access in the LHC
  precision era},''
  \href{http://dx.doi.org/10.1140/epjc/s10052-015-3318-8}{{\em Eur. Phys. J. C}
  {\bfseries 75} (2015) 132}, \href{http://arxiv.org/abs/1412.7420}{{\ttfamily
  arXiv:1412.7420 [hep-ph]}}.

\bibitem{Butterworth:2015oua}
J.~Butterworth {\em et~al.}, ``{PDF4LHC recommendations for LHC Run II},''
  \href{http://dx.doi.org/10.1088/0954-3899/43/2/023001}{{\em J. Phys.}
  {\bfseries G43} (2016) 023001},
\href{http://arxiv.org/abs/1510.03865}{{\ttfamily arXiv:1510.03865 [hep-ph]}}.

\bibitem{Jimenez-Delgado:2014twa}
P.~Jimenez-Delgado and E.~Reya, ``{Delineating parton distributions and the
  strong coupling},'' \href{http://dx.doi.org/10.1103/PhysRevD.89.074049}{{\em
  Phys. Rev.} {\bfseries D89} no.~7, (2014) 074049},
\href{http://arxiv.org/abs/1403.1852}{{\ttfamily arXiv:1403.1852 [hep-ph]}}.

\bibitem{Dulat:2015mca}
S.~Dulat, T.-J. Hou, J.~Gao, M.~Guzzi, J.~Huston, P.~Nadolsky, J.~Pumplin,
  C.~Schmidt, D.~Stump, and C.~P. Yuan, ``{New parton distribution functions
  from a global analysis of quantum chromodynamics},''
  \href{http://dx.doi.org/10.1103/PhysRevD.93.033006}{{\em Phys. Rev.}
  {\bfseries D93} no.~3, (2016) 033006},
\href{http://arxiv.org/abs/1506.07443}{{\ttfamily arXiv:1506.07443 [hep-ph]}}.

\bibitem{Harland-Lang:2014zoa}
L.~A. Harland-Lang, A.~D. Martin, P.~Motylinski, and R.~S. Thorne, ``{Parton
  distributions in the LHC era: MMHT 2014 PDFs},''
  \href{http://dx.doi.org/10.1140/epjc/s10052-015-3397-6}{{\em Eur. Phys. J.}
  {\bfseries C75} no.~5, (2015) 204},
\href{http://arxiv.org/abs/1412.3989}{{\ttfamily arXiv:1412.3989 [hep-ph]}}.

\bibitem{Ball:2017nwa}
{\bfseries NNPDF} Collaboration, R.~D. Ball {\em et~al.}, ``{Parton
  distributions from high-precision collider data},''
  \href{http://dx.doi.org/10.1140/epjc/s10052-017-5199-5}{{\em Eur. Phys. J.}
  {\bfseries C77} no.~10, (2017) 663},
\href{http://arxiv.org/abs/1706.00428}{{\ttfamily arXiv:1706.00428 [hep-ph]}}.

\bibitem{Flett:2020duk}
C.~A. Flett, A.~D. Martin, M.~G. Ryskin, and T.~Teubner, ``{Very low $x$ gluon
  density determined by LHCb exclusive $J/\psi$ data},''
\href{http://arxiv.org/abs/2006.13857}{{\ttfamily arXiv:2006.13857 [hep-ph]}}.

\bibitem{Bertone:2013vaa}
V.~Bertone, S.~Carrazza, and J.~Rojo, ``{APFEL: A PDF Evolution Library with
  QED corrections},'' \href{http://dx.doi.org/10.1016/j.cpc.2014.03.007}{{\em
  Comput. Phys. Commun.} {\bfseries 185} (2014) 1647--1668},
  \href{http://arxiv.org/abs/1310.1394}{{\ttfamily arXiv:1310.1394 [hep-ph]}}.

\bibitem{Carrazza:2014gfa}
S.~Carrazza, A.~Ferrara, D.~Palazzo, and J.~Rojo, ``{APFEL Web}: {a web-based
  application for the graphical visualization of parton distribution
  functions},'' \href{http://dx.doi.org/10.1088/0954-3899/42/5/057001}{{\em J.
  Phys. G} {\bfseries 42} no.~5, (2015) 057001},
  \href{http://arxiv.org/abs/1410.5456}{{\ttfamily arXiv:1410.5456 [hep-ph]}}.

\bibitem{Ball:2013bra}
R.~D. Ball, M.~Bonvini, S.~Forte, S.~Marzani, and G.~Ridolfi, ``{Higgs
  production in gluon fusion beyond NNLO},''
  \href{http://dx.doi.org/10.1016/j.nuclphysb.2013.06.012}{{\em Nucl. Phys. B}
  {\bfseries 874} (2013) 746--772},
  \href{http://arxiv.org/abs/1303.3590}{{\ttfamily arXiv:1303.3590 [hep-ph]}}.

\bibitem{Bonvini:2014jma}
M.~Bonvini, R.~D. Ball, S.~Forte, S.~Marzani, and G.~Ridolfi, ``{Updated Higgs
  cross section at approximate N$^3$LO},''
  \href{http://dx.doi.org/10.1088/0954-3899/41/9/095002}{{\em J. Phys. G}
  {\bfseries 41} (2014) 095002},
  \href{http://arxiv.org/abs/1404.3204}{{\ttfamily arXiv:1404.3204 [hep-ph]}}.

\bibitem{Bonvini:2016frm}
M.~Bonvini, S.~Marzani, C.~Muselli, and L.~Rottoli, ``{On the Higgs cross
  section at N$^{3}$LO+N$^{3}$LL and its uncertainty},''
  \href{http://dx.doi.org/10.1007/JHEP08(2016)105}{{\em JHEP} {\bfseries 08}
  (2016) 105},
\href{http://arxiv.org/abs/1603.08000}{{\ttfamily arXiv:1603.08000 [hep-ph]}}.

\bibitem{Bonciani:2007ex}
R.~Bonciani, G.~Degrassi, and A.~Vicini, ``{Scalar particle contribution to
  Higgs production via gluon fusion at NLO},''
  \href{http://dx.doi.org/10.1088/1126-6708/2007/11/095}{{\em JHEP} {\bfseries
  11} (2007) 095},
\href{http://arxiv.org/abs/0709.4227}{{\ttfamily arXiv:0709.4227 [hep-ph]}}.

\bibitem{Anastasiou:2015ema}
C.~Anastasiou, C.~Duhr, F.~Dulat, F.~Herzog, and B.~Mistlberger, ``{Higgs Boson
  Gluon-Fusion Production in QCD at Three Loops},''
  \href{http://dx.doi.org/10.1103/PhysRevLett.114.212001}{{\em Phys. Rev.
  Lett.} {\bfseries 114} (2015) 212001},
\href{http://arxiv.org/abs/1503.06056}{{\ttfamily arXiv:1503.06056 [hep-ph]}}.

\bibitem{Anastasiou:2016cez}
C.~Anastasiou, C.~Duhr, F.~Dulat, E.~Furlan, T.~Gehrmann, F.~Herzog,
  A.~Lazopoulos, and B.~Mistlberger, ``{High precision determination of the
  gluon fusion Higgs boson cross-section at the LHC},''
  \href{http://dx.doi.org/10.1007/JHEP05(2016)058}{{\em JHEP} {\bfseries 05}
  (2016) 058},
\href{http://arxiv.org/abs/1602.00695}{{\ttfamily arXiv:1602.00695 [hep-ph]}}.

\bibitem{Aaij:2014bga}
{\bfseries LHCb} Collaboration, R.~Aaij {\em et~al.}, ``{Measurement of the
  $\eta_c (1S)$ production cross-section in proton-proton collisions via the
  decay $\eta_c (1S) \rightarrow p \bar{p}$},''
  \href{http://dx.doi.org/10.1140/epjc/s10052-015-3502-x}{{\em Eur. Phys. J.}
  {\bfseries C75} no.~7, (2015) 311},
\href{http://arxiv.org/abs/1409.3612}{{\ttfamily arXiv:1409.3612 [hep-ex]}}.

\bibitem{Aaij:2019gsn}
{\bfseries LHCb} Collaboration, R.~Aaij {\em et~al.}, ``{Measurement of the
  $\eta_c(1S)$ production cross-section in $pp$ collisions at $\sqrt{s} = 13$
  TeV},'' \href{http://dx.doi.org/10.1140/epjc/s10052-020-7733-0}{{\em Eur.
  Phys. J.} {\bfseries C80} no.~3, (2020) 191},
\href{http://arxiv.org/abs/1911.03326}{{\ttfamily arXiv:1911.03326 [hep-ex]}}.

\bibitem{Aubert:2008ba}
{\bfseries BaBar} Collaboration, B.~Aubert {\em et~al.}, ``{Observation of the
  bottomonium ground state in the decay $\upsilon_{3S} \to \gamma eta_b$},''
  \href{http://dx.doi.org/10.1103/PhysRevLett.102.029901,
  10.1103/PhysRevLett.101.071801}{{\em Phys. Rev. Lett.} {\bfseries 101} (2008)
  071801}, \href{http://arxiv.org/abs/0807.1086}{{\ttfamily arXiv:0807.1086
  [hep-ex]}}.
[Erratum: Phys. Rev. Lett.102,029901(2009)].

\bibitem{Mizuk:2012pb}
{\bfseries Belle} Collaboration, R.~Mizuk {\em et~al.}, ``{Evidence for the
  $\eta_b(2S)$ and observation of $h_b(1P) \to \eta_b(1S) \gamma$ and $h_b(2P)
  \to \eta_b(1S) \gamma$},''
  \href{http://dx.doi.org/10.1103/PhysRevLett.109.232002}{{\em Phys. Rev.
  Lett.} {\bfseries 109} (2012) 232002},
\href{http://arxiv.org/abs/1205.6351}{{\ttfamily arXiv:1205.6351 [hep-ex]}}.

\bibitem{Aubert:2009as}
{\bfseries BaBar} Collaboration, B.~Aubert {\em et~al.}, ``{Evidence for the
  eta(b)(1S) Meson in Radiative Upsilon(2S) Decay},''
  \href{http://dx.doi.org/10.1103/PhysRevLett.103.161801}{{\em Phys. Rev.
  Lett.} {\bfseries 103} (2009) 161801},
\href{http://arxiv.org/abs/0903.1124}{{\ttfamily arXiv:0903.1124 [hep-ex]}}.

\bibitem{Bonvicini:2009hs}
{\bfseries CLEO} Collaboration, G.~Bonvicini {\em et~al.}, ``{Measurement of
  the eta(b)(1S) mass and the branching fraction for Upsilon(3S) ---> gamma
  eta(b)(1S)},'' \href{http://dx.doi.org/10.1103/PhysRevD.81.031104}{{\em Phys.
  Rev.} {\bfseries D81} (2010) 031104},
\href{http://arxiv.org/abs/0909.5474}{{\ttfamily arXiv:0909.5474 [hep-ex]}}.

\bibitem{Tamponi:2015xzb}
{\bfseries Belle} Collaboration, U.~Tamponi {\em et~al.}, ``{First observation
  of the hadronic transition $ \Upsilon(4S) \to \eta h_{b}(1P)$ and new
  measurement of the $h_b(1P)$ and $\eta_b(1S)$ parameters},''
  \href{http://dx.doi.org/10.1103/PhysRevLett.115.142001}{{\em Phys. Rev.
  Lett.} {\bfseries 115} no.~14, (2015) 142001},
\href{http://arxiv.org/abs/1506.08914}{{\ttfamily arXiv:1506.08914 [hep-ex]}}.

\bibitem{Kou:2018nap}
{\bfseries Belle-II} Collaboration, W.~Altmannshofer {\em et~al.}, ``{The Belle
  II Physics Book},'' \href{http://dx.doi.org/10.1093/ptep/ptz106,
  10.1093/ptep/ptaa008}{{\em PTEP} {\bfseries 2019} no.~12, (2019) 123C01},
  \href{http://arxiv.org/abs/1808.10567}{{\ttfamily arXiv:1808.10567
  [hep-ex]}}.
[Erratum: PTEP2020,no.2,029201(2020)].

\bibitem{Feng:2017hlu}
F.~Feng, Y.~Jia, and W.-L. Sang, ``{Next-to-Next-to-Leading-Order QCD
  Corrections to the Hadronic width of Pseudoscalar Quarkonium},''
  \href{http://dx.doi.org/10.1103/PhysRevLett.119.252001}{{\em Phys. Rev.
  Lett.} {\bfseries 119} no.~25, (2017) 252001},
\href{http://arxiv.org/abs/1707.05758}{{\ttfamily arXiv:1707.05758 [hep-ph]}}.

\bibitem{Zyla:2020zbs}
{\bfseries Particle Data Group} Collaboration, P.~A. Zyla {\em et~al.},
  ``{Review of Particle Physics},''
\href{http://dx.doi.org/10.1093/ptep/ptaa104}{{\em PTEP} {\bfseries 2020}
  no.~8, (2020) 083C01}.

\bibitem{Braaten:2000cm}
E.~Braaten, S.~Fleming, and A.~K. Leibovich, ``{NRQCD analysis of bottomonium
  production at the Tevatron},''
  \href{http://dx.doi.org/10.1103/PhysRevD.63.094006}{{\em Phys. Rev.}
  {\bfseries D63} (2001) 094006},
\href{http://arxiv.org/abs/hep-ph/0008091}{{\ttfamily arXiv:hep-ph/0008091
  [hep-ph]}}.

\bibitem{Maltoni:2004hv}
F.~Maltoni and A.~D. Polosa, ``{Observation potential for eta(b) at the
  Tevatron},'' \href{http://dx.doi.org/10.1103/PhysRevD.70.054014}{{\em Phys.
  Rev.} {\bfseries D70} (2004) 054014},
\href{http://arxiv.org/abs/hep-ph/0405082}{{\ttfamily arXiv:hep-ph/0405082
  [hep-ph]}}.

\bibitem{Hao:2006nf}
G.~Hao, Y.~Jia, C.-F. Qiao, and P.~Sun, ``{Hunting eta(b) through radiative
  decay into J/psi},''
  \href{http://dx.doi.org/10.1088/1126-6708/2007/02/057}{{\em JHEP} {\bfseries
  02} (2007) 057},
\href{http://arxiv.org/abs/hep-ph/0612173}{{\ttfamily arXiv:hep-ph/0612173
  [hep-ph]}}.

\bibitem{Gong:2008ue}
B.~Gong, Y.~Jia, and J.-X. Wang, ``{Exclusive eta(b) decay to double J / psi at
  next-to-leading order in alpha(s)},''
  \href{http://dx.doi.org/10.1016/j.physletb.2008.10.063}{{\em Phys. Lett.}
  {\bfseries B670} (2009) 350--355},
\href{http://arxiv.org/abs/0808.1034}{{\ttfamily arXiv:0808.1034 [hep-ph]}}.

\bibitem{Santorelli:2007xg}
P.~Santorelli, ``{Long-distance contributions to the etab--->J/psiJ/psi
  decay},'' \href{http://dx.doi.org/10.1103/PhysRevD.77.074012}{{\em Phys.
  Rev.} {\bfseries D77} (2008) 074012},
\href{http://arxiv.org/abs/hep-ph/0703232}{{\ttfamily arXiv:hep-ph/0703232
  [HEP-PH]}}.

\bibitem{Aaij:2020fnh}
{\bfseries LHCb} Collaboration, R.~Aaij {\em et~al.}, ``{Observation of
  structure in the $J/\psi$-pair mass spectrum},''
  \href{http://dx.doi.org/10.1016/j.scib.2020.08.032}{{\em Sci. Bull.}
  {\bfseries 2020} (2020) 65},
\href{http://arxiv.org/abs/2006.16957}{{\ttfamily arXiv:2006.16957 [hep-ex]}}.

\bibitem{Jia:2006rx}
Y.~Jia, ``{Which hadronic decay modes are good for eta(b) searching: double
  J/psi or something else?},''
  \href{http://dx.doi.org/10.1103/PhysRevD.78.054003}{{\em Phys. Rev.}
  {\bfseries D78} (2008) 054003},
\href{http://arxiv.org/abs/hep-ph/0611130}{{\ttfamily arXiv:hep-ph/0611130
  [hep-ph]}}.

\bibitem{Hao:2007rb}
G.~Hao, C.-F. Qiao, and P.~Sun, ``{Investigate the Bottomonium Ground State
  eta(b) via Its Inclusive Charm Decays},''
  \href{http://dx.doi.org/10.1103/PhysRevD.76.125013}{{\em Phys. Rev.}
  {\bfseries D76} (2007) 125013},
\href{http://arxiv.org/abs/0710.3339}{{\ttfamily arXiv:0710.3339 [hep-ph]}}.

\bibitem{Czarnecki:2001zc}
A.~Czarnecki and K.~Melnikov, ``{Charmonium decays: J / psi ---> e+ e- and
  eta(c) ---> gamma gamma},''
  \href{http://dx.doi.org/10.1016/S0370-2693(01)01129-7}{{\em Phys. Lett.}
  {\bfseries B519} (2001) 212--218},
\href{http://arxiv.org/abs/hep-ph/0109054}{{\ttfamily arXiv:hep-ph/0109054
  [hep-ph]}}.

\bibitem{Feng:2015uha}
F.~Feng, Y.~Jia, and W.-L. Sang, ``{Can Nonrelativistic QCD Explain the
  $\gamma\gamma^* \to \eta_c$ Transition Form Factor Data?},''
  \href{http://dx.doi.org/10.1103/PhysRevLett.115.222001}{{\em Phys. Rev.
  Lett.} {\bfseries 115} no.~22, (2015) 222001},
\href{http://arxiv.org/abs/1505.02665}{{\ttfamily arXiv:1505.02665 [hep-ph]}}.

\bibitem{Lansberg:2006sy}
J.~P. Lansberg and T.~N. Pham, ``{Two-photon width of eta(b), eta-prime(b) and
  eta-prime-prime(b) from Heavy-Quark Spin Symmetry},''
  \href{http://dx.doi.org/10.1103/PhysRevD.75.017501}{{\em Phys. Rev.}
  {\bfseries D75} (2007) 017501},
\href{http://arxiv.org/abs/hep-ph/0609268}{{\ttfamily arXiv:hep-ph/0609268
  [hep-ph]}}.

\bibitem{Qiao:2008mw}
C.-F. Qiao, J.~Wang, J.~X. Wang, and Y.~Zheng, ``{EtabFDC: An eta(b) event
  generator in hadroproduction at LHC},''
  \href{http://dx.doi.org/10.1016/j.cpc.2008.08.003}{{\em Comput. Phys.
  Commun.} {\bfseries 180} (2009) 61--68},
\href{http://arxiv.org/abs/0804.1288}{{\ttfamily arXiv:0804.1288 [hep-ph]}}.

\bibitem{Brodsky:2012vg}
S.~J. Brodsky, F.~Fleuret, C.~Hadjidakis, and J.~P. Lansberg, ``{Physics
  Opportunities of a Fixed-Target Experiment using the LHC Beams},''
  \href{http://dx.doi.org/10.1016/j.physrep.2012.10.001}{{\em Phys. Rept.}
  {\bfseries 522} (2013) 239--255},
\href{http://arxiv.org/abs/1202.6585}{{\ttfamily arXiv:1202.6585 [hep-ph]}}.

\bibitem{Lansberg:2012kf}
J.~P. Lansberg, S.~J. Brodsky, F.~Fleuret, and C.~Hadjidakis, ``{Quarkonium
  Physics at a Fixed-Target Experiment using the LHC Beams},''
  \href{http://dx.doi.org/10.1007/s00601-012-0445-8}{{\em Few Body Syst.}
  {\bfseries 53} (2012) 11--25},
\href{http://arxiv.org/abs/1204.5793}{{\ttfamily arXiv:1204.5793 [hep-ph]}}.

\bibitem{Massacrier:2015qba}
L.~Massacrier, B.~Trzeciak, F.~Fleuret, C.~Hadjidakis, D.~Kikola, J.~P.
  Lansberg, and H.~S. Shao, ``{Feasibility studies for quarkonium production at
  a fixed-target experiment using the LHC proton and lead beams (AFTER@LHC)},''
  \href{http://dx.doi.org/10.1155/2015/986348}{{\em Adv. High Energy Phys.}
  {\bfseries 2015} (2015) 986348},
\href{http://arxiv.org/abs/1504.05145}{{\ttfamily arXiv:1504.05145 [hep-ex]}}.

\bibitem{Hadjidakis:2018ifr}
C.~Hadjidakis {\em et~al.}, ``{A Fixed-Target Programme at the LHC: Physics
  Case and Projected Performances for Heavy-Ion, Hadron, Spin and Astroparticle
  Studies},''
\href{http://arxiv.org/abs/1807.00603}{{\ttfamily arXiv:1807.00603 [hep-ex]}}.

\bibitem{arbuzov2020physics}
A.~Arbuzov {\em et~al.}, ``On the physics potential to study the gluon content
  of proton and deuteron at {NICA SPD},''
  \href{http://arxiv.org/abs/2011.15005}{{\ttfamily arXiv:2011.15005
  [hep-ex]}}.

\bibitem{Zenaiev:2015rfa}
{\bfseries PROSA} Collaboration, O.~Zenaiev {\em et~al.}, ``{Impact of
  heavy-flavour production cross sections measured by the LHCb experiment on
  parton distribution functions at low x},''
  \href{http://dx.doi.org/10.1140/epjc/s10052-015-3618-z}{{\em Eur. Phys. J.}
  {\bfseries C75} no.~8, (2015) 396},
\href{http://arxiv.org/abs/1503.04581}{{\ttfamily arXiv:1503.04581 [hep-ph]}}.

\bibitem{Zenaiev:2019ktw}
{\bfseries PROSA} Collaboration, O.~Zenaiev, M.~V. Garzelli, K.~Lipka, S.~O.
  Moch, A.~Cooper-Sarkar, F.~Olness, A.~Geiser, and G.~Sigl, ``{Improved
  constraints on parton distributions using LHCb, ALICE and HERA heavy-flavour
  measurements and implications for the predictions for prompt
  atmospheric-neutrino fluxes},''
  \href{http://dx.doi.org/10.1007/JHEP04(2020)118}{{\em JHEP} {\bfseries 04}
  (2020) 118},
\href{http://arxiv.org/abs/1911.13164}{{\ttfamily arXiv:1911.13164 [hep-ph]}}.

\bibitem{Gauld:2016kpd}
R.~Gauld and J.~Rojo, ``{Precision determination of the small-$x$ gluon from
  charm production at LHCb},''
  \href{http://dx.doi.org/10.1103/PhysRevLett.118.072001}{{\em Phys. Rev.
  Lett.} {\bfseries 118} no.~7, (2017) 072001},
\href{http://arxiv.org/abs/1610.09373}{{\ttfamily arXiv:1610.09373 [hep-ph]}}.

\bibitem{Kusina:2017gkz}
A.~Kusina, J.-P. Lansberg, I.~Schienbein, and H.-S. Shao, ``{Gluon Shadowing in
  Heavy-Flavor Production at the LHC},''
  \href{http://dx.doi.org/10.1103/PhysRevLett.121.052004}{{\em Phys. Rev.
  Lett.} {\bfseries 121} no.~5, (2018) 052004},
\href{http://arxiv.org/abs/1712.07024}{{\ttfamily arXiv:1712.07024 [hep-ph]}}.

\bibitem{Lansberg:2016deg}
J.-P. Lansberg and H.-S. Shao, ``{Towards an automated tool to evaluate the
  impact of the nuclear modification of the gluon density on quarkonium, D and
  B meson production in proton-nucleus collisions},''
  \href{http://dx.doi.org/10.1140/epjc/s10052-016-4575-x}{{\em Eur. Phys. J.}
  {\bfseries C77} no.~1, (2017) 1},
\href{http://arxiv.org/abs/1610.05382}{{\ttfamily arXiv:1610.05382 [hep-ph]}}.

\end{thebibliography}\endgroup

\appendix
\onecolumn 
\section{Splitting Functions}
\label{sec:split}

In this Appendix we define the splitting functions used in the text,
\begin{equation}
    P_{gg}\left(z\right)=2C_A \left(\frac{1-z}{z}+\frac{z}{\left(1-z\right)}_{+}+z\left(1-z\right)\right)+b_0\delta\left(1-z\right),
\end{equation}
\begin{equation}
    P_{gq}\left(z\right)=C_F\frac{1+(1-z)^2}{z},
\end{equation}
where $C_A=N_c$ and $C_F=\frac{N_c^2-1}{2N_c}$ are the Casimirs of the adjoint and the fundamental representation. In order to apply the splitting functions we need to define the plus distribution $\frac{1}{\left(1-z\right)}_{+}$ that regulates the pole at $z=1$. This distribution can be applied to any arbitrary function $f\left(z\right)$ that is finite at $z=1$,
\begin{equation}
\begin{split}
    \int_{0}^1 dz\, \frac{1}{\left(1-z\right)}_{+}\, f\left(z\right)=\int_{0}^1 dz\, \frac{f\left(z\right)-f\left(1\right)}{1-z}.
\end{split}
\end{equation}
In cross-section computations, one usually integrates $z$ from a non-zero value that is bounded by for example the center of mass energy $\tau_0=\frac{M_Q^2}{s}$. The plus distributions are however defined for the integral over the entire region from $[0,1]$. In order to deal with a modified bound $[\tau_0,1]$ however, we need to make a modification.

Defining the following modified plus distribution as,
\begin{equation}
    \int_{\tau_0}^1 dz\, \frac{1}{\left(1-z\right)}_{\tau_0}\,f\left(z\right)=\int_{\tau_0}^1 dz\, \frac{f\left(z\right)-f\left(1\right)}{1-z},
\end{equation}
we can thus make the replacement,
\begin{equation}
     \frac{1}{\left(1-z\right)}_{+}= \frac{1}{\left(1-z\right)}_{\tau_0} + \log{\left(1-\tau_0\right)}\,\delta\left(1-z\right).
\end{equation}

\section{Analytical Results for $\sigma$}
\label{sec:sigma}

For completeness and for the sake of the discussion we present here the result for $\sigma$ as well. To obtain these, one can for instance take the expressions for the partonic cross sections from Refs~\cite{Petrelli:1997ge,Kuhn:1992qw,Schuler:1994hy} and fold these with the PDFs.  Recalling that $z=\tau_0/\tau$ and $\tau_0=4m_Q^2/s$, one gets
\begin{equation}\label{eq:sig-NLO-gg}
\begin{split}
\sigma_{gg}&=\frac{\alpha_s^2 \pi^2 |R_0|^2}{96 m_c^5}\left[\frac{\partial\mathcal{L}_{gg}}{\partial\tau}(\tau_0)\,\left(\tau_0\,+\frac{\alpha_s}{\pi}\frac{\tau_0}{12} \left(-44+7 \pi^2+12 b_0 \log{\left(\frac{\mu_R^2}{\mu_F^2}\right)}+72 \log^2{\left(1-\tau_0\right)} -72 \log{\left(1-\tau_0\right)}\log{\left(\frac{\mu_F^2}{4 m_c^2}\right)}\right)\right)\right.
\\
&+\frac{\alpha_s}{\pi}\frac{1}{2}\left[\int_{\tau_0}^{1}d\tau\,\frac{\partial\mathcal{L}_{gg}}{\partial\tau}(\tau)\,\left(24\log{\left(1-z\right)}\left(\left(1-z\right) z^2-2 z\right) +\left(\frac{1}{1-z}\right)\left(\frac{12\log{z}}{(1-z)(1+z)^3}\left(1-z^2\left(5+z \left(2+z+3z^3+2z^4\right)\right)\right)\right.\right.\right.
\\
&\left.\left.-\frac{1}{(1+z)^2} \left(12+z^2 \left(23+z \left(24+2 z+11 z^3\right)\right)+12 \left(1+z^3\right)^2 \log{z}\right)\right)\right)+24\int_{\tau_0}^{1}d\tau\,\left[\left(\frac{\log{\left(1-z\right)}}{1-z}\right) \, \left(\frac{\partial\mathcal{L}_{gg}}{\partial\tau}(\tau)-z^2 \frac{\partial\mathcal{L}_{gg}}{\partial\tau}(\tau_0)\right)\right]
\\
&\left.\left.-12\int_{\tau_0}^{1}d\tau\, \left[\log{\left(\frac{\mu_F^2}{4 m_c^2}\right)}\left(\frac{1}{1-z}\right)\,\left(\left(1-z+z^2\right)^2\,\frac{\partial\mathcal{L}_{gg}}{\partial\tau}(\tau)-z^2\, \frac{\partial\mathcal{L}_{gg}}{\partial\tau}(\tau_0)\right)\right]\right]\right],
\end{split}
\end{equation}
\begin{equation}\label{eq:sig-NLO-qq}
\sigma_{q\overbar{q}}=\int_{\tau_0}^{1}d\tau\,\frac{\partial\mathcal{L}_{q\bar{q}}}{\partial\tau}(\tau)\,\frac{\alpha_s^3 \pi |R_0|^2}{81 m_c^5}z^2\left(1-z\right),
\end{equation}
\begin{equation}\label{eq:sig-NLO-gq}
\begin{split}
\sigma_{qg}=&\int_{\tau_0}^{1}d\tau\,\frac{\partial\mathcal{L}_{qg}}{\partial\tau}(\tau)\,\left[\frac{\alpha_s^3 \pi |R_0|^2}{72 m_c^5}\left(\frac{1}{2} z^2+z-1+2 \left(\frac{1}{2} z^2-z+1\right) \log{\left(1-z\right)}-\left(\frac{1}{2} z^2-z+1\right)\log{\left(\frac{\mu_F^2}{4m_c^2}\right)}-\frac{1}{2} z^2\log{z}\right)\right],
\end{split}
\end{equation}
where have defined
\begin{equation}
\begin{split}
\frac{\partial\mathcal{L}_{gg}}{\partial\tau}(\tau)&=\int_{1/2\log{\tau}}^{-1/2\log{\tau}}dy\,f_{g}\left(\sqrt{\tau}e^{y},\mu_F\right)f_{g}\left(\sqrt{\tau}e^{-y},\mu_F\right),
\\
\frac{\partial\mathcal{L}_{q\bar{q}}}{\partial\tau}(\tau)&=\sum_{q=u,d,s}\int_{1/2\log{\tau}}^{-1/2\log{\tau}}dy\,\left(f_{q}\left(\sqrt{\tau}e^{y},\mu_F\right)f_{\overbar{q}}\left(\sqrt{\tau}e^{-y},\mu_F\right)+f_{\overbar{q}}\left(\sqrt{\tau}e^{y},\mu_F\right)f_{q}\left(\sqrt{\tau}e^{-y},\mu_F\right)\right),
\\
\frac{\partial\mathcal{L}_{qg}}{\partial\tau}(\tau)&=\sum_{q=u,d,s,\overbar{u},\overbar{d},\overbar{s}} \int_{1/2\log{\tau}}^{-1/2\log{\tau}}dy\,\left(f_{q}\left(\sqrt{\tau}e^{y},\mu_F\right)f_{g}\left(\sqrt{\tau}e^{-y},\mu_F\right)+f_{g}\left(\sqrt{\tau}e^{y},\mu_F\right)f_{q}\left(\sqrt{\tau}e^{-y},\mu_F\right)\right).
\end{split}  
\end{equation}

\section{Analytical Results for $d\sigma/dy$}
\label{sec:dsigdy}

In this appendix,  we provide the analytical expressions in terms of convolution of PDFs for the rapidity-differential cros section for $\eta_c$ hadron-production. These are not available in the literature.
As discussed above, 3 channels should be considered. The formulae below hold for $y\geq 0$ in order to perform the integration-boundary decomposition. For symmetric collisions,
$d\sigma/dy$ is just symmetric. For asymmetric  hadron $A$ - hadron $B$  collisions,  one can obtain $d\sigma/dy$ for $y<0$, by assigning the PDF depending on $x_1$ to hadron $B$ and conversely. 
%
%
%
%
%
We start with the $gg$-channel:
\begin{equation}\label{eq:dsidy-NLO-gg}
\begin{split}
&\frac{d\sigma_{NLO, gg}}{dy}=\frac{\alpha_s^2 \pi^2 R_0^2}{96 m_c^5}
\Bigg[\tilde{\mathcal{L}}_{gg}(\tau_0,y)\,\,\tau_0\,\left(1+\frac{\alpha_s}{12 \pi}\left(-44+7 \pi^2+12 b_0 L_{RF}
+36 \Big\{\log (1-\eta_1) \Big(L_{MF}+\log{(1-\eta_1})\Big) + \eta_1 \leftrightarrow \eta_2\Big\}\right)\right)
\\&+\frac{3 \alpha_s}{\pi}\Bigg(
\int_{\eta_1}^1 d\tau \int_{t_1}^{t_2}\!\!\!\!dw\, \tilde{\mathcal{L}}_{gg}(\tau,y_3)\,\frac{2a_1}{1-w^2}\,  
+\int_{\eta_2}^{\eta_1} \!\!\!\!d\tau \int_{t_1}^{1}\!\!\!\!dw\,\Big[\frac{\tilde{\mathcal{L}}_{gg}(\tau,y_3)\, a_1-\tilde{\mathcal{L}}_{gg}(\tau,y_4)\, a_2}{1-w}
+
\frac{\tilde{\mathcal{L}}_{gg}(\tau,y_3)\, a_1}{1+w} \Big]
\\&
+\Bigg\{\int_{\tau_0}^{\eta_1}\!\!\!\! d\tau\Big[\frac{\tilde{\mathcal{L}}_{gg}(\tau,y_1)\, c_1\; L_{MxF}-z^2
\tilde{\mathcal{L}}_{gg}(\tau_0,y)L_{MF}}{1-z}
+2\frac{\log{\left(1-z\right)}}{1-z}\left(\tilde{\mathcal{L}}_{gg}(\tau,y_1)\, c_1-z^2 \tilde{\mathcal{L}}_{gg}(\tau_0,y)\right)\Big]
+ (\eta_1,y_1) \leftrightarrow (\eta_2,y_2)\Bigg\}
\\&
+\int_{\eta_2}^{\eta_1} \!\!\!\!d\tau\; a_2\; \tilde{\mathcal{L}}_{gg}(\tau,y_4)\log{\Big(\frac{1-t_1}{2}\Big)}
+\int_{\tau_0}^{\eta_2} \!\!\!\!d\tau \int_{-1}^{1}\!\!\!\!dw
\Big[\frac{\tilde{\mathcal{L}}_{gg}(\tau,y_3)\, a_1-\tilde{\mathcal{L}}_{gg}(\tau,y_4)\, a_2}{1-w} 
+
\frac{\tilde{\mathcal{L}}_{gg}(\tau,y_3)\, a_1-\tilde{\mathcal{L}}_{gg}(\tau,y_5)\, a_2}{1+w} \Big]\Bigg)\Bigg]
\end{split}
\end{equation}
\noindent where $z=\tau_0/\tau$ and $\tau_0=4m_c^2/s$. The definition for
$\tilde{\mathcal{L}}_{gg}\left(\tau,\tilde{y}\right)=\frac{\partial\mathcal{L}_{gg}}{\partial\tau \partial y}\left(\tau,\tilde{y}\right)=f_{g}(\sqrt{\tau}e^{\tilde{y}},\mu_F)f_{g}(\sqrt{\tau}e^{-\tilde{y}},\mu_F)$,
where $f_{g}(x_1,\mu_F)$ is the gluon PDF with the factorisation scale $\mu_F$ and $x_1$-value.
We  now turn to the $qg$-channel:
\begin{equation}\label{eq:dsidy-NLO-qg}
\begin{split}
&\frac{d\sigma_{NLO, gq+qg}}{dy}=\frac{\alpha_s^3 \pi R_0^2}{144 m_c^5}\left[\int_{\tau_0}^{\eta_1}\!\!\!\! d\tau\,\tilde{\mathcal{L}}_{gq}\left(\tau,y_1\right)\,\left(c_2 \left(L_{MxF}+2\log{\left(1-x\right)}\right)+z^2\right)\right.
+\int_{\tau_0}^{\eta_2}\!\!\!\! d\tau\,\tilde{\mathcal{L}}_{qg}\left(\tau,y_2\right)\,\left(c_2 \left(L_{MxF}+2\log{\left(1-z\right)}\right)+z^2\right)
\\
&+\int_{\eta_1}^1\!\!\!\! d\tau \int_{t_1}^{t_2}\!\!\!\!dw\frac{2\left(\tilde{\mathcal{L}}_{gq}\left(\tau,y_3\right)\, a_4+\tilde{\mathcal{L}}_{qg}\left(\tau,y_3\right)\, a_5\right)}{1-w^2}
+\int_{\eta_2}^{\eta_1}\!\!\!\! d\tau \; c_2 \; \tilde{\mathcal{L}}_{gq}\left(\tau,y_4\right)\log{\left(\frac{1-t_1}{2}\right)}
\\
&+\int_{\eta_2}^{\eta_1}\!\!\!\! d\tau \int_{t_1}^{1}\!\!\!\!dw\left[\frac{\tilde{\mathcal{L}}_{gq}\left(\tau,y_3\right)\, a_4-\tilde{\mathcal{L}}_{gq}\left(\tau,y_4\right)\, c_2+\tilde{\mathcal{L}}_{qg}\left(\tau,y_3\right)\, a_5}{1-w}
+
\frac{\left(\tilde{\mathcal{L}}_{gq}\left(\tau,y_3\right)\, a_4+\tilde{\mathcal{L}}_{qg}\left(\tau,y_3\right)\, a_5\right)}{1+w}\right]
\\
&+\int_{\tau_0}^{\eta_2}\!\!\!\! d\tau \int_{-1}^{1}\!\!\!\!dw\left[\frac{\tilde{\mathcal{L}}_{gq}\left(\tau,y_3\right)\, a_4-\tilde{\mathcal{L}}_{gq}\left(\tau,y_4\right)\, c_2+\tilde{\mathcal{L}}_{qg}\left(\tau,y_3\right)\, a_5}{1-w}
\left.+
\frac{\tilde{\mathcal{L}}_{gq}\left(\tau,y_3\right)\, a_4+\tilde{\mathcal{L}}_{qg}\left(\tau,y_3\right)\, a_5-\tilde{\mathcal{L}}_{qg}\left(\tau,y_5\right)\, c_2}{1+w}\right]\right]
\end{split}
\end{equation}
where
$\tilde{\mathcal{L}}_{gq}\left(\tau,\tilde{y}\right)=\frac{\partial\mathcal{L}_{gq}}{\partial\tau \partial y}\left(\tau,\tilde{y}\right)=\sum_{q=u,d,s,\overbar{u},\overbar{d},\overbar{s}}f_{g}(\sqrt{\tau}e^{\tilde{y}},\mu_F)f_{q}(\sqrt{\tau}e^{-\tilde{y}},\mu_F),$
$\tilde{\mathcal{L}}_{qg}\left(\tau,\tilde{y}\right)=\frac{\partial\mathcal{L}_{qg}}{\partial\tau \partial y}\left(\tau,\tilde{y}\right)=\sum_{q=u,d,s,\overbar{u},\overbar{d},\overbar{s}}f_{q}(\sqrt{\tau}e^{\tilde{y}},\mu_F)f_{g}(\sqrt{\tau}e^{-\tilde{y}},\mu_F)$,
with $f_{q}(x_1,\mu_F)$ being quark PDF and $f_{g}(x_1,\mu_F)$ the gluon PDF with the factorisation scale $\mu_F$ and $x_1$-value.
Finally, we have for $q\overbar{q}$-channel:
\begin{equation}\label{eq:dsidy-NLO-qq}
\begin{split}
\frac{d\sigma_{NLO, q\overbar{q}}}{dy}&=\frac{\alpha_s^3 \pi R_0^2}{216 m_c^5}\left(\int_{\eta_1}^1 d\tau \int_{t_1}^{t_2}dw\, \, \tilde{\mathcal{L}}_{q\overbar{q}}\left(\tau,y_3\right)\, a_3 \right.
+\int_{\eta_2}^{\eta_1} d\tau \int_{t_1}^{1}dw \, \, \tilde{\mathcal{L}}_{q\overbar{q}}\left(\tau,y_3\right)\, a_3
\left.+\int_{\tau_0}^{\eta_2} d\tau \int_{-1}^{1}dw\, \, \tilde{\mathcal{L}}_{q\overbar{q}}\left(\tau,y_3\right)\, a_3\right)
\end{split}
\end{equation}
where
$\tilde{\mathcal{L}}_{q\overbar{q}}\left(\tau,\tilde{y}\right)=\frac{\partial\mathcal{L}_{q\overbar{q}}}{\partial\tau \partial y}\left(\tau,\tilde{y}\right)=\sum_{q=u,d,s,\overbar{u},\overbar{d},\overbar{s}}f_{q}(\sqrt{\tau}e^{\tilde{y}},\mu_F)f_{\overbar{q}}(\sqrt{\tau}e^{-\tilde{y}},\mu_F)$,
with $f_{q}(x_1,\mu_F)$ being the quark PDF at the factorisation scale $\mu_F$ and $x_1$ value.
In the above expressions, we have adopted the following definitions:
\begin{equation}
\begin{split}
b_0&=C_A \frac{11}{6}-n_l T_F \frac{2}{3}; \;\;
L_{RF}=\log{\left(\frac{\mu_R^2}{\mu_F^2}\right)}; \;\;
L_{MF}=\log{\left(\frac{4 m_c^2}{\mu_F^2}\right)}; \;\;
L_{MxF}=\log{\left(\frac{4 m_c^2}{\mu_F^2 z}\right)}; \;\;
\eta_{1,2}=\sqrt{\tau_0}e^{\pm y}; \;\;
\\
y_{1,2}&=y\pm\frac{1}{2}\log{z}; \;\;
t_{1,2}=\left(\frac{1+z}{1-z}\right)\tanh{\left(y\pm\frac{1}{2}\log{\tau}\right)}; \;\;
y_3=y-\textrm{arctanh}\left(\left(\frac{1-z}{1+z}\right)w\right); \;\;
y_{4,5}=y\mp\textrm{arctanh}\left(\frac{1-z}{1+z}\right); \;\;
\\
a_1&=\left(\frac{z^2 \left(-z w^2+z+w^2+3\right)^2 \left(9 z^4-4 z^3+6 z^2+(z-1)^4 w^4+6 (z-1)^4 w^2-4 z+9\right)}{16 (1-z) \left((z+1)^2-(z-1)^2 w^2\right)^2}\right)
\\
a_2&=\frac{((z-1) z+1)^2}{1-z}; \;\;
a_3=\left(\left(1-z\right) z^2 \left(1+w^2\right)\right); \;\;
c_1=\left(z^2-z+1\right)^2; \;\;
c_2=\left(\left(z-2\right) z+2\right); \;\;
\end{split}
\end{equation}
\begin{equation}
\begin{split}
a_4&=\frac{z^2 (w+1) \left(z^2 (w+1)^2-2 z (w+1)^2+w (w+2)+5\right)}{2 ((z-1) w+z+1)^2}; \;\;
a_5=\frac{z^2 (1-w) \left(z^2 (w-1)^2-2 z (w-1)^2+(w-2) w+5\right)}{2 (z (-w)+z+w+1)^2}
\end{split}
\end{equation}

\end{document}